\documentclass[%
 reprint,
superscriptaddress,
 amsmath,amssymb,
 aps,
]{revtex4-2}

\usepackage{graphicx}
\usepackage[caption=false]{subfig}
\usepackage{dcolumn}
\usepackage{bm}
\usepackage{hyperref}
\hypersetup{
  colorlinks   = true, 
  urlcolor     = blue, 
  linkcolor    = blue, 
  citecolor   = blue 
}

\usepackage{siunitx}
\newcommand{\mmicro}{\si\micro} 
\usepackage{float}
\usepackage{longtable}
\newcolumntype{P}[1]{>{\centering\arraybackslash}p{#1}}
\newcolumntype{L}[1]{>{\raggedright\arraybackslash}p{#1}}
\usepackage{multirow}
\usepackage{boldline}

\usepackage{etoolbox}
\makeatletter
\appto\abstract{%
  \let\latexlist\list
  \def\list{\edef\keeprightskip{\the\rightskip}\latexlist}%
  \patchcmd\latexlist{\ignorespaces}{\rightskip\keeprightskip\ignorespaces}{}{}%
}
\makeatother

\begin{document}


\title{Investigating High-Energy Proton-Induced Reactions on Spherical Nuclei: Implications for the Pre-Equilibrium Exciton Model}

\author{Morgan B. Fox}
\email{Email Address: morganbfox@berkeley.edu}
\affiliation{Department of Nuclear Engineering, University of California, Berkeley, Berkeley, California 94720, USA}
\author{Andrew S. Voyles}
\email{asvoyles@lbl.gov}
\affiliation{Department of Nuclear Engineering, University of California, Berkeley, Berkeley, California 94720, USA}
\affiliation{Lawrence Berkeley National Laboratory, Berkeley, California 94720, USA}
\author{Jonathan T. Morrell}
\affiliation{Department of Nuclear Engineering, University of California, Berkeley, Berkeley, California 94720, USA}
\author{Lee A. Bernstein}
\affiliation{Department of Nuclear Engineering, University of California, Berkeley, Berkeley, California 94720, USA}
\affiliation{Lawrence Berkeley National Laboratory, Berkeley, California 94720, USA}
\author{Amanda M. Lewis}
\affiliation{Department of Nuclear Engineering, University of California, Berkeley, Berkeley, California 94720, USA}
\author{Arjan J. Koning}
\affiliation{International Atomic Energy Agency, P.O. Box 100, A-1400 Vienna, Austria}
\author{Jon C. Batchelder}
\affiliation{Department of Nuclear Engineering, University of California, Berkeley, Berkeley, California 94720, USA}
\author{Eva R. Birnbaum}
\affiliation{Los Alamos National Laboratory, Los Alamos, New Mexico 87544, USA}
\author{Cathy S. Cutler}
\affiliation{Brookhaven National Laboratory, Upton, New York 11973, USA}
\author{Dmitri G. Medvedev}
\affiliation{Brookhaven National Laboratory, Upton, New York 11973, USA}
\author{Francois M. Nortier}
\affiliation{Los Alamos National Laboratory, Los Alamos, New Mexico 87544, USA}
\author{Ellen M. O'Brien}
\affiliation{Los Alamos National Laboratory, Los Alamos, New Mexico 87544, USA}
\author{Christiaan Vermeulen}
\affiliation{Los Alamos National Laboratory, Los Alamos, New Mexico 87544, USA}

\date{\today}

\begin{abstract}
\begin{description}
\item[Background] A number of accelerator-based isotope production facilities utilize 100- to 200-MeV proton beams due to the high production rates enabled by high-intensity beam capabilities and the greater diversity of isotope production brought on by the long range of high-energy protons. However, nuclear reaction modeling at these energies can be challenging because of the interplay between different reaction modes and a lack of existing guiding cross section data.
\item[Purpose] A Tri-lab collaboration has been formed among the Lawrence Berkeley, Los Alamos, and Brookhaven National Laboratories to address these complexities by characterizing charged-particle nuclear reactions relevant to the production of established and novel radioisotopes.
\item[Method]  In the inaugural collaboration experiments, stacked-targets of niobium foils were irradiated at the Brookhaven Linac Isotope Producer ($E_p=200$\,MeV) and the Los Alamos Isotope Production Facility (${E_p=100}$\,MeV) to measure $^{93}$Nb(p,x) cross sections between 50 and 200\,MeV. First measurements of the $^{93}$Nb(p,4n)$^{90}$Mo beam monitor reaction beyond 100\,MeV are reported in this work, as part of the broadest energy-spanning dataset for the reaction to date. $^{93}$Nb(p,x) production cross sections are additionally reported for 22 other measured residual products. The measured cross section results were compared with literature data as well as the default calculations of the nuclear model codes TALYS, CoH, EMPIRE, and ALICE.
\item[Results] The default code predictions largely failed to reproduce the measurements, with consistent underestimation of the pre-equilibrium emission. Therefore, we developed a standardized procedure that determines the reaction model parameters that best reproduce the most prominent reaction channels in a physically justifiable manner. The primary focus of the procedure was to determine the best parameterization for the pre-equilibrium two-component exciton model via a comparison to the energy-dependent $^{93}$Nb(p,x) data, as well as previously published $^{139}$La(p,x) cross sections.
\item[Conclusions] This modeling study revealed a trend toward a relative decrease for internal transition rates at intermediate proton energies ($E_p=20-60$\,MeV) in the current exciton model as compared to the default values. The results of this work are instrumental for the planning, execution, and analysis essential to isotope production.
\end{description}
\end{abstract}

\maketitle


\section{\label{Introduction}Introduction}
The continued rise of nuclear medicine to study physiological processes, diagnose, and treat diseases requires improved production routes for existing radionuclides, as well as new production pathways for entirely novel radioisotopes \cite{NucMedGeneral}. The implementation of these new methodologies or products in nuclear medicine relies on accurate and precise nuclear reaction cross section data in order to properly inform and optimize large scale creation for clinical use \cite{Voyles2018:Nb,Becker2020:protonsLa,Tarkanyi2019:MedicalIsotopesDataNeeds,Khandaker2009:ProtonsTi,Morrell2020,Bernstein2019:AnnualReview}. A primary component in obtaining these data is a suitable reaction monitor, defined as a long-lived radionuclide with a well-known cross section as a function of incident beam energy, that can accurately describe beam properties during a production irradiation \cite{Voyles2018:Nb,VoylesThesis,Khandaker2009:ProtonsTi,Graves2016:StackTarget,Marus2015}.

In the case of high-energy proton-induced reactions, which are important production routes at national accelerator facilities on account of the high beam intensities and large projectile range in targets \cite{Tarkanyi2019:MedicalIsotopesDataNeeds,Bernstein2019:AnnualReview,Khandaker2009:ProtonsTi}, the $^{93}$Nb(p,4n)$^{90}$Mo reaction is emerging as a valuable new monitor candidate as evidenced by \textcite{Voyles2018:Nb}.

In this work, proton-induced reaction cross sections for $^{93}$Nb were measured for energies 50--200 MeV using the stacked-target activation technique. The results include the first cross section measurements for $^{93}$Nb(p,4n)$^{90}$Mo beyond 100 MeV within the most comprehensive dataset for the reaction to date, spanning over the broadest energy range. 

In addition to the (p,4n) channel, production cross sections were extracted for 22 additional reaction products. This extensive body of data forms a valuable tool to study nuclear reaction modeling codes and assess the predictive capabilities for proton reactions on spherical nuclei up to 200 MeV \cite{Morrell2020,Koning2004:GlobalPEMexcitonOMP,Singh2006:ProtonAlphaModelExcitationFunc,Blann1975:PreqOverview,Babu2012:ProtonTalysParams,Mills1992:ProtonsCu}, which have been studied less than neutron-induced reactions \cite{Alhassan2019:ProtonsBestFitChiSquared}. It was demonstrated that default modeling predictions from TALYS, CoH, EMPIRE, and ALICE codes failed to reproduce the measured niobium data and required modifications to improve \cite{TalysManual,Kawano2019:CoHOverview,EmpireOverview,Blann1982}. In this manuscript, we set forth a systematic algorithm to determine the set of reaction model input parameters, in a scientifically justifiable manner, that best reproduce the most prominent reaction channels. The algorithm is built in the TALYS modeling framework and sets a premier focus on determining the best parameterization of the two-component exciton model in order to gain insight into high-energy pre-equilibrium reaction dynamics \cite{Koning2004:GlobalPEMexcitonOMP,KoningRochman2012:MethodTALYSEval,TalysManual}. The algorithm was then further applied to existing high-energy $^{139}$La(p,x) data. Taken together, this work suggests that the default internal transition rates of the exciton model must be modified as a function of exciton number and total system energy when considering residual product data from high-energy proton-induced reactions.

The fitting methodology proposed in this work aims to improve an accepted approach in cross section measurement literature where too few observables are used to guide modeling parameter adjustments, thereby potentially subjecting the modeling to compensating errors.

The results of this work should benefit the experimental and theoretical calculations central to isotope production planning and execution, as well as help inform the physical basis of the exciton model.

\section{\label{2}Experimental Methods and Materials}
The charged-particle irradiations in this work were performed as part of a Tri-lab collaboration between Lawrence Berkeley National Laboratory (LBNL), Los Alamos National Laboratory (LANL), and Brookhaven National Laboratory (BNL). The associated experimental facilities were the 88-Inch Cyclotron at LBNL for proton energies of $E_p<$ 55 MeV, the Isotope Production Facility (IPF) at LANL for 50 $<E_p<$ 100 MeV, and the Brookhaven Linac Isotope Producer (BLIP) at BNL for 100 $<E_p<$ 200 MeV. 

\subsection{\label{2.}Stacked-Target Design}
The stacked-target activation technique was employed in this work, where three separate target stacks were constructed and irradiated, each at a different accelerator facility. The stacked-target approach requires a layered ensemble of thin foils such that induced activation on these foils by a well-characterized incident charged-particle beam allows for the measurement of multiple energy-separated cross section values per reaction channel. Monitor foils are included among the thin foil targets in order to properly assess the beam intensity and energy reduction throughout the depth of the stack. Degraders are additionally interleaved throughout the stack to reduce and selectively control the primary beam energy incident upon each target foil \cite{Voyles2018:Nb,Morrell2020,Graves2016:StackTarget}.

\subsubsection{\label{2.1}LBNL Stack and Irradiation}
The initial primary motivation for these Tri-lab stacked-target experiments was to determine residual nuclide production cross sections for $^{75}$As(p,x) from threshold to 200 MeV, with a specific focus on the production of $^{68}$Ge and $^{72}$Se for PET imaging. However, the $^{76}$Se compound system is non-spherical, which could necessitate the use of coupled-channel calculations in the reaction modeling. Deformed systems may also require the use of a modified Hauser-Feshbach code that extends angular momentum and level density considerations to include nuclei spin projections on the symmetry axis. This modification is presented in \textcite{Grimes2013:ModHF} and suggests an increased accuracy for deformed nuclei calculations versus the assumption of spherical symmetry inherent to the standard Hauser-Feshbach formalism. Yet these deformation aspects lie beyond the scope of this current paper and in turn, the results from the $^{75}$As(p,x) measurements will be presented in a separate publication. 

Consequently, the LBNL stack in this campaign focused only on arsenic targets and did not contain niobium foils. The experimental setup and procedure at this site will therefore not be discussed in this work.

\subsubsection{\label{2.2}LANL Stack and Irradiation}
The IPF stack utilized 25 $\mmicro$m $^{\mathrm{nat}}$Cu foils (99.999\%, LOT: U02F019, Part: 10950, Alfa Aesar, Tewksbury, MA 01876,
USA), 25 $\mmicro$m $^{\mathrm{nat}}$Al foils (99.999\%, LOT: Q26F026, Part: 44233, Alfa Aesar), 25 $\mmicro$m $^{\mathrm{nat}}$Nb foils (99.8\%, LOT: T23A035, Alfa Aesar), and thin metallic $^{75}$As layers electroplated onto 25 $\mmicro$m $^{\mathrm{nat}}$Ti foil backings (99.6\%, TI000205/TI000290, Goodfellow Metals). $^{\mathrm{nat}}$Nb is 100\% $^{\mathrm{93}}$Nb isotopic abundance.

Ten copper, niobium, and aluminum foils each were cut into 2.5 cm $\times$ 2.5 cm squares and their physical dimensions were characterized by taking four length and width measurements using a digital caliper (Mitutoyo America Corp.) and four thickness measurements taken at different locations using a digital micrometer (Mitutoyo America Corp.). Multiple mass measurements at 0.1\,mg precision were taken after cleaning the foils with isopropyl alcohol. Ten titanium foils were cut to the same approximate sizes but the same dimensioning and weighting techniques could not be used due to the chemical and mechanical constraints of the collaboration-developed electroplating process. Instead, the nominal manufacturer thickness and density were accepted for the titanium, with confidence and uncertainties gathered from separate physical measurements of extra titanium foils not used in the stack. The creation and characterization of the accompanying 2.25\,cm diameter arsenic depositions used in this stack will be described in detail in a future publication dedicated to the arsenic irradiation products. This characterization involved dimensional measurements, electron transmission, and reactor-based neutron activation analysis.

The electroplated arsenic targets, as well as the niobium foils, were sealed using LINQTAPE PIT0.5S-UT Series Kapton polyimide film tape composed of 12 $\mmicro$m of silicone adhesive on 13 $\mmicro$m of polyimide backing (total nominal 3.18 mg/cm$^2$). The copper and aluminum foils were not encapsulated in any tape.

\begin{figure}[t]
{\includegraphics[width=1.0\columnwidth]{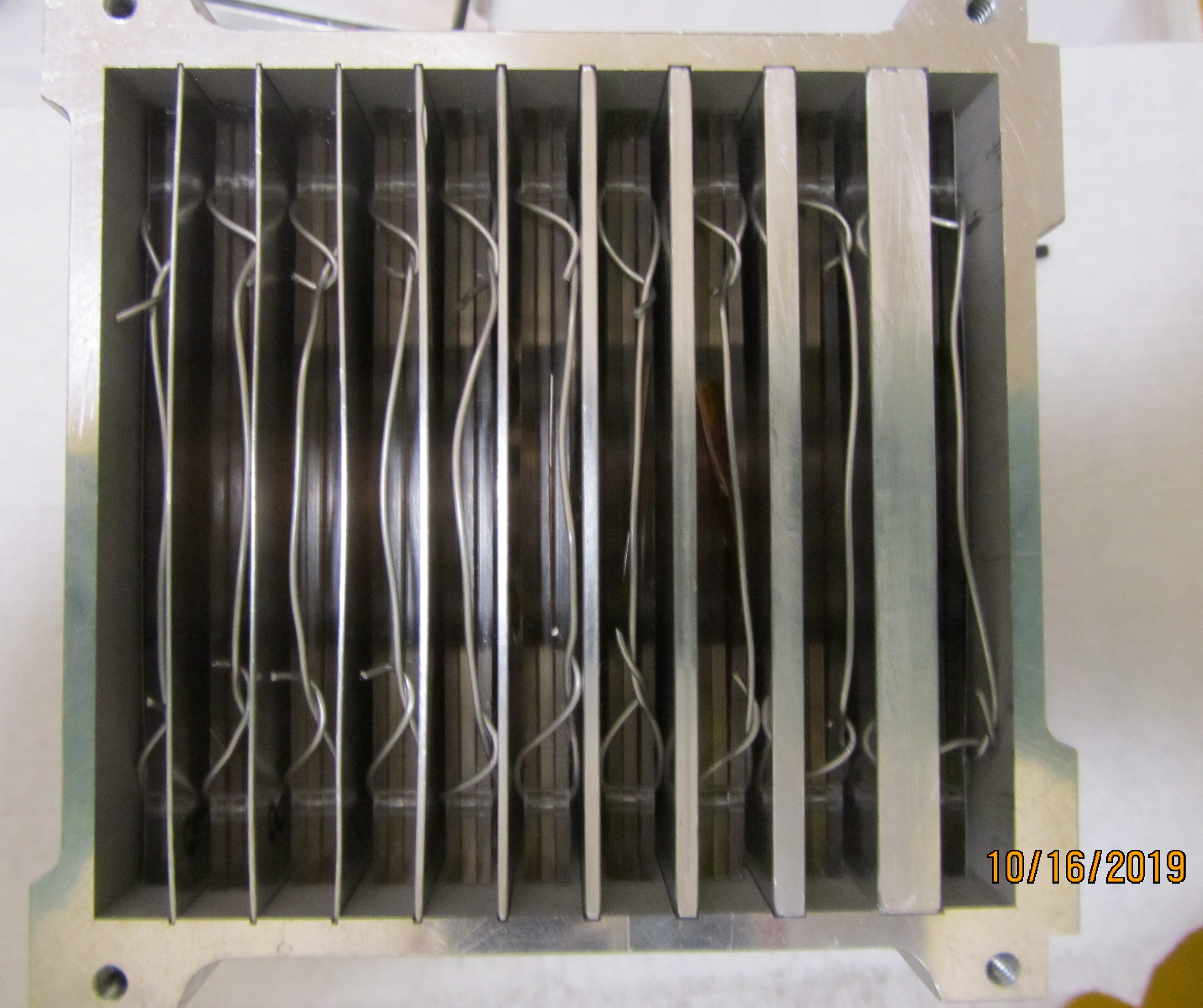}}
\vspace{-0.55cm}
\caption{A top view of the assembled LANL target stack showing the ten target ``compartments" separated by aluminum degraders. The beam enters through a 0.411\,mm aluminum entrance window on the right hand side of the target box.}
\vspace{-0.25cm}
\label{LANL_TargetBox}
\end{figure}

The electroplated arsenic foils were attached to ten acrylic frames (1.5\,mm in thickness), which protected the foils during handling and centered them in the bombardment position after the stack was fully arranged. The ten copper foils were treated in an identical manner. The aluminum and niobium foils were paired up and mounted on the front and back of the same frames due to physical space limitations of the machined 6061-T6 aluminum IPF target box. Nine aluminum 1100 series degraders were characterized in the same manner as the Cu, Nb, and Al foils and included in the stack to yield ten different beam energy ``compartments'' for cross section measurements. In each compartment, one $^{93}$Nb+$^{\mathrm{nat}}$Al target, one $^{75}$As+$^{\mathrm{nat}}$Ti target, and one $^{\mathrm{nat}}$Cu target were placed and bundled together using baling wire. The baling wire, attached at the top of the frames and not obstructing any target material, was necessary to aid the removal of the foils from the target box following irradiation using the hot cell's tele-manipulators. The assembled stack in the IPF target box can be seen in Figure \ref{LANL_TargetBox}, where it is also noted that the box has a 0.411\,mm aluminum beam entrance window and is specially designed to be watertight since the IPF target station is located underwater. Additionally, stainless steel plates (approximately 100\,mg/cm$^2$) were placed in the front and back of the stack. Post-irradiation dose mapping of the activated stainless plates using radiochromic film (Gafchromic EBT3) was used to determine the spatial profile of the beam entering and exiting the stack \cite{Voyles2018:Nb,Morrell2020}.

The upstream beamline components at IPF have a significant effect on beam energy that must be taken into account \cite{Ellen2020:IPFUpstream}. Two materials exist upstream of the target box entrance window: the beam window separating beamline vacuum from the target chamber and a single cooling water channel defined by the distance between the beam window and the aluminum target box window during operation. The installed beam window is 0.381\,mm thick Inconel alloy 718 and it is pre-curved toward the vacuum side of the beamline by 1.3\,mm. However, under the hydrostatic and vacuum loading pressures experienced during operation, the beam window further elastically deforms toward the vacuum side. During operation at low beam currents, typical of this work, the beam window elastically deforms toward the vacuum side by approximately 0.12\,mm. Given the geometry of the target box, this information implies that the proton beam travels through a cooling water channel 7.414\,mm thick \cite{Ellen2020:IPFUpstream}. The combined upstream effects total an approximate effective degrader areal density of 1165\,mg/cm$^2$.

The full detailed target stack ordering and properties for the LANL irradiation are given in Table \ref{LANLStack} in Appendix \ref{StackTables}. The stack was irradiated for 7203 seconds with an H$^+$ beam of 100 nA nominal current. The beam current, measured using an inductive pickup, remained stable under these conditions for the duration of the irradiation. The mean beam energy extracted was 100.16\,MeV at a 0.1\% uncertainty.

\subsubsection{\label{2.3}BNL Stack and Irradiation}
The target stack for the BNL irradiation was composed of 25 $\mmicro$m $^{\mathrm{nat}}$Cu foils (99.95\%, CU000420,  Goodfellow Metals, Coraopolis, PA 15108-9302, USA), 25 $\mmicro$m $^{\mathrm{nat}}$Nb foils (99.8\%, LOT: T23A035, Alfa Aesar), and thin metallic $^{75}$As layers electroplated onto 25 $\mmicro$m $^{\mathrm{nat}}$Ti foil backings (99.6\%, TI000205/TI000290, Goodfellow Metals). The arsenic targets were again produced by members of this collaboration and characterized similarly to the arsenic targets created for the LANL experiment. The copper, niobium, and titanium foils for BNL were prepared according to the process outlined for the same foils in Section \ref{2.2}.

Seven targets of each material were prepared for this irradiation and six copper degraders were in turn characterized to create seven energy compartments within the stack.

The electroplated arsenic targets were sealed using the same LINQTAPE PIT0.5S-UT Series Kapton polyimide film tape described in Section \ref{2.2}. The copper and niobium foils were encapsulated with DuPont Kapton polyimide film tape of 43.2\,$\mmicro$m of silicone adhesive on 50.8\,$\mmicro$m of polyimide backing (total nominal 11.89\,mg/cm$^2$). The foils were mounted to plastic frames, with copper and niobium foils paired due to space limitations of the BLIP target box. Similar to the LANL irradiation, baling wire was used to secure one $^{\mathrm{nat}}$Cu+$^{93}$Nb target and one $^{75}$As+$^{\mathrm{nat}}$Ti target together in each energy compartment of the stack between degraders. The BNL target box, also specially designed to be watertight since the BLIP target station is located underwater, has a 0.381\,mm aluminum beam entrance window. A single stainless steel plate could only be included at the beginning of the stack in this experiment to assess the physical beam profile post-irradiation due to space constraints.

BLIP facility upstream beamline components that influence beam properties were also included into the stack considerations. Beryllium and AlBeMet windows exist to facilitate the beamline vacuum connections; two stainless steel windows and two water cooling channels are also in place \cite{Cutler2018:BLIPMedicalIsotopes}. Together, these components give an approximately 1820 mg/cm$^2$ system that the proton beam must traverse before reaching the target box's aluminum window. Unlike IPF, possible deformation of the BLIP upstream windows under hydrostatic and vacuum loading conditions are not measured and may introduce unknown uncertainties to the stack characterization. Though the effect of these uncertainties is expected to be small due to the lower stopping power at a higher beam energy, corrections for potential changes to these upstream conditions are considered in the stack transport calculations in Section \ref{Current_Calc}.

The BNL target stack (Table \ref{BNLStack} in Appendix \ref{StackTables}) was irradiated for 3609 seconds with an H$^+$ beam of 200 nA nominal current. The beam current during operation was recorded using toroidal beam transformers and remained stable under these conditions for the duration of the irradiation. The mean beam energy extracted was 200\,MeV at a 0.2\% uncertainty \cite{Degraffenreid2019:BNLAs}.

\subsection{\label{3.}Gamma Spectroscopy and Measurement of Foil Activities}
The collaborative nature of this work prompted the use of different types of germanium detectors and data acquisition systems to measure the induced activities of target foils.

\subsubsection{\label{3.1}LANL}
The LANL counting took place at two locations. One ORTEC IDM-200-VTM High-Purity Germanium (HPGe) detector and one ORTEC GEM p-type coaxial HPGe detector (model GEM20P-PLUS) were used to capture short- and intermediate-lived activation species directly at the IPF site of target irradiation. The IDM is a mechanically-cooled coaxial p-type HPGe with a single, large-area 85\,mm diameter $\times$ 30\,mm length crystal and built-in spectroscopy electronics. The energy and absolute photopeak efficiency of the detectors were calibrated using standard $^{152}$Eu, $^{207}$Bi, and $^{241}$Am sources as well as a mixed gamma source containing $^{57}$Co, $^{60}$Co, $^{109}$Cd, and $^{137}$Cs. The efficiency model used in this work is taken from the physical model presented by \textcite{Gallagher1974:EffModel}. The LANL countroom was further commissioned to perform longer counts over a multi-week period, which was not possible at IPF. The countroom uses p-type ORTEC GEM series HPGes with aluminum windows.

Following the irradiation, the IPF target box was removed from the beamline and raised into the IPF hot cell. Tele-manipulators were used to disassemble the stack and extract the foils. The radiochromic film showed that an $\approx$1\,cm diameter proton beam was fully inscribed within the samples throughout the stack. All target frames were wrapped in one layer of Magic Cover Clear Vinyl Self-Adhesive to fix any surface contamination. Due to elevated dose rates, only the arsenic, titanium, and copper targets were made available for counting on the day of irradiation. Initial data were acquired from 10--20 minute counts of the targets starting approximately 2 hours after the end-of-bombardment (EoB) at distances of 15\,cm and 17\,cm from the GEM detector face and 55\,cm and 60\,cm from the IDM face. One day post-irradiation, within 19 hours of EoB, the aluminum and niobium targets were accessible and counted multiple times along with the other targets throughout the day at positions of 15, 17, 25, 55, and 60\,cm from the detector faces. Once appropriate statistics had been acquired to either establish necessary decay curves for induced products or characterize monitor reaction channels, all targets were packaged and shipped to the LANL countroom.

In the dedicated counting lab, the 40 available targets were first repeatedly cycled in front of detectors at 10--15\,cm capturing 1 hour counts over the course of a week. The countroom curators varied the foil distance from the detector face on a regular basis to optimize count rate and dead time. The calibration data for each detector used, at each counting position, were collected each day and made available with the foil data. Over the following 6 weeks, cycling of the target foils in front of the detectors continued and count times were increased to 6--8 hours to capture the longest-lived activation products.

\subsubsection{\label{3.2}BNL}
The BNL gamma spectroscopy setup incorporated two EURISYS MESURES 2 Fold Segmented ``Clover" detectors in addition to two GEM25P4-70 ORTEC GEM coaxial p-type HPGe detectors and an ORTEC GAMMA-X n-type Coaxial HPGe detector (model GMX-13180). All detector efficiencies were calculated using a combination of $^{54}$Mn, $^{60}$Co, $^{109}$Cd, $^{137}$Cs, $^{133}$Ba, $^{152}$Eu, and $^{241}$Am calibrated point sources, with the \textcite{Gallagher1974:EffModel} physical model. One GEM detector was situated in the BLIP facility at the irradiation site while the remaining detectors were in a counting lab in a neighbouring building. 

Within 2 hours of EoB at BLIP, the copper foils and electroplated arsenic targets were removed from the hot cell and counted for over 10 minutes each using the GEM detector in the facility. The observed beam spot size on targets was $\approx$1\,cm in diameter.  Once the niobium foils had been pulled from the BLIP hot cell, all targets were transported to the nearby counting lab. There, the copper and arsenic foils were cycled first through 10--30 minute counts, followed by hour long counts, on the Clovers and GEM at 10--15\,cm from the detector faces. The niobium foils were assigned a similar counting scheme starting approximately 20 hours after EoB. Cycling and counting of the foils continued for an additional 24 hours.

Within two weeks of EoB, all targets were shipped back to LBNL. The subsequent gamma-spectroscopy at the 88-Inch Cyclotron utilized an ORTEC GMX series (model GMX-50220-S) HPGe, which is a nitrogen-cooled coaxial n-type HPGe with a 0.5\,mm beryllium window and a 64.9\,mm diameter $\times$ 57.8\,mm long crystal. Multi-day to week-long counts of the copper, arsenic, and niobium foils were performed with the LBNL GMX over the course of 2+ months to ensure that all observable long-lived products could be quantified.

\subsubsection{\label{Activation_Analysis}Activation Analysis}
While the specifications of counting equipment and procedure varied between irradiations, the data analysis for the measurement of induced target activities and cross sections followed a standardized approach. The procedure is well-described in \textcite{Voyles2018:Nb} and \textcite{Morrell2020} but is included here for clarity and completeness.

\begin{figure*}[t]
{\includegraphics[width=1.0\textwidth]{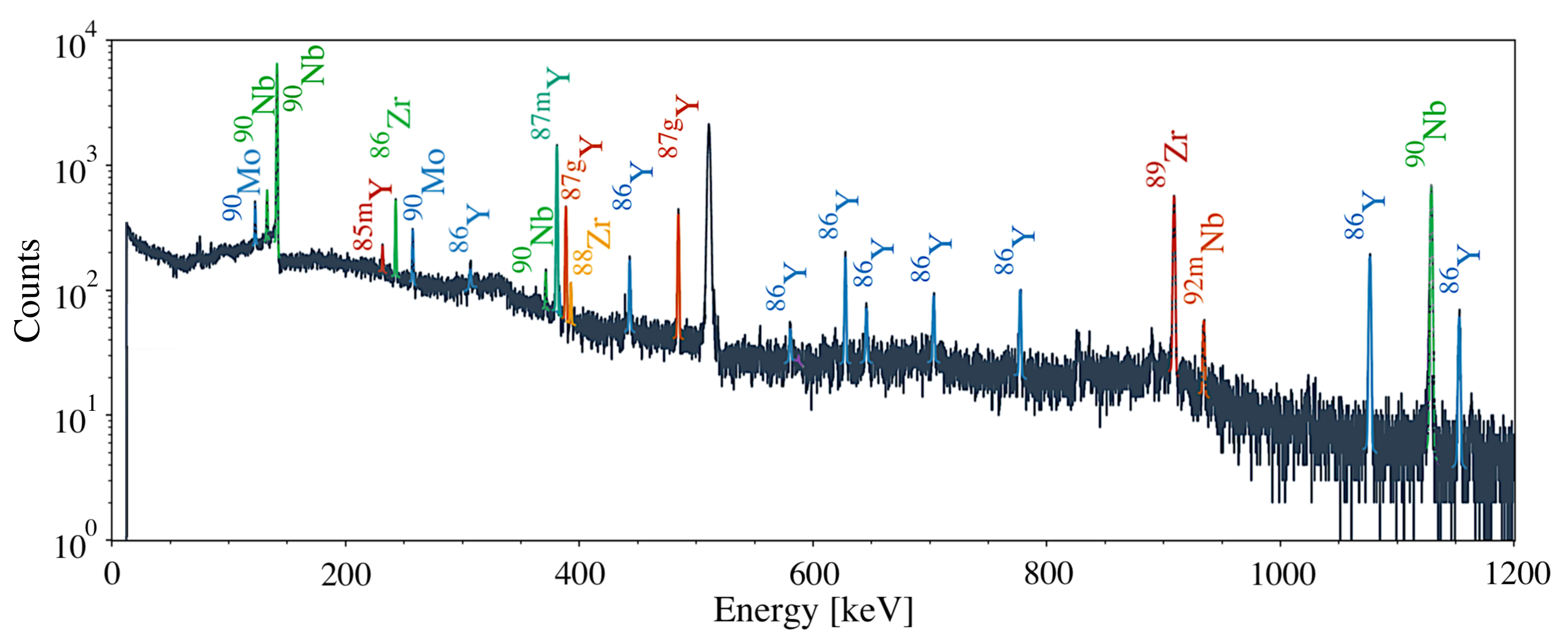}}
\vspace{-0.55cm}
\caption{Example gamma-ray spectrum from the induced activation of a niobium target in the LANL stack at approximately $E_p=91$\,MeV. The spectrum was taken approximately 20 hours after EoB, and the smooth fits to the peaks of interest shown are produced by the NPAT package \cite{Morrell2019}.}
\label{fitspectrum_ex}
\end{figure*}

The gamma emission peaks from decaying activation products were identified from the previously described gamma-ray spectra. These photopeaks were fit using the NPAT code package developed at UC Berkeley \cite{Morrell2019}. Example fits are shown in Figure \ref{fitspectrum_ex} for a spectrum collected from the LANL Nb-SN1 target of the stack in Table \ref{LANLStack} (see Appendix \ref{StackTables}).

The activity $A$ for each activation product of interest at a delay time $t_d$ since the end-of-bombardment to the start of counting was then determined from the net counts found $N_c$ after corrections for gamma intensity $I_\gamma$, detector efficiency $\epsilon$, dead time, counting time, and self-attenuation within the foils according to:
\begin{gather}\label{ApparentActivity}
A(t_d)=\frac{N_c \lambda}{(1-e^{-\lambda t_{real}})I_\gamma \epsilon}\frac{t_{real}}{t_{live}} F_{att},
\end{gather}
where $\lambda$ is the decay constant for the radionuclide of interest, $t_{real}$ and $t_{live}$ describe the real and live time for detector acquisition, respectively, and $F_{att}$ is the photon self-attenuation correction factor. $F_{att}$ is calculated using photon attenuation cross sections retrieved from the XCOM database \cite{XCOM2010} and takes the convention that all activity is assumed to be made at the midplane of the foils.

The EoB activity $A_0$ for a given radionuclide was subsequently found from a fit to the relevant Bateman equation. Moreover, the benefit of repeated foil counts in this work and the use of multiple gamma-rays is evidenced here by providing multiple radionuclide activities at numerous $t_d$, which establish a consistent decay curve. Through a regression analysis of decay curves, it is possible to extract the $A_0$ for each activation product in a more accurate manner than simply basing its calculation on a single time point and a single gamma-ray observation. 

If an activation product of interest is populated without contribution from the decay of a parent radionuclide, the EoB activity is found from a fit to the first order Bateman equation:
\begin{gather}\label{Bateman1}
A(t_d)=A_0e^{-\lambda t_d}.
\end{gather}

Typically, if it is needed to calculate EoB activities within a feeding chain in this work, the required calculation is only second order. This is the case for isomeric to ground state conversions as well as two-step beta-decay chains. In these circumstances, the decay curve is given by:
\begin{gather}\label{Bateman2}
A_2(t_d)=A_{0,1}B_r\frac{\lambda_2}{\lambda_2-\lambda_1}(e^{-\lambda_1 t_d}-e^{-\lambda_2 t_d})+A_{0,2}e^{-\lambda_2 t_d},
\end{gather}
where $A_2(t_d)$ is still found from Equation (\ref{ApparentActivity}), $B_r$ is the decay branching ratio, and the 1 and 2 subscripts denote the parent and daughter nuclides, respectively, in the two-step decay chain. This two-step fit to calculate $A_{0,2}$ uses the independently determined $A_{0,1}$ from Equation (\ref{Bateman1}) when possible, but otherwise both variables are fit together. The decay curve regressions in this work were additionally performed with the NPAT code package \cite{Morrell2019}.

\begin{figure}
{\includegraphics[width=1.0\columnwidth]{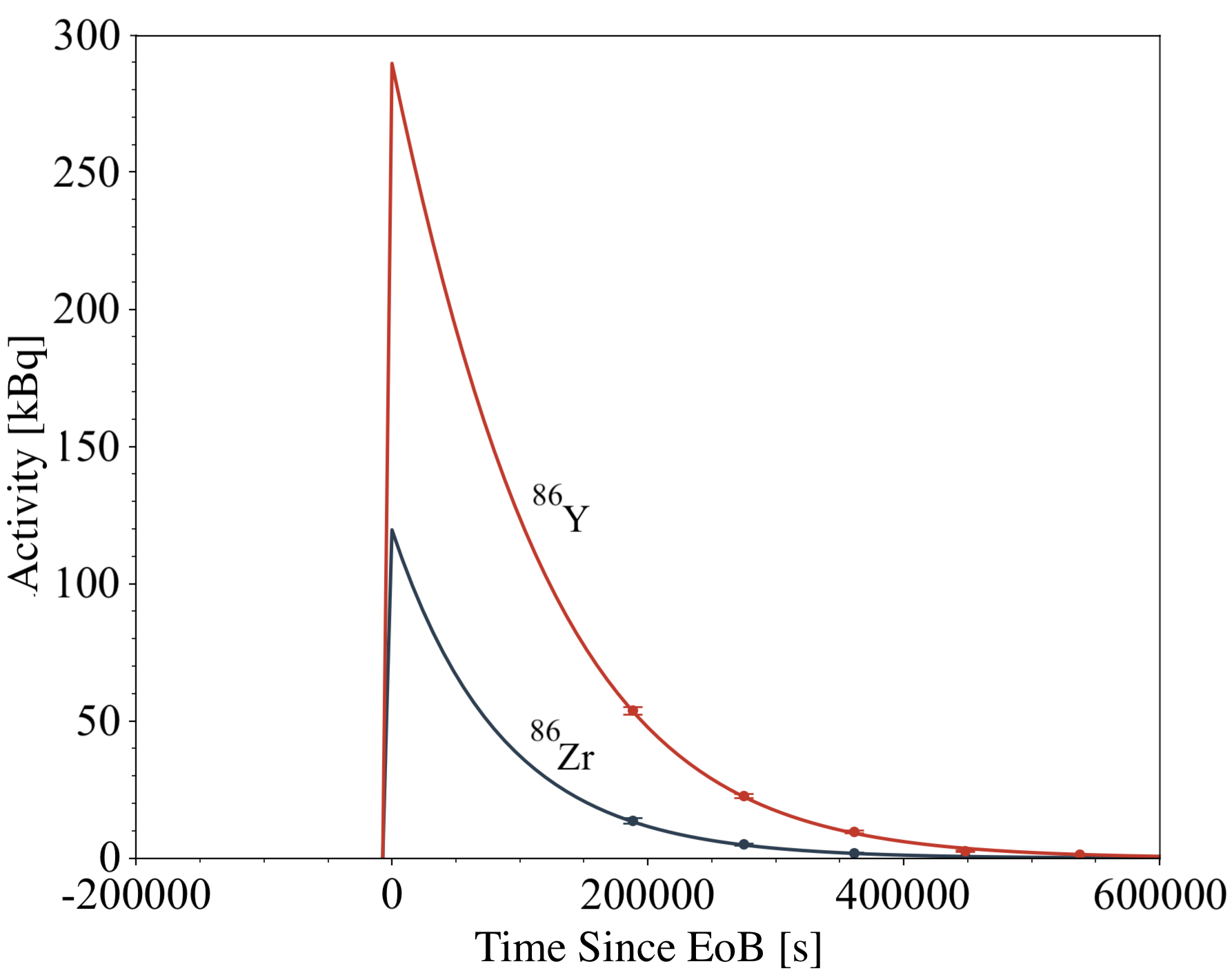}}
\vspace{-0.55cm}
\caption{Example of initial activity fitting for two-step beta-decay chain of $^{86}$Zr feeding $^{86}$Y as residual products in the niobium irradiations.}
\vspace{-0.15cm}
\label{FeedingRegressionEx}
\end{figure}

A regression example for the $^{86}$Zr$\rightarrow ^{86}$Y decay chain is shown in Figure \ref{FeedingRegressionEx}.

The total uncertainties in the determined EoB activities had contributions from uncertainties in fitted peak areas, evaluated half-lives and gamma intensities, and detector efficiency calibrations. Each contribution to the total uncertainty was assumed to be independent and was added in quadrature. The impact of calculated $A_0$ uncertainties on final cross section results is detailed in Section \ref{xscalc}.

\subsection{\label{Current_Calc}Stack Current and Energy Properties}
The methods of current monitoring during beam operation discussed in Sections \ref{2.2} and \ref{2.3} provide valuable information for the experimental conditions, their output is not sufficient to precisely describe the beam energy and intensity evolution throughout a target stack \cite{Voyles2018:Nb,Morrell2020,Graves2016:StackTarget,Marus2015}. Instead, more detailed calculations must be retrieved from monitor foil activation analysis, where known reaction cross sections can be used to measure beam current in the multiple energy positions of a stack.
\newpage

The relevant proton fluence monitor reactions used in the irradiations were:\\

\noindent LANL
\begin{itemize}
\item $^{\mathrm{nat}}$Cu(p,x)$^{56}$Co, $^{58}$Co, $^{62}$Zn
\item $^{\mathrm{nat}}$Ti(p,x)$^{48}$V
\item $^{\mathrm{nat}}$Al(p,x)$^{22}$Na
\end{itemize}
BNL
\begin{itemize}
\item $^{\mathrm{nat}}$Cu(p,x)$^{58}$Co
\end{itemize}
where only reactions with IAEA-recommended data in the relevant proton energy ranges have been considered \cite{IAEARefCrossSections}.

In the BNL irradiation, the lack of reliable data for high proton energy reactions precluded the use of most monitor channels and as a result only the $^{58}$Co activation product was taken to extract the beam current. However, $^{\mathrm{nat}}$Cu(p,x)$^{56}$Co has significant data in this high-energy region and was preliminarily used as a validation of the beam current derived from the $^{58}$Co calculations.

The $A_0$ for the monitor reaction products were calculated according to the formalism presented in Section \ref{Activation_Analysis}. Since the beam was constant throughout the irradiation period, the proton beam current $I_p$ was calculated at each monitor foil position by the relation:
\begin{gather}\label{beamcurrent}
I_p=\frac{A_0}{(\rho_N \Delta r)(1-e^{-\lambda t_{irr}})\bar{\sigma}},
\end{gather}
where $I_p$ is output in units of protons per second, $(1-e^{-\lambda t_{irr}})$ corrects for decay that occurred during the beam-on irradiation time $t_{irr}$, $\rho_N \Delta r$ is the relevant measured areal number density calculated from Tables \ref{LANLStack} and \ref{BNLStack} (see Appendix \ref{StackTables}), and $\bar{\sigma}$ is the flux-weighted production cross section.

\begin{figure}[H]
{\includegraphics[width=1.0\columnwidth]{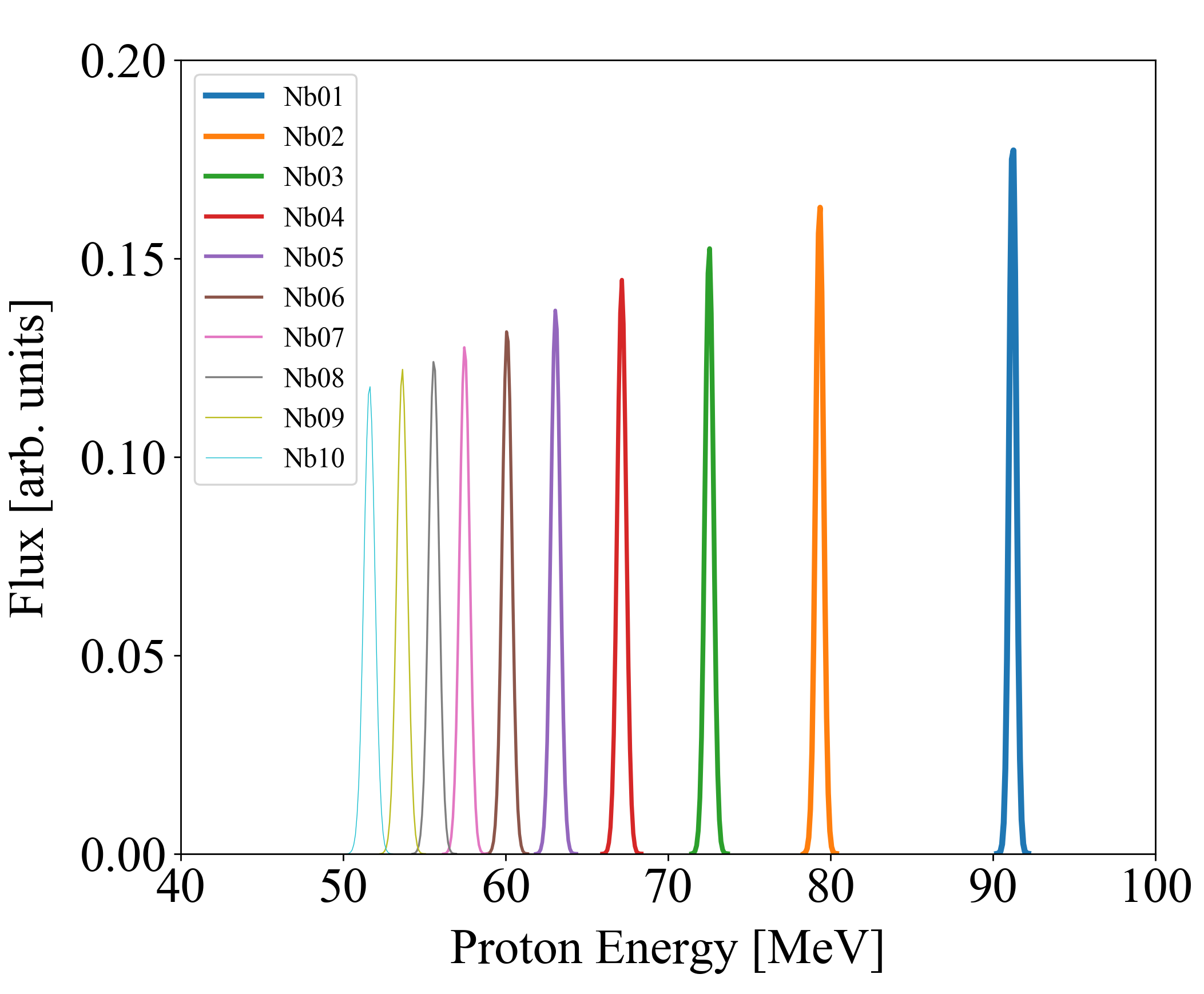}}
\vspace{-0.55cm}
\caption{Visualization of the calculated proton energy spectrum for each niobium foil in the LANL stack.}
\label{AZ_Flux_Output}
\end{figure}

The $\bar{\sigma}$ formalism is needed to account for the energy width broadening resulting from energy straggle of the beam as it is propagated toward the back of the stack \cite{Voyles2018:Nb,Morrell2020,Graves2016:StackTarget,Marus2015}. Using the IAEA-recommended cross section data $\sigma(E)$ for the relevant monitor reactions \cite{IAEARefCrossSections}, the flux-weighted cross section is calculated from:
\begin{gather}\label{sigma_bar}
\bar{\sigma}=\frac{\int \sigma(E)\phi(E)dE}{\int \phi(E)dE},
\end{gather}
where $\phi(E)$ is the proton flux energy spectrum. $\phi(E)$ was determined here using an Anderson \& Ziegler-based Monte Carlo code, as implemented in NPAT \cite{SRIM,Morrell2019}. The calculated energy spectrum resulting from the Anderson \& Ziegler calculation in the LANL irradiation is shown in Figure \ref{AZ_Flux_Output} as an example.

\vspace{-0.25cm}
\vspace{-0.3cm}

\begin{figure}[t]
{\includegraphics[width=1.0\columnwidth]{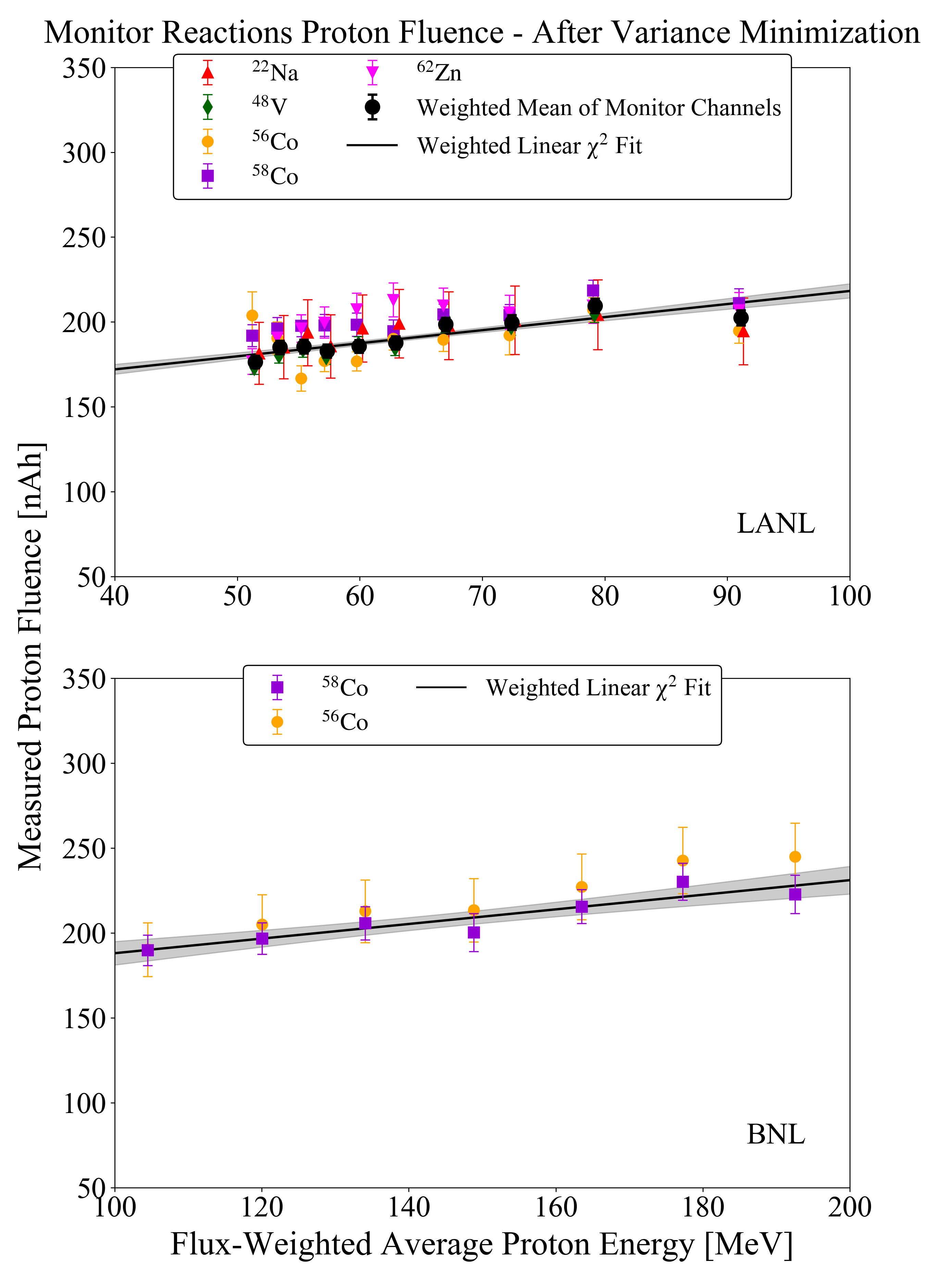}}
\vspace{-0.65cm}
\caption{Plots of the proton beam current measured by monitor reactions in the LANL and BNL stacks following adjustments made by the variance minimization technique. The $^{\mathrm{nat}}$Cu(p,x)$^{56}$Co monitor reaction is plotted for BNL but its data were not used for any of the BNL fluence calculations or the variance minimization.}
\label{3Fluences}
\end{figure}

The implementation of this monitor foil deduced current, following Equations (\ref{beamcurrent}) and (\ref{sigma_bar}), is shown for each irradiation site in Figure \ref{3Fluences}. Included in Figure \ref{3Fluences} are weighted averages of all the available monitor foils for the fluence at each stack position. The weighted averages account for data and measurement correlations between the reaction channels in each compartment. An uncertainty-weighted linear fit is also included for each site as a global model to impose a smooth and gradual fleunce depletion.

Included in the results of Figure \ref{3Fluences} is a reduction in systematic uncertainty using the ``variance minimization" technique presented in \textcite{Graves2016:StackTarget}, \textcite{Voyles2018:Nb}, and \textcite{Morrell2020}. This technique was applied, as partial disagreement between the initial proton fluence predictions from each monitor channel in each energy compartment of the stack at each experiment site was observed. The disagreement was most noticeable near the rear of each stack where contributions of poor stopping power characterization, straggling, and systematic uncertainties from upstream components became most compounded. The independent measurements of proton fluence from the monitor reactions should all theoretically be consistent at each energy position given accurate monitor reaction cross sections and foil energy assignments. The variance minimization technique is a corrective tool applied to the stopping power in simulations to address this discrepancy through the treatment of the effective density of the Al/Cu degraders in each stack as a free parameter. This is reasonable because the majority of the stopping power for the beam occurs in the thick degraders. The free parameter can then be optimized by a reduced $\chi^2$ minimization technique for the global linear fit of the monitor fluence data.

\begin{figure}[t]
{\includegraphics[width=1.0\columnwidth]{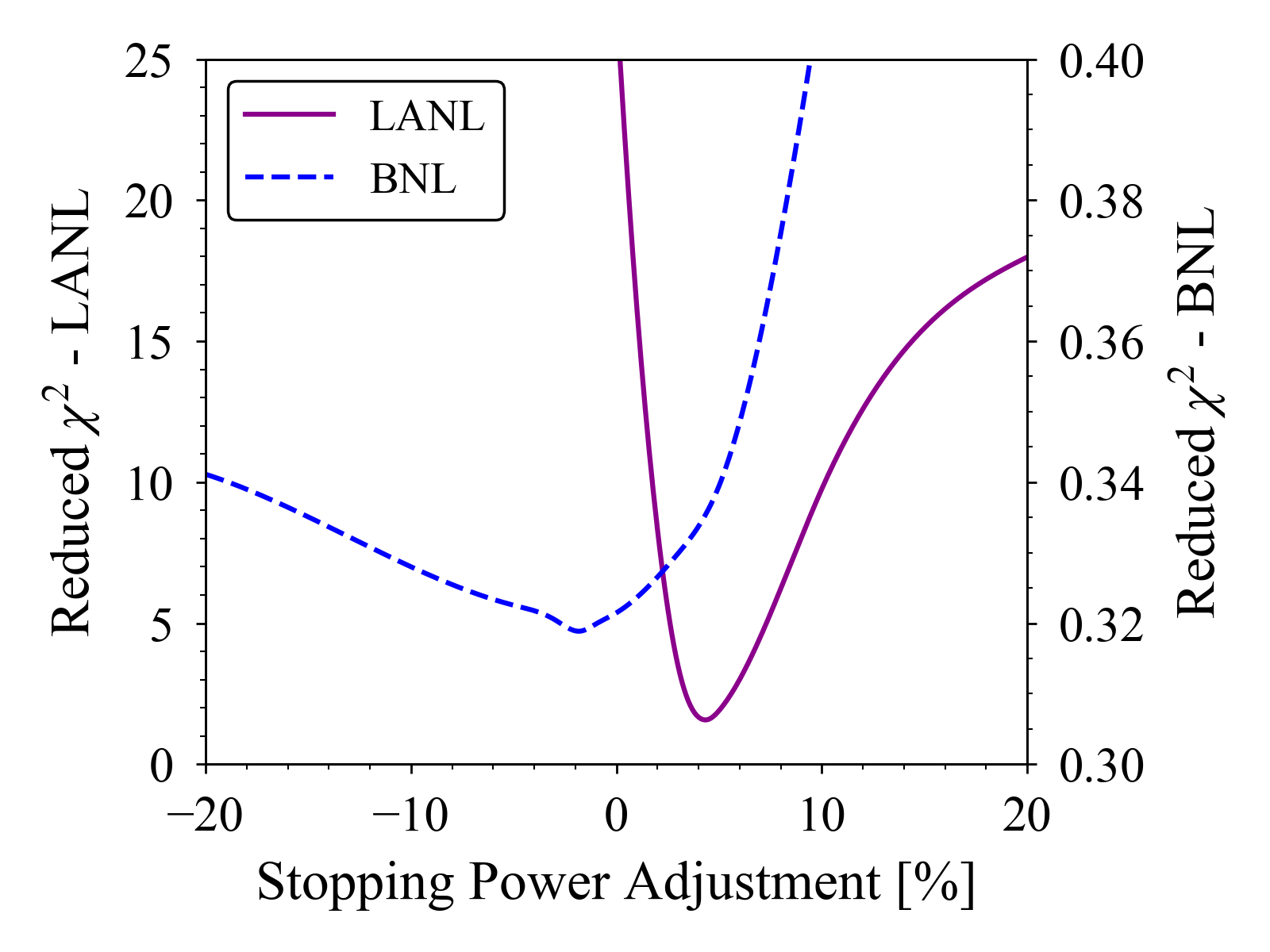}}
\vspace{-0.65cm}
\caption{Result of $\chi^2$ analysis used in the variance minimization technique to determine the required adjustment to stopping power within the proton energy spectrum calculations per stack.}
\label{VarMinChi}
\end{figure}

For both stacks, the degraders' effective densities were varied uniformly in the stopping power simulations by a factor of up to $\pm$25\% of nominal values. The resulting reduced $\chi^2$ in each case is given in Figure \ref{VarMinChi}. Figure \ref{VarMinChi} indicates that a change in degrader density, which is equivalent to a linear change in stopping power, of $+4.35\%$, and $-1.84\%$ compared to nominal measurements for the LANL and BNL stacks, respectively, minimizes the monitor foil disagreement in each case. Previous stacked-target work has always shown a modest positive enhancement to the stopping power of $+$2--5\%, which makes the BNL optimization interesting \cite{Voyles2018:Nb,Morrell2020}. It is likely that the negative adjustment in the BNL case is mostly due to compensation for the less well-known characterizations of the upstream cooling water channel and window deformation. It is also possible that some of this effect may be attributed to the use of copper degraders at BNL versus the aluminum degraders used at LANL and LBNL.

Monitor reactions that threshold in the energy region of the stack, such as $^{56}$Co near the LANL stack rear, are extremely valuable in this minimization approach as they are most sensitive to changes in stopping power and energy assignment thereby providing physical limits for the problem. The relative shallowness of the BNL $\chi^2$ curve is most likely due to the limitation of minimizing using just one monitor reaction. Note that this degrader density variation procedure is a computation tool to correct for poorly-characterized stopping power at these energies and does not mean that the actual degrader density was physically different than what was measured \cite{Morrell2020}. 

The minimized reduced $\chi^2$ also provides optimized beam energy assignments for each foil in a stack from the corrected transport simulation. The energy assignments are the flux-averaged energies using $\phi(E)$ with uncertainties per foil taken as the full-width at half maximum. These energy assignments for the niobium targets are provided in Table \ref{Nb_xsValues}.

In the BNL fluence results, the optimized global linear model provides an interpolation to each individual niobium foil with a better accuracy and uncertainty than just utilizing the sole $^{58}$Co fluence prediction in each compartment. In the LANL fluence results, the linear fit was used for the variance minimization but the correlation-weighted-average values in each compartment were directly used for calculating production cross sections. This is possible without any need for interpolation or worry of model selection influence because of the contributions from multiple available monitor reactions.

\vspace{-0.55cm}

\subsection{\label{xscalc}Cross Section Determination}
Given the activity, weighted-average beam current and energy, timing, and areal density factors previously discussed, the flux-averaged cross sections for products of interest in this work were calculated using Equation (\ref{CrossSectionCalc}):
\begin{gather}\label{CrossSectionCalc}
\sigma=\frac{A_0}{I_p(\rho_N\Delta r)(1-e^{-\lambda t_{irr}})}.
\end{gather}

The $^{93}$Nb(p,x) cross section results are given in Table \ref{Nb_xsValues}, which reports measurements for $^{93\textnormal{m}}$Mo, $^{92\textnormal{m}}$Nb, $^{91\textnormal{m}}$Nb, $^{90}$Mo, $^{90}$Nb, $^{89}$Zr, $^{88}$Zr, $^{88}$Y, $^{87\textnormal{m}}$Y, $^{87}$Y, $^{86}$Zr, $^{86}$Y,  $^{86}$Rb, $^{85\textnormal{m}}$Y, $^{84}$Rb, $^{83}$Sr, $^{83}$Rb, $^{82\textnormal{m}}$Rb, $^{81}$Rb, $^{75}$Se, $^{74}$As, $^{73}$As, and $^{72}$Se. The $^{75}$As(p,x) data in addition to the $^{\mathrm{nat}}$Cu(p,x) and $^{\mathrm{nat}}$Ti(p,x) results will be detailed in a future publication.

A distinction is made in this work between cumulative, $(c)$, and independent, $(i)$, cross section values. Numerous reaction products in these irradiations were produced both directly and from decay feeding. Where the decay of any precursors could be measured and the in-growth contribution separated, or where no decay precursors exist, independent cross sections for direct production of a nucleus are reported. Where the in-growth due to parent decay could not be deconvolved, due to timing or decay property limitations, cumulative cross sections are reported.

The final uncertainty contributions to the cross section measurements include uncertainties in evaluated half-lives (0.1--0.8\%), foil areal density measurements (0.05--0.4\%), proton current determination calculated from monitor fluence measurements and variance minimization (2--4\%), and $A_0$ quantification that accounts for efficiency uncertainty in addition to other factors listed in Section \ref{Activation_Analysis} (2--10\%). These contributions were added in quadrature to give uncertainty in the final results at the 3--6\% level on average (Table \ref{Nb_xsValues}).

\begin{table*}
\caption{Summary of cross sections measured in this work. Subscripts $(i)$ and $(c)$ indicate independent and cumulative cross sections, respectively. Uncertainties are listed in the least significant digit, that is, 119.8 (10)\,MeV means 119.8 $\pm$ 1.0\,MeV.}
\label{Nb_xsValues}
\begin{ruledtabular}
\begin{tabular}{L{1.8cm}L{1.61cm}L{1.61cm}L{1.61cm}L{1.61cm}L{1.61cm}L{1.61cm}L{1.61cm}L{1.61cm}L{1.61cm}}
\multicolumn{10}{c}{\bfseries $^{93}$Nb(p,x) Production Cross Sections [mb]}\\[+0.1cm]
\hline\\[-0.22cm]
E$\mathrm{_p}$ [MeV] & 192.38 (73) & 177.11 (77) & 163.31 (81) & 148.66 (86) & 133.87 (92) & 119.8 (10) & 104.2 (11) & 91.21 (52) & 79.32 (58) \\[+0.1cm]
\hline\\[-0.22cm]
$^{72}$Se$_{(c)}$& 0.066 (13) & 0.0193 (26) & - & - & - & - & - & - & - \\[0.1cm]

$^{73}$As$_{(c)}$& 1.15 (30) & 0.77 (18) & - & - & - & - & - & - & - \\[0.1cm]

$^{74}$As$_{(i)}$& 0.182 (12) & 0.1071 (71) & - & - & - & - & - & - & - \\[0.1cm]

$^{75}$Se$_{(c)}$& 1.443 (76) & 0.963 (25) & 0.603 (21) & 0.200 (24) & - & - & - & - & - \\[0.1cm]

$^{81}$Rb$_{(c)}$& - & - & - & - & - & - & - & 2.99 (55) & - \\[0.1cm]

$^{82\textnormal{m}}$Rb$_{(i)}$& 10.55 (36) & 9.28 (27) & 8.39 (22) & 6.86 (24) & 4.93 (18) & 3.65 (20) & 3.49 (18) & 3.07 (13) & 1.06 (15) \\[0.1cm]

$^{83}$Rb$_{(c)}$& 40.0 (22) & 36.8 (17) & 35.0 (19) & 30.9 (19) & 27.0 (19) & 15.97 (71) & 5.59 (41) & 6.27 (47) & 7.12 (53) \\[0.1cm]

$^{83}$Sr$_{(c)}$& 32.3 (20) & 29.1 (17) & 27.1 (15) & 25.0 (16) & 20.5 (13) & 13.2 (11) & 3.64 (42)  & 3.88 (61) & 5.13 (75) \\[0.1cm]

$^{84}$Rb$_{(i)}$& 3.11 (17) & 2.89 (16) & 2.64 (14) & 2.32 (13) & 2.06 (11) & 1.701 (94) & 0.699 (40) & 0.563 (37) & 0.436 (31) \\[0.1cm]

$^{85\textnormal{m}}$Y$_{(c)}$& - & - & - & - & - & - & - & 26.1 (28) & 18.8 (24) \\[0.1cm]

$^{86}$Rb$_{(i)}$& - & 0.256 (21) & - & - & - & - & - & - & - \\[0.1cm]

$^{86}$Y$_{(i)}$& 45.2 (11) & 43.88 (93) & 44.77 (84) & 44.21 (84) & 42.64 (80) & 38.67 (88) & 29.31 (78) & 33.4 (13) & 42.7 (15) \\[0.1cm]

$^{86}$Zr$_{(c)}$& 20.3 (18) & 21.5 (19) & 22.3 (19) & 22.5 (19) & 23.0 (19) & 18.4 (16)  & 9.91 (90) & 16.4 (15) & 23.5 (20) \\[0.1cm]

$^{87}$Y$_{(c)}$& 106.5 (27) & 110.3 (26) & 112.9 (24) & 115.7 (24) & 120.2 (26) & 123.7 (30) & 103.2 (30) & 106.1 (48) & 56.2 (25) \\[0.1cm]

$^{87\textnormal{m}}$Y$_{(c)}$& 86.5 (57) & 89.4 (58) & 92.5 (59) & 94.6 (61) & 98.4 (63) & 99.2 (65) & 82.4 (55) & 87.9 (41) & 47.1 (21) \\[0.1cm]

$^{88}$Y$_{(i)}$& 18.36 (52) & 18.71 (46) & 18.63 (40) & 18.39 (38) & 18.22 (39) & 17.84 (41) & 17.18 (47) & 19.07 (62) & 14.86 (48) \\[0.1cm]

$^{88}$Zr$_{(c)}$& 85.9 (48) & 91.5 (50) & 95.9 (51) & 101.1 (54) & 109.0 (58) & 117.6 (64) & 136.5 (77) & 159 (12) & 141.5 (95) \\[0.1cm]

$^{89}$Zr$_{(c)}$& 108.6 (36) & 114.4 (35) & 125.2 (43) & 136.2 (52) & 145.5 (50) & 159.5 (59) & 177.3 (63) & 196 (15) & 249 (16) \\[0.1cm]

$^{90}$Nb$_{(i)}$& 69.4 (22) & 76.2 (21) & 84.7 (21) & 90.4 (24) & 102.8 (25) & 110.5 (31) & 131.2 (39) & 155.1 (46) & 174.4 (49)\\[0.1cm]

$^{90}$Mo$_{(i)}$& 4.54 (33) & 5.01 (34) & 5.46 (32) & 6.55 (59) & 7.70 (70) & 9.64 (88) & 12.3 (11) & 17.9 (11) & 22.8 (14) \\[0.1cm]

$^{91\textnormal{m}}$Nb$_{(c)}$& 14.1 (22) & 14.7 (23) & 14.7 (23) & 17.3 (27) & 17.3 (27) & 20.5 (32) & 22.0 (34) & 25.8 (40) & 27.3 (42) \\[0.1cm]

$^{92\textnormal{m}}$Nb$_{(i)}$& 25.9 (12) & 29.5 (13) & 30.9 (13) & 32.4 (14) & 35.4 (15) & 37.8 (16) & 41.4 (19) & 45.4 (24) & 47.8 (26) \\[0.1cm]

$^{93\textnormal{m}}$Mo$_{(i)}$& - & - & - & - & - & - & - & 1.069 (71) & 0.75 (10)\\[0.1cm]

\hline\\[-0.22cm]
E$\mathrm{_p}$ [MeV] & 72.52 (62) & 67.14 (65) & 63.06 (68) & 60.08 (71) & 57.47 (73) & 55.58 (75) & 53.62 (77) & 51.61 (80)\\[+0.1cm]
\hline\\[-0.22cm]

$^{83}$Rb$_{(c)}$& 5.32 (39) & 2.31 (19) & 0.71 (11) & 0.19 (11) & - & - & - & - \\[0.1cm]

$^{83}$Sr$_{(c)}$& 4.31 (68) & 1.40 (59) & 1.04 (55) & - & - & - & - & -\\[0.1cm]

$^{84}$Rb$_{(i)}$& 0.625 (43) & 0.637 (44) & 0.533 (39) & 0.368 (31) & 0.250 (25) & 0.143 (21) & 0.078 (14) & -\\[0.1cm]

$^{85\textnormal{m}}$Y$_{(c)}$& 5.8 (13) & - & - & - & - & - & - & -\\[0.1cm]

$^{86}$Y$_{(i)}$& 43.5 (15) & 32.7 (12) & 21.8 (10) & 10.02 (61) & 4.38 (46) & - & - & -\\[0.1cm]

$^{86}$Zr$_{(c)}$& 28.0 (23) & 22.1 (18) & 12.3 (13) & 5.9 (10) & 2.50 (64) & 1.58 (72) & - & -\\[0.1cm]

$^{87}$Y$_{(c)}$& 61.5 (23) & 78.3 (26) & 101.1 (32) & 115.3 (43) & 116.2 (56) & 109.3 (41) & 97.3 (31) & 86.9 (36)\\[0.1cm]

$^{87\textnormal{m}}$Y$_{(c)}$& 50.6 (23) & 64.7 (30) & 83.6 (39) & 93.8 (43) & 96.5 (45) & 90.6 (43) & 80.3 (38) & 69.7 (36)\\[0.1cm]

$^{88}$Y$_{(i)}$& 11.82 (41) & 9.60 (35) & 9.15 (34) & 9.55 (36) & 10.93 (60) & 10.53 (40) & 11.45 (42) & 13.34 (47)\\[0.1cm]

$^{88}$Zr$_{(c)}$& 92.0 (75) & 45.2 (56) & 27.3 (41) & 24.0 (41) & 25.4 (70) & 27.6 (42) & 31.9 (42) & 41.0 (47)\\[0.1cm]

$^{89}$Zr$_{(c)}$& 309 (21) & 328 (17) & 296 (21) & 205 (15) & 171 (23) & 136 (14) & 80.3 (86) & 54.6 (77)\\[0.1cm]

$^{90}$Nb$_{(i)}$& 201.0 (58) & 225.0 (62) & 271.2 (79) & 307.2 (85) & 350.7 (97) & 369 (10) & 394 (11) & 429 (12)\\[0.1cm]

$^{90}$Mo$_{(i)}$& 28.5 (17) & 36.2 (22) & 48.9 (36) & 63.7 (37) & 83.3 (46) & 91.7 (51) & 103.3 (57) & 118.9 (63)\\[0.1cm]

$^{91\textnormal{m}}$Nb$_{(c)}$& 30.7 (47) & 31.0 (48) & 34.0 (53) & 36.3 (56) & 37.0 (62) & - & 36.9 (57) & 40.6 (63)\\[0.1cm]

$^{92\textnormal{m}}$Nb$_{(i)}$& 51.3 (28) & 51.2 (32) & 54.7 (30) & 58.3 (30) & 58.2 (31) & 56.6 (30) & 57.7 (29) & 61.7 (32)\\[0.1cm]

$^{93\textnormal{m}}$Mo$_{(i)}$& 1.19 (12) & 1.11 (14) & 1.33 (15) & 1.59 (20) & 1.45 (24) & 1.25 (19) & 1.86 (25) & 1.76 (18)\\[0.1cm]
\end{tabular}
\end{ruledtabular}
\end{table*}

\section{\label{Results}Results and Discussion}
The experimentally extracted cross sections are compared with the predictions of nuclear reaction modeling codes TALYS-1.9 \cite{TalysManual}, CoH-3.5.3 \cite{Kawano2019:CoHOverview}, EMPIRE-3.2.3 \cite{EmpireOverview}, and ALICE-20 \cite{Blann1982}, each using default settings and parameters, to initially explore variations between the codes and their sensitivity to pre-equilibrium reaction dynamics. Where measured cumulative cross sections are plotted, the corresponding code calculations shown also include the necessary parent production to estimate cumulative yields. Note, however, that ALICE-20 is not suited to calculate independent isomer or ground state production due to a lack of detailed angular momentum modeling.

Furthermore, in the code comparisons, the TALYS and ALICE codes account for potential deuteron, $^{3}$He, and triton emissions at all incident proton energies. Default EMPIRE limits these emissions and CoH ignore these effects altogether. The TALYS output provides total production cross sections for these emission channels that can be used to estimate their influence. In TALYS, the cumulative deuteron, $^{3}$He, and triton cross section is calculated as 3.1\%, 3.5\%, and 11.8\% of the combined proton and neutron production at 50\,MeV, 100\,MeV, and 200\,MeV, respectively. At each energy, the deuteron production dominates over $^{3}$He and triton emissions. Therefore, while the inclusion of these more complex emission types accounts for mostly a small effect, it is a point of difference between the code calculations.

A summary of the key default models implemented in each code is given in Table \ref{DefaultCodeModelSum}.

Comparisons with the TENDL-2019 library \cite{KoningRochman2012:MethodTALYSEval} are also made. Additionally, the cross section measurements in this work are compared to the existing body of literature data, retrieved from EXFOR \cite{Michel1997:ProtonsTiCuNb,Titarenko2011,Voyles2018:Nb,Ditroi2008,Steyn2011,Ditroi2009,Albouy1963,Korteling1964,AvilaRodriguez2008,Levkovskii1991,Kiselev1974,Rizvi2012,Lawriniang2018,Parashari2018,Singh2006:ProtonAlphaModelExcitationFunc,James1954,Forsthoff1953,Blaser1951}.

\begin{table*}
\caption{Default models implemented in reaction codes}
\label{DefaultCodeModelSum}
\begin{ruledtabular}
\begin{tabular}{L{2.3cm}L{3cm}L{2.5cm}L{4.75cm}L{3.75cm}}
Reaction Code & Proton/Neutron Optical Model & Alpha Optical Model & Level Density & Pre-Equilibrium\\[0.1cm]
\hline\\[-0.25cm]
TALYS-1.9 & Koning-Delaroche \cite{KD2003:OMP} & Avrigeanu(2014) \cite{Avrigeanu2014} & Gilbert-Cameron constant temperature and Fermi gas model \cite{TalysManual} & Two-component exciton model \cite{Koning2004:GlobalPEMexcitonOMP}\\[0.2cm]
CoH-3.5.3 & Koning-Delaroche & Avrigeanu(1994) \cite{Avrigeanu1994} & Gilbert-Cameron constant temperature and Fermi gas model & Two-component exciton model\\[0.2cm]
EMPIRE-3.2.3 & Koning-Delaroche & Avrigeanu(2009) \cite{Avrigeanu2009} & Enhanced Generalized Superfluid Model \cite{EmpireOverview} & PCROSS one-component exciton model \cite{EmpireOverview}\\[0.2cm]
ALICE-20 & Becchetti-Greenlees \cite{Blann1982,BGreenOMP} & Igo(1959) \cite{IgoOMP} & Shell-dependent Kataria-Ramamurthy model \cite{Blann1982} & Hybrid Monte-Carlo Simulation pre-compound decay \cite{Blann1982}\\[0.1cm]
\end{tabular}
\end{ruledtabular}
\end{table*}

The cross sections and code comparisons for four residual products of interest are described in detail below. The remaining cross section figures are given in Appendix \ref{Appendix_Plots} (Figures \ref{Nb_72Se}--\ref{Nb_93mMo}).

\subsection{\label{90MoExFunction}{$^{93}$Nb(p,4n)$^{90}$Mo Cross Section}}
As presented in \textcite{Voyles2018:Nb}, the $^{93}$Nb(p,4n)$^{90}$Mo reaction is compelling as a new, higher energy proton monitor reaction standard. The $^{93}$Nb(p,4n) reaction channel is independent of any (n,x) contaminant production that could be due to secondary neutrons stemming from (p,xn) reactions and requires no corrections for precursor decays. $^{90}$Mo decays with seven intense gamma lines ranging from near 100\,keV to 1300\,keV that allow for easy delineation on most detectors \cite{DataSheetsA90}. Further, the $^{90}$Mo 5.56 $\pm$ 0.09\,hr half-life is fairly flexible for a monitor reaction \cite{DataSheetsA90}, as the isotope can still be readily quantified more than one day post-irradiation, as was done in the counting for these experiments.

\begin{figure}[H]
{\includegraphics[width=1.0\columnwidth]{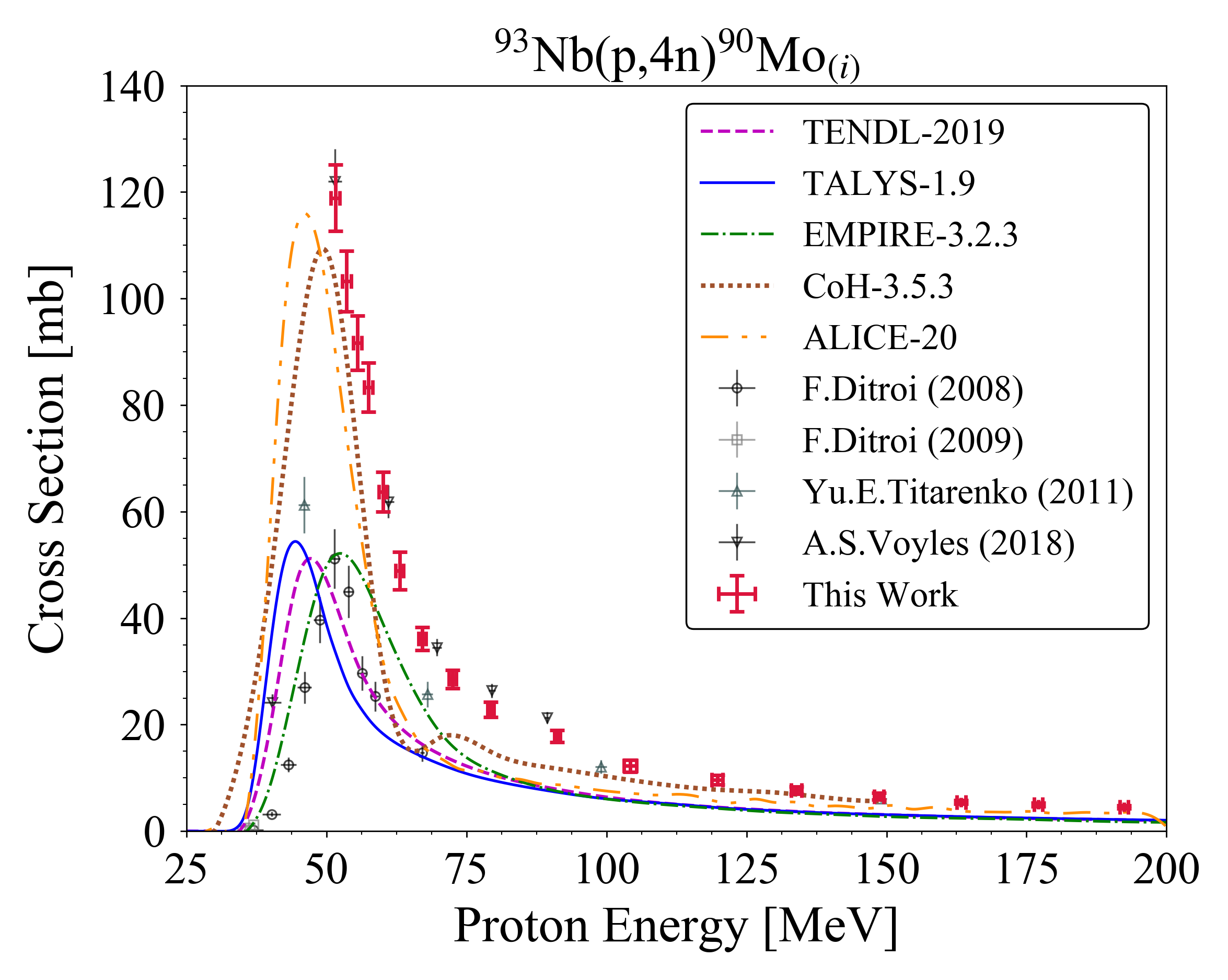}}
\vspace{-0.65cm}
\caption{Experimental and theoretical cross sections for $^{90}$Mo production, peaking near 120 mb around 50\,MeV.}
\label{Nb_90Mo}
\end{figure}

The cross section results here, shown in Figure \ref{Nb_90Mo}, align very well with the \textcite{Voyles2018:Nb} measurements in predicting a peak cross section of approximately 120 mb near 50 MeV.

\vspace{-0.15cm}
\vspace{-0.3cm}

The \textcite{Ditroi2008} data in Figure \ref{Nb_90Mo} predicts a compound peak of less than half the magnitude observed in this work and \textcite{Voyles2018:Nb}. This underprediction appears as a trend across numerous reaction products and can be seen in the remaining excitation function plots shown in Appendix \ref{Appendix_Plots}. The \textcite{Titarenko2011} dataset is also slightly inconsistent with this work, as it too implies a smaller peak, though not as small as that put forth by \textcite{Ditroi2008}.

Only CoH and ALICE reproduce the peak magnitude of the cross section, while TALYS, EMPIRE, and TENDL predict a smaller magnitude similar to \textcite{Ditroi2008}. Further, the TALYS and EMPIRE default calculations misplace the compound peak centroid relative to the other calculations. Although CoH and ALICE perform best, neither properly accounts for the increased production on the peak's high-energy falling edge due to a pre-equilibrium ``tail" contribution.

This work gives the first measurements of $^{93}$Nb(p,4n)$^{90}$Mo above 100\,MeV and is the broadest energy-spanning dataset for the reaction to date. A recent proton irradiation with niobium targets was conducted in a separate experiment at LBNL for energies from 55\,MeV to threshold in order to fully characterize the remaining low-energy side of the compound peak. These results will be discussed in a subsequent publication.

\subsection{\label{90NbExFunction}{$^{93}$Nb(p,p3n)$^{90}$Nb Cross Section}}
$^{90}$Nb is the most strongly-fed observed residual product stemming from proton reactions on niobium in this investigation, accounting for $\approx$30\% of the total non-elastic reaction value at its peak. The $^{90}$Nb cross section data in this work were measured independently through a two-step beta-decay chain fit that accounted for contributions from its $^{90}$Mo parent.

\begin{figure}[H]
{\includegraphics[width=1.0\columnwidth]{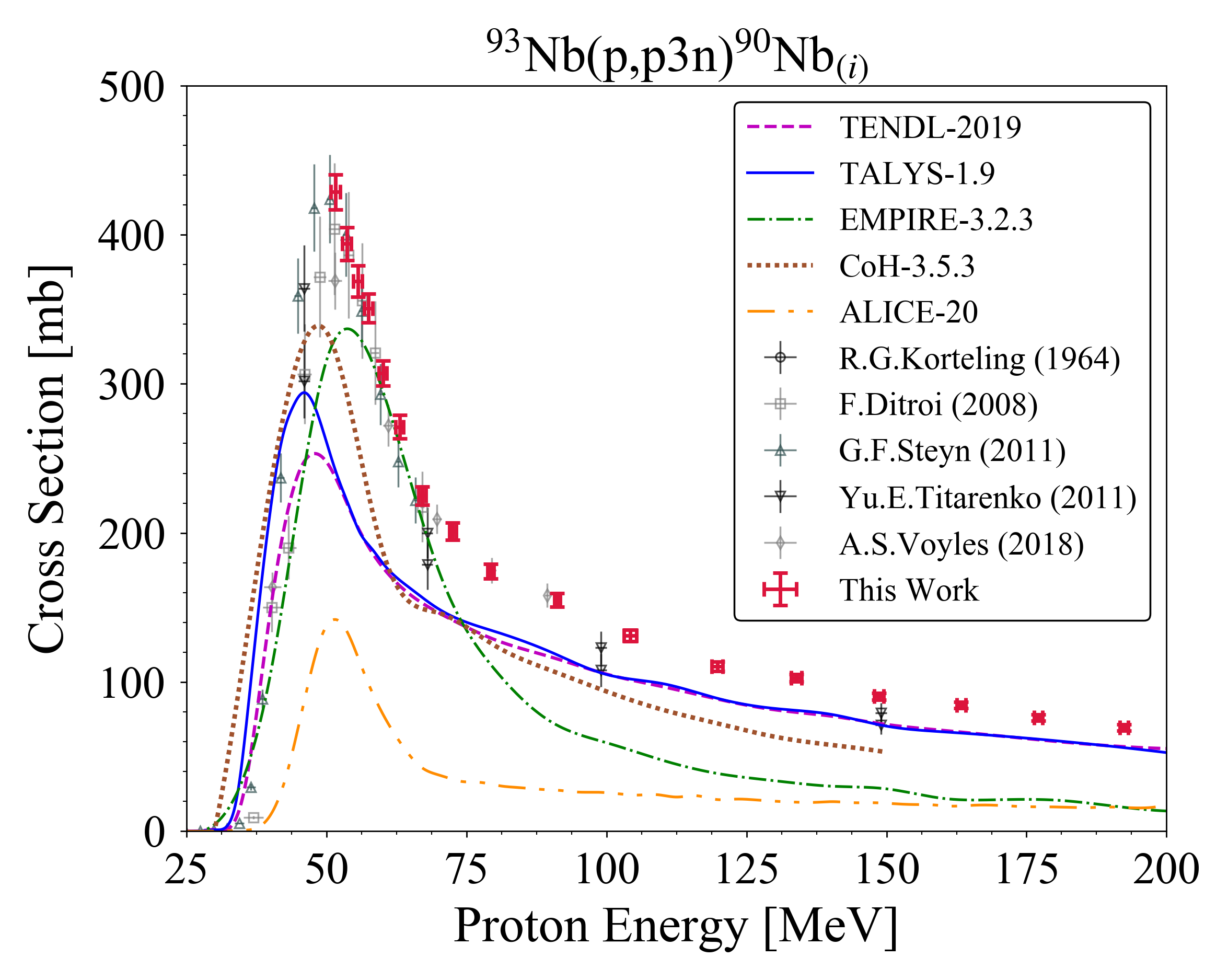}}
\vspace{-0.65cm}
\caption{Experimental and theoretical cross sections for $^{90}$Nb production, peaking near 425 mb around 50\,MeV.}
\label{Nb_90Nb}
\end{figure}

The $^{93}$Nb(p,p3n)$^{90}$Nb results of this work (Figure \ref{Nb_90Nb}) agree very well with the prior literature data and provide a well-characterized, significant extension beyond 75\,MeV.

\vspace{-0.15cm}
\vspace{-0.3cm}

No code matches the large compound peak magnitude of the experimental data. CoH and EMPIRE come the closest but suffer from their misplacement of the peak's energy by approximately 5\,MeV. The shapes of default TALYS, TENDL, and CoH show some affinity for the very pronounced high-energy pre-equilibirum tail in $^{90}$Nb production whereas default ALICE and EMPIRE lack in this regard. The misprediction from ALICE here is in stark contrast to its close prediction of the neighbouring (p,4n) reaction.

It is particularly concerning for the global predictive power of $^{93}$Nb(p,x) modeling that no code adequately reproduces this dominant reaction channel. Moreover, the proton emitted in the (p,p3n) channel is likely to result from pre-equilibrium emission at higher energies due to its suppression from the Coulomb barrier. The poor default predictions of this channel thereby suggest a systematic issue in the pre-equilibrium modeling of these codes.

\subsection{\label{89ZrExFunction}{$^{93}$Nb(p,x)$^{89}$Zr Cross Section}}
\begin{figure}[H]
{\includegraphics[width=1.0\columnwidth]{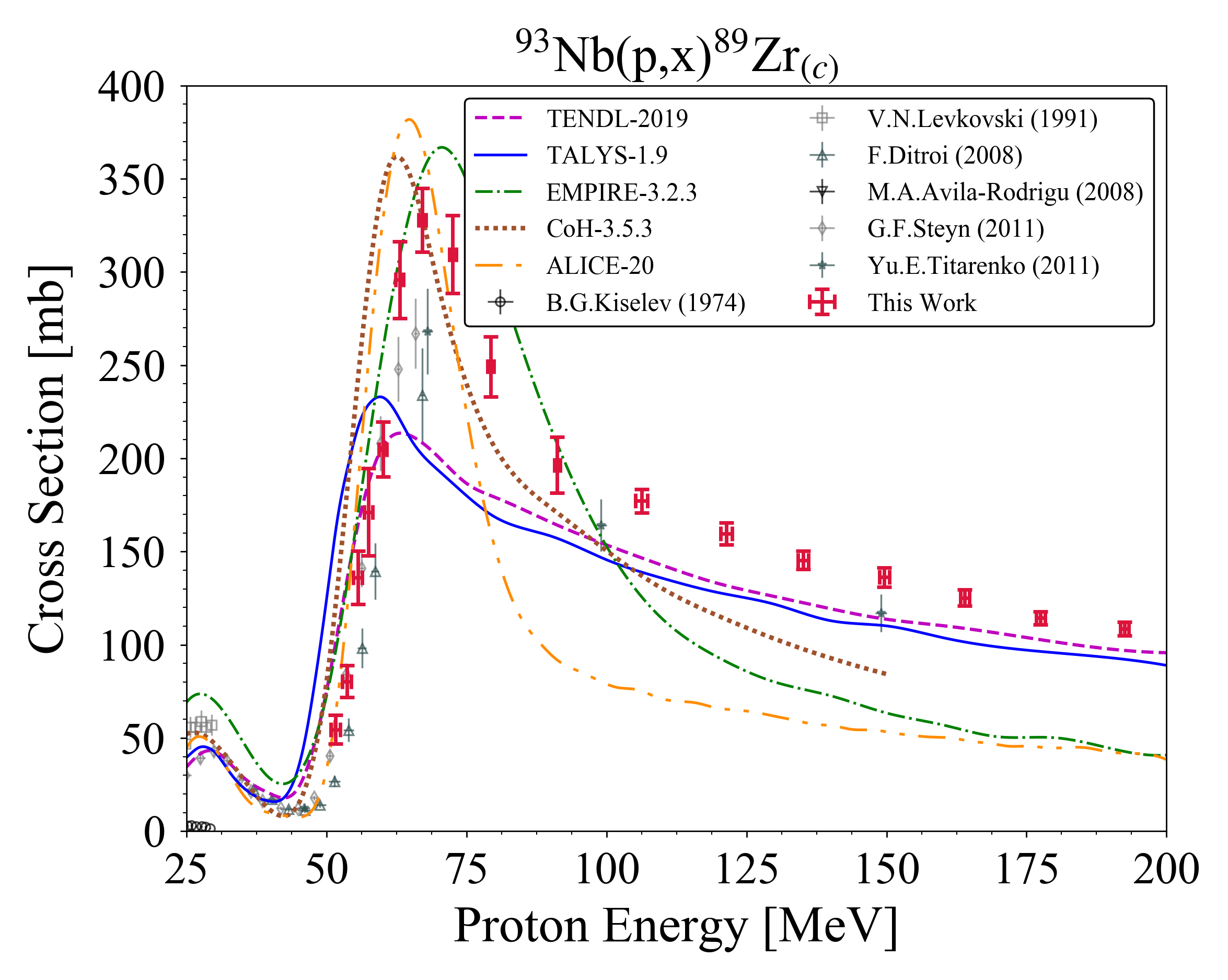}}
\vspace{-0.65cm}
\caption{Experimental and theoretical cross sections for cumulative $^{89}$Zr production, showing peaks for both $^{93}$Nb(p,$\alpha$n) and $^{93}$Nb(p,2p3n) formation mechanisms.}
\label{Nb_89Zr}
\end{figure}

The lifetimes of $^{89}$Zr precursor feeding nuclei ($^{89}$Mo, $^{89\textnormal{m}}$Nb, $^{89}$Nb, $^{89\textnormal{m}}$Zr) were too short to be able to quantify their production in these irradiations given the counting procedures described in Sections \ref{3.1} and \ref{3.2} \cite{DataSheetsA89}. As a result, the measurement of $^{93}$Nb(p,x)$^{89}$Zr, provided in Figure \ref{Nb_89Zr}, is cumulative and includes contributions from all of these precursors as well as the ground state of $^{89}$Zr.

$^{89\textnormal{g}}$Zr is a useful positron emitting isotope for radiolabelling monoclonal antibodies to provide an accurate picture of dose distribution and targeting effectiveness in immunoPET \cite{89ZrPET,NucMedGeneral,Sadeghi2012:89ZrDifferentModels}. Its 78.41 $\pm$ 0.12\,hr half-life meshes nicely with the typical 2--4 day pharmacokinetic properties of monoclonal antibodies in tumours \cite{DataSheetsA89,89ZrPET}. Further, zirconium is especially attractive for this application because of existing commercially available chelating agents for labelling, which have been proven to remain bound in-vivo. Production of $^{89\textnormal{g}}$Zr via $^{93}$Nb(p,x) using 200\,MeV protons may offer an attractive alternative to the established $^{89}$Y(p,n)$^{89}$Zr route used in low-energy cyclotrons, potentially facilitating $^{89}$Zr production in locations such as IPF and BLIP \cite{89ZrPET}. However, the co-production of $^{88}$Zr (t$_{1/2}=83.4$ $\pm$ 0.3\,d \cite{DataSheetsA88}) in the $^{93}$Nb(p,x) path may make the low-energy (p,n) route more viable.

This work gives the most complete description of the cumulative higher-energy production peak near 67\,MeV and greatly extends the cross section information beyond 75 MeV, where only two prior data points existed. The larger higher-energy peak is indicative of independent $^{89}$Zr formation through the $^{93}$Nb(p,2p3n) mechanism in contrast to the lower-energy compound peak around 25\,MeV, denoting formation by $^{93}$Nb(p,$\alpha$n). The measured values agree well with \textcite{Steyn2011} on the higher-energy peak rising edge, but predict a peak value of approximately 325 mb, which is larger than both \textcite{Steyn2011} and \textcite{Titarenko2011}. The \textcite{Ditroi2008} magnitude discrepancy is noticeable in this measurement where the dataset underpredicts both the rising edge and peak relative to all the other literature.

\vspace{-0.3cm}

It is difficult to comment on the performance of the codes here due to the feeding from the three nuclei, and multiple isomeric states, involved in the calculations. It can be noted that there is still the persistent difficulty in properly modeling the pre-equilibrium effect throughout these nuclei though, which manifests in these codes as both a shift in the centroids for the higher-energy peak and a missing high-energy tail.

\subsection{\label{86YExFunction}{$^{93}$Nb(p,x)$^{86}$Y Cross Section}}
The LANL and BNL irradiations in this investigation allowed for a measurement of $^{86}$Y production from reaction threshold to near 200 MeV. As specifically referenced in Figure \ref{FeedingRegressionEx}, the cumulative $^{86}$Zr production could be directly determined, which then enabled an independent quantification of $^{86}$Y. The 33\% $\beta^+$ decay mode of $^{86}$Y along with its 14.74 $\pm$ 0.02\,hr half-life make it a promising surrogate for imaging the biodistribution and studying the absorbed dose of $^{90}$Y (100\% $\beta^-$) for bone palliative treatments \cite{DataSheetsA86,86YPET1}. However, compared to the established $^{86}$Y production routes using strontium targets, a niobium target based pathway introduces long-lived $^{88}$Y (t$_{1/2}=106.626$ $\pm$ 0.021\,d \cite{DataSheetsA88}) isotopic impurities and suffers a lower yield, making it less advantageous \cite{86YPET2}.

\begin{figure}[H]
{\includegraphics[width=1.0\columnwidth]{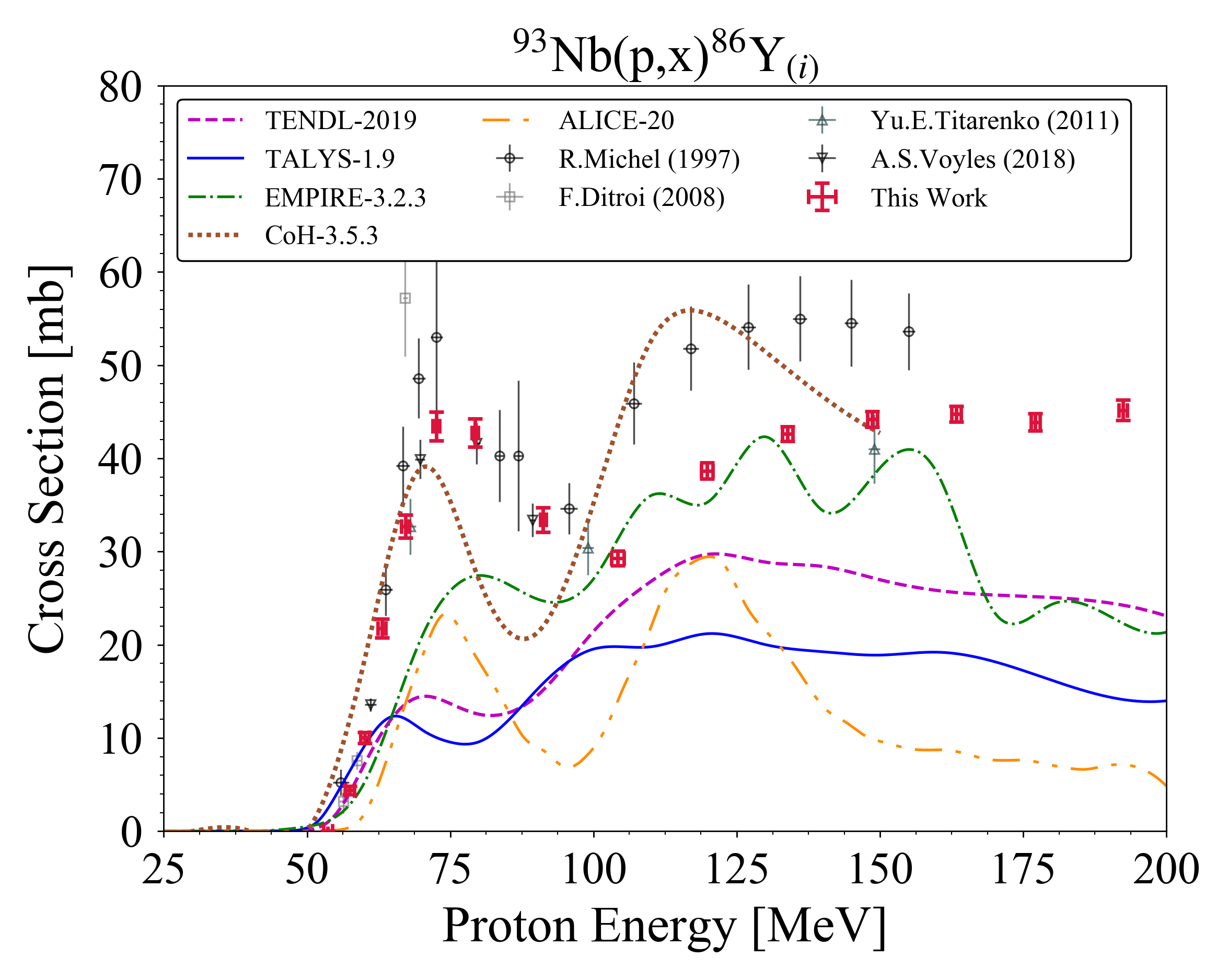}}
\vspace{-0.65cm}
\caption{Experimental and theoretical cross sections for $^{86}$Y production, spanning from reaction threshold to near 200 MeV.}
\label{Nb_86Y}
\end{figure}

The extracted excitation function (Figure \ref{Nb_86Y}) is in excellent agreement with the measurements of \textcite{Voyles2018:Nb} and \textcite{Titarenko2011}. This wide-spanning dataset, similar to the \textcite{Michel1997:ProtonsTiCuNb} work, characterizes the full compound behaviour as well as the high-energy pre-equilibrium component. However, where there is good agreement to the \textcite{Michel1997:ProtonsTiCuNb} work below 100 MeV, our dataset predicts lower values for the remainder of the pre-equilibrium tail by 10--15 mb.

\vspace{-0.3cm}

$^{86}$Y is not a strongly-fed residual product channel, which gives some explanation to the variation between different code calculations. The theoretical predictions are sensitive to compensating effects from miscalculations in more dominant reaction channels. As a result, no code properly reproduces both the experimentally determined magnitude and shape of the excitation function using default parameters. CoH predicts the compound peak with the closest magnitude, though the peak centroid, falling edge, and pre-equilibrium shape are incorrect. TALYS and TENDL perhaps best represent the overall shape but are far lower in magnitude than the experimental data.

Other notable cross section results in this work include $^{82\textnormal{m}}$Rb, $^{83}$Sr, and $^{84}$Rb production, where data had been extremely sparse but now have their excitation functions well-characterized beginning from threshold. These cross section results, along with the measurements of all other observed nuclei, are detailed in Appendix \ref{Appendix_Plots}.

\section{\label{}High-Energy Proton Reaction Modeling}
The large body of data measured here, in addition to the existing $^{93}$Nb(p,x) literature data, presents a good opportunity to study high-energy proton reaction modeling on spherical nuclei. Our approach is to follow the procedure established for modeling high-energy (n,x) reactions by comprehensively fitting the most prominent residual product channels first, followed by the weaker channels. A critical focus in developing a consistent fitting procedure is to gain insight into pre-equilibrium reaction dynamics in an attempt to isolate shortcomings in the current theoretical understanding.

As a note, the fitting work presented here is based in the TALYS reaction code. TALYS has widespread use in the nuclear community and is an accessible code-of-choice for reaction cross section predictions. Further, TALYS incorporates the widely employed two-component exciton model for pre-equilibrium physics, which means that any outcomes derived in this work can be applied broadly by the nuclear reaction data evaluation community \cite{TalysManual,Griffin1966:Exciton,Kalbach1986:Exciton,Cline1971:PEMexciton}.

\subsection{\label{PEM_talys}Pre-Equilibrium in TALYS-1.9}
The currently-used two-component exciton model in TALYS-1.9 was constructed through an extensive global pre-equilibrium study by \textcite{Koning2004:GlobalPEMexcitonOMP}. Their work relied on virtually all existing angle-integrated experimental continuum emission spectra for (p,xp), (p,xn), (n,xn), and (n,xp) reactions for A$\geq$24 spanning incident energies between 7--200 MeV. No double-differential or residual product cross sections were included in the semi-classical two-component model development, but these results were expected to fall out naturally from globally fitting the emission spectra. The decision to adopt the exciton model over other potential pre-equilibrium calculation methods is detailed by \textcite{Koning2004:GlobalPEMexcitonOMP}.

The significant updates made by \textcite{Koning2004:GlobalPEMexcitonOMP} to previous two-component models include using a more recent optical model potential (OMP) for neutrons and protons, a new and improved determination of collision probabilities for intranuclear scattering to more or less complex particle-hole states, surface interactions specific to projectiles and targets, and greater detail applied to multiple pre-equilibrium emission. The most noteworthy of these changes is the collision probabilities, which use a new parameterization of the phenomenological squared matrix element for the effective exciton residual interaction applicable across the entire 7--200 MeV energy range \cite{Koning2004:GlobalPEMexcitonOMP,Kalbach1986:Exciton,DobesBetak1983:Exciton}.

Moreover, in the two-component exciton master equation used by \textcite{Koning2004:GlobalPEMexcitonOMP}, which describes the temporal development of the composite system for projectile-target interaction in terms of exciton states characterized by proton and neutron particle and hole numbers, internal transition rates are defined to model particle-hole creation ($\lambda^+$), conversion ($\lambda^0$), and annihilation ($\lambda^-$). These transition rates govern the evolution of the total exciton state and are critical pieces for the overall pre-equilibrium energy-differential cross section calculation \cite{Kalbach1986:Exciton,DobesBetak1983:Exciton,Cline1971:PEMexciton}. Formally, the model is approximated to disregard pair annihilation where it has been shown that decay rates to less complex exciton states are small compared to other processes in the pre-equilibrium part of the reaction and can be neglected \cite{Koning2004:GlobalPEMexcitonOMP,Kalbach1986:Exciton}. Transition rates are calculated from collision probabilities, determined using time-dependent perturbation theory and Fermi's golden rule to give expressions such as Equation (\ref{collisionprob}) for a proton ($\pi$)-proton ($\pi$) collision $\lambda_{\pi\pi}$, leading to an additional proton particle-hole pair $(1p)$ \cite{TalysManual}:
\begin{gather}\label{collisionprob}
\lambda_{\pi\pi}^{1p}=\frac{2\pi}{\hbar}M^2_{\pi\pi}\omega.
\end{gather}

In the collision probability definition given in Equation\,(\ref{collisionprob}), $\omega$ is the particle-hole state density as a function of the exciton state configuration and excitation energy, as formulated by \textcite{DobesBetak1983:Exciton}. An exciton state configuration is defined by $(p_\pi,h_\pi,p_\nu,h_\nu)$ with the proton (neutron) particle number as $p_\pi$ $(p_\nu)$ and the proton (neutron) hole number as $h_\pi$ $(h_\nu)$. $M^2_{\pi\pi}$, and the other corresponding proton and neutron ($\nu$) permutations ($M^2_{\pi\nu}$ etc.), are average squared matrix elements of the residual interaction inside the nucleus that depend only on the total energy of the composite nucleus to describe two-body scattering to exciton states of different complexity \cite{TalysManual}. In TALYS-1.9, the matrix element variations for like and unlike nucleons can be cast in terms of a total average $M^2$ by:
\begin{gather}
M^2_{xy}=R_{xy}M^2,
\end{gather}
with $x$ and $y$ denoting some combination of $\pi$ and $\nu$. $R_{xy}$ is a free parameter with default values in TALYS-1.9 such as $R_{\pi\nu}=1.0$ \cite{TalysManual}.

Given the complete body of experimental emission spectra data, the following semi-empirical expression for the total average squared matrix element is implemented in TALYS-1.9 for incident energies 7--200 MeV \cite{TalysManual}:
\begin{gather}\label{M2eqn}
M^2=\frac{C_1A_p}{A^3}\left[7.48C_2+\frac{4.62\times10^5}{\left(\frac{E^{tot}}{nA_p}+10.7C_3\right)^3}\right],
\end{gather}
where $C_1$, $C_2$, and $C_3$ are adjustable parameters, $A$ is the target mass, $A_p$ is the mass number of the projectile, $n$ is the total exciton number, and $E^{tot}$ is the total energy of the composite system. In particle-hole creation, the change in state exciton number is ${\Delta n=+2}$, while in a conversion transition $\Delta n=0$.

\begin{figure}[H]
	{\includegraphics[width=1.0\columnwidth]{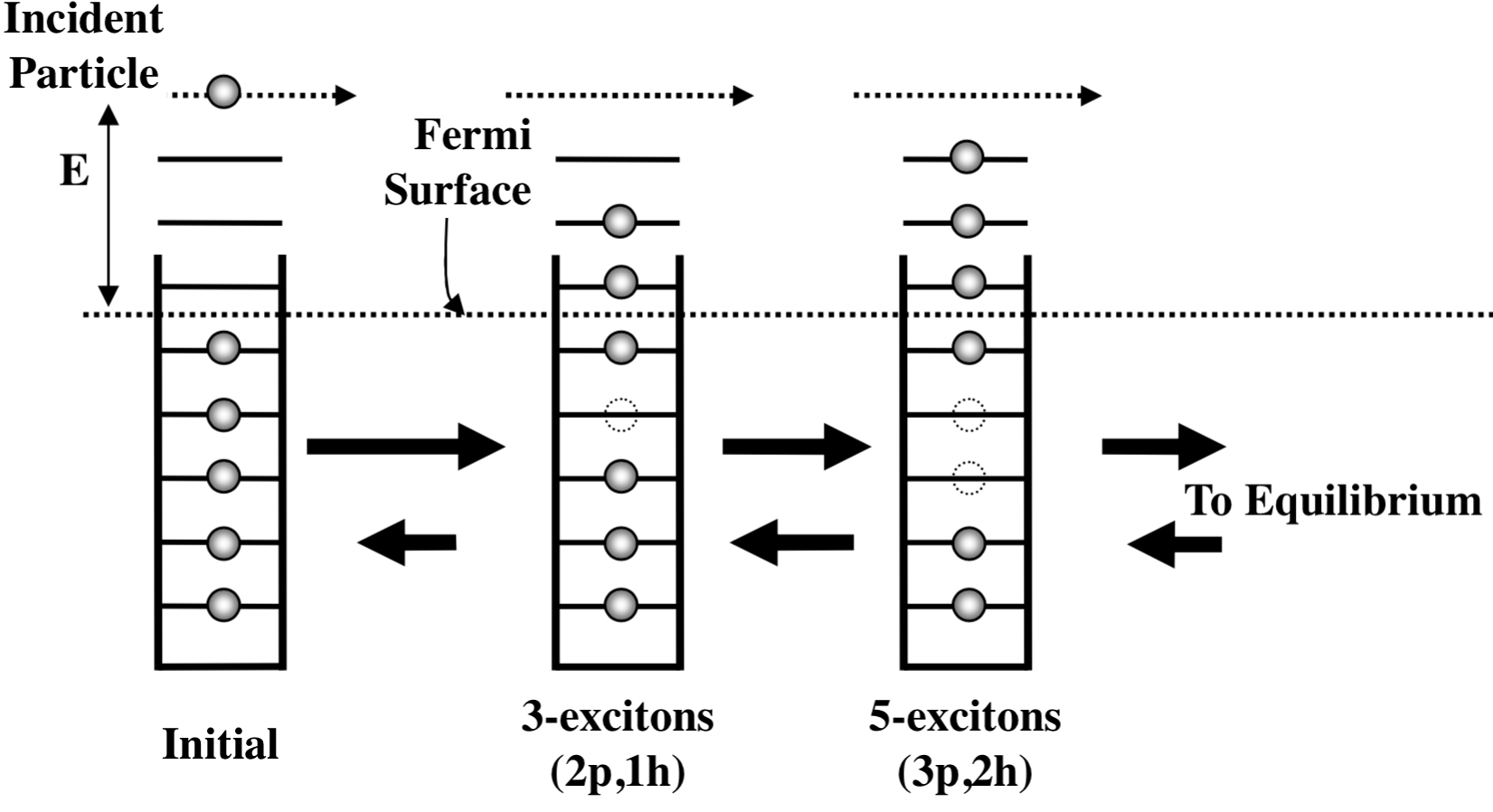}}
\vspace{-0.65cm}
\caption{Illustration of the initial stages of reaction in the pre-equilibrium exciton model from \textcite{Selman2009:Thesis}. Solid horizontal lines are representative of single particle states in a potential well. Particles are shown as solid circles while holes are empty dashed circles
 \cite{Cline1971:PEMexciton}.}\label{ScatteringGraphic}
 \vspace{0.15cm}
	{\includegraphics[width=1.0\columnwidth]{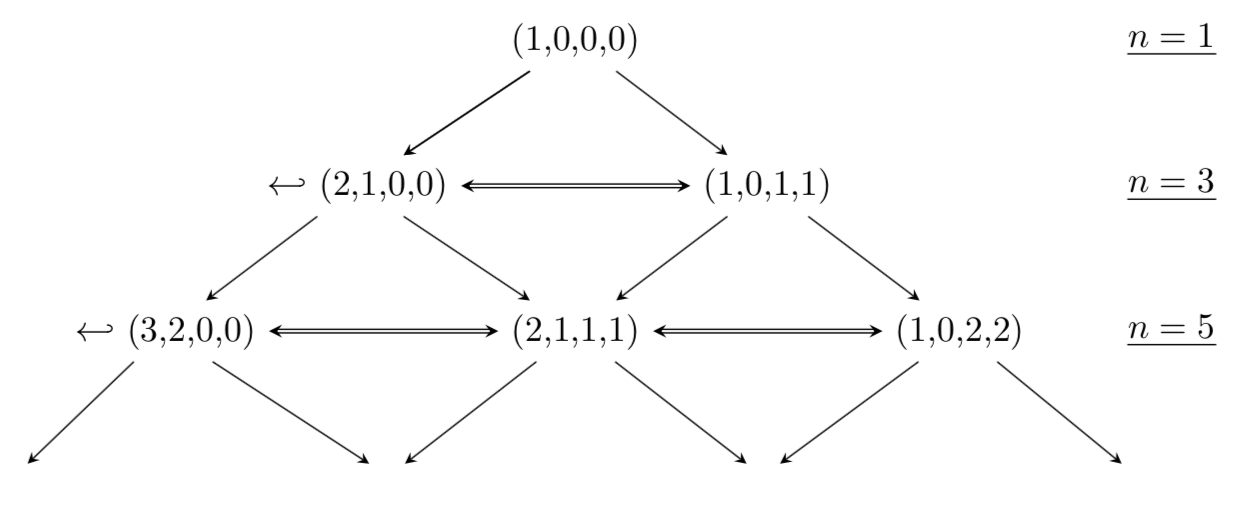}}
\vspace{-0.65cm}
\caption{Scheme of the two-body interaction pathways in the two-component exciton model where individual exciton states are characterized by $(p_\pi,h_\pi,p_\nu,h_\nu)$. The particle-hole annihilation pathways to less complex states are neglected here. The single arrows represent particle-hole creation transitions and the double arrows represent conversion transitions. The hooked arrows represent the chance for particle emission to the continuum at the given exciton number $n$, where $n$ is the sum of all present particles and holes in a configuration.}\label{ExcitonGraphic}
\end{figure}

For an incident proton projectile, a simplified visualization of the scattering with target nucleons defined by the exciton model is shown in Figure \ref{ScatteringGraphic}. Additionally, a schematic of the two-component transitions from an initial exciton state configuration of $(1,0,0,0)$ to more complex states is given in Figure \ref{ExcitonGraphic} \cite{DobesBetak1983:Exciton}.

\vspace{-0.35cm}
\vspace{-0.1cm}

Each state in Figure \ref{ExcitonGraphic} has an associated mean lifetime $\tau(p_\pi,h_\pi,p_\nu,h_\nu)$ defined as the inverse sum of the various internal transition rates and the total emission rate \cite{TalysManual}. As a result, the parameterization of $M^2$ is an essential component of the state lifetime calculation. Moreover, it can be noted from the representation in Figure \ref{ExcitonGraphic} that in order to calculate the overall energy differential pre-equilibrium cross section, the exciton model calculation must keep track of all emissions in addition to the part of the pre-equilibrium flux that has survived emission and now passes through new configurations. This survival population is generally denoted by $P(p_\pi,h_\pi,p_\nu,h_\nu)$ and is also calculated on the basis of the $M^2$ parameterization. The total emission rate $W$ for an ejectile $k$ of emission energy $E_k$ is not a function of $M^2$ but is instead calculated from the optical model and $\omega$ \cite{TalysManual}.

Given these considerations, the energy differential pre-equilibrium cross section can be calculated by \cite{TalysManual}:
\begin{gather}
\begin{split}
\frac{d\sigma_k^{PE}}{dE_k}&=\sigma^{CF}\sum_{p_\pi=p^0_\pi}^{p_\pi^{max}}\sum_{p_\nu=p^0_\nu}^{p_\nu^{max}}W_k(p_\pi,h_\pi,p_\nu,h_\nu,E_k)\\ &\times \tau(p_\pi,h_\pi,p_\nu,h_\nu)P(p_\pi,h_\pi,p_\nu,h_\nu),
\end{split}
\end{gather}
where $\sigma^{CF}$ is the compound nucleus formation cross section, also calculated from the optical model. $p_\pi^{max}$ and $p_\nu^{max}$ are particle numbers representing the equilibration limit for the scattering interactions at which point the Hauser-Feshbach mechanism handles the reaction calculations. In the case of multiple pre-equilibrium emissions, additional proton and neutron number dependencies are introduced into the exciton model, though $M^2$ and the internal transition rates play similar critical roles \cite{Koning2004:GlobalPEMexcitonOMP}.

Ultimately, given that the level density and optical model parameters at high energies are well-characterized compared to the relative paucity of information surrounding pre-equilibrium dynamics, it can be argued that an exploration of pre-equilibrium emission resulting from the exciton model in TALYS is centrally an exploration of the effective squared matrix element parameterization. TALYS's abundance of adjustable keywords related to $M^2$ make it an ideal tool to investigate this parameterization using measured residual product excitation function data. However, it will not be possible to entirely neglect the effects of level density and optical model adjustments on reaction observables and it is necessary to be cognizant of these additional degrees of freedom in any attempt to isolate $M^2$ effects \cite{Koning2004:GlobalPEMexcitonOMP}.

\subsection{Residual Product Based Standardized Fitting Procedure}
\begin{figure*}[t]
{\includegraphics[width=1.0\textwidth]{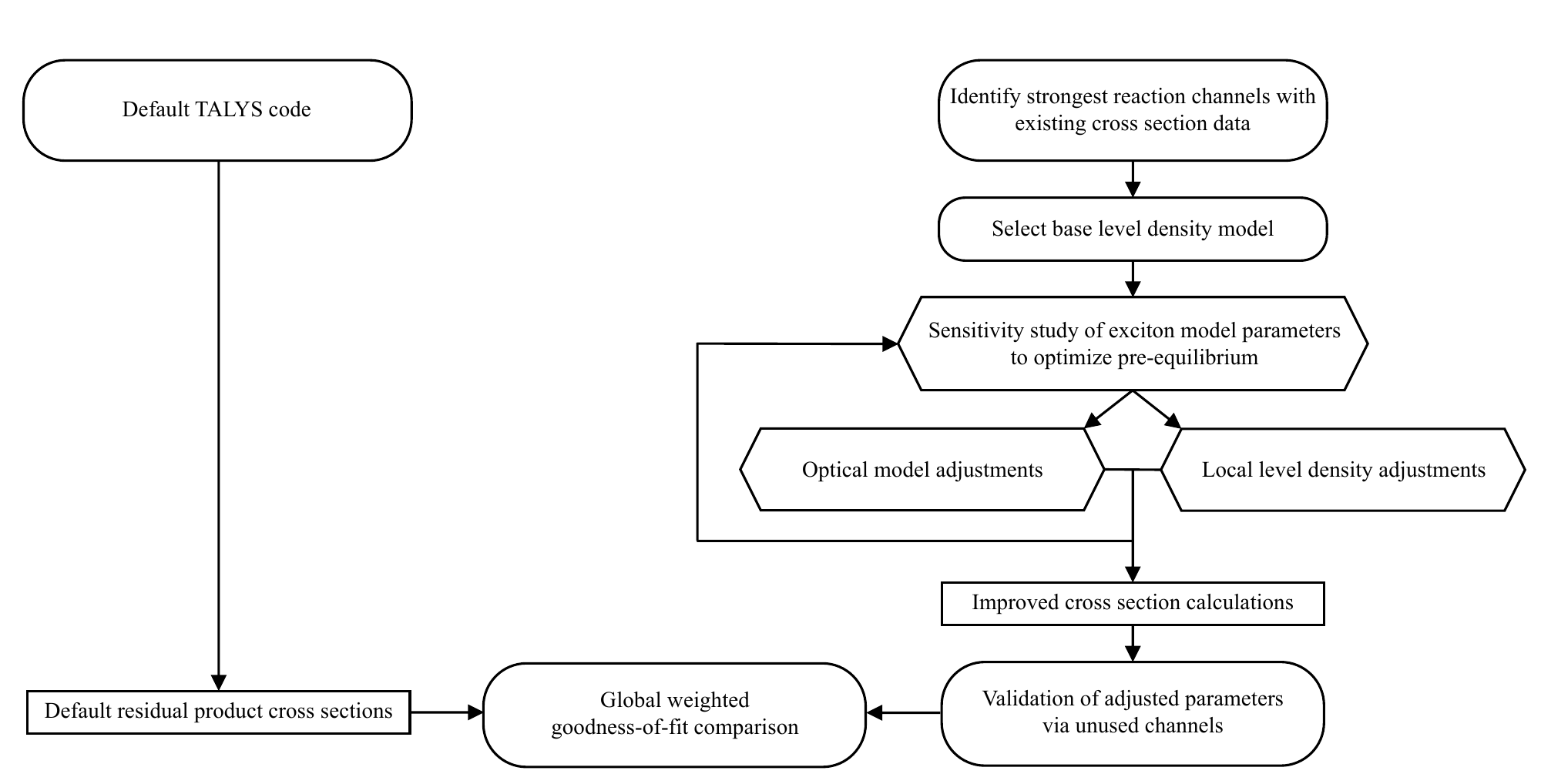}}
\vspace{-0.45cm}
\caption{Proposed standardized reaction modeling code parameter adjustment procedure, reliant on residual product excitation function data, built to best fit multiple dominant reaction channels and gain justified insight into the pre-equilibrium mechanism.}
\label{ProcedureFlowChart}
\end{figure*}

The approach pursued in this work to accurately reproduce production probabilities for high-energy proton-induced reactions on spherical nuclei using TALYS and its associated adjustable parameters is outlined in the flow chart of Figure \ref{ProcedureFlowChart}. This fitting procedure prioritizes an examination of exciton model physics to help identify trends and biases within the current calculation technique.

A further motivation of this procedure is to avoid the compensating errors caused by current non-evaluation fitting methods that utilize too few experimental data and/or too simplistic parameter changes, which may ultimately hinder modeling as a whole. Particularly, simplistic or arbitrary parameter adjustments in TALYS, tuned to provide a better fit for a singular reaction channel of interest, are non-unique and may not hold a global physical basis because neighbouring reaction channels can suffer from the fit choice \cite{Babu2012:ProtonTalysParams,Fuladvand2013:AsParametersPEM,Kakavand2010:AliceParameters89Zr,Sadeghi2012:89ZrDifferentModels,Lawriniang2018,Parashari2018,Koning2013:TalysBestParams,Koning2015:TENDLstatusBestFitEvals,Singh2006:ProtonAlphaModelExcitationFunc,Alhassan2019:ProtonsBestFitChiSquared,Brodovitch1976:ProtonsPEM}. Nevertheless, these adjustment methods are representative of a norm in non-evaluation modeling work and can have real-world implications such as incorrect predicted yields during medical radioisotope production, high level co-production of an unwanted contaminant, or poor particle transport calculations. Even with a foundational understanding of the level density, OMP, and exciton model parameter adjustments, the interplay between the permutations and combinations of changes in each component is not well understood \cite{Koning2004:GlobalPEMexcitonOMP}. In turn, it is difficult to determine the most physically justifiable modeling parameters if the data from every open reaction channel is not known.

For example, consider the numerous modeling possibilities for the large residual product channel $^{93}$Nb(p,p3n)$^{90}$Nb, as shown in Figure \ref{NbTALYSModels}. The list of parameter adjustments in each modeling case is described in Appendix \ref{10Models} (Table \ref{NbParameterDetails}). It is qualitatively seen that ten different models, with arbitrary choices of which simplistic or complex parameters are adjusted, can reproduce similar improvement over the default prediction. 

\begin{figure}
{\includegraphics[width=1.0\columnwidth]{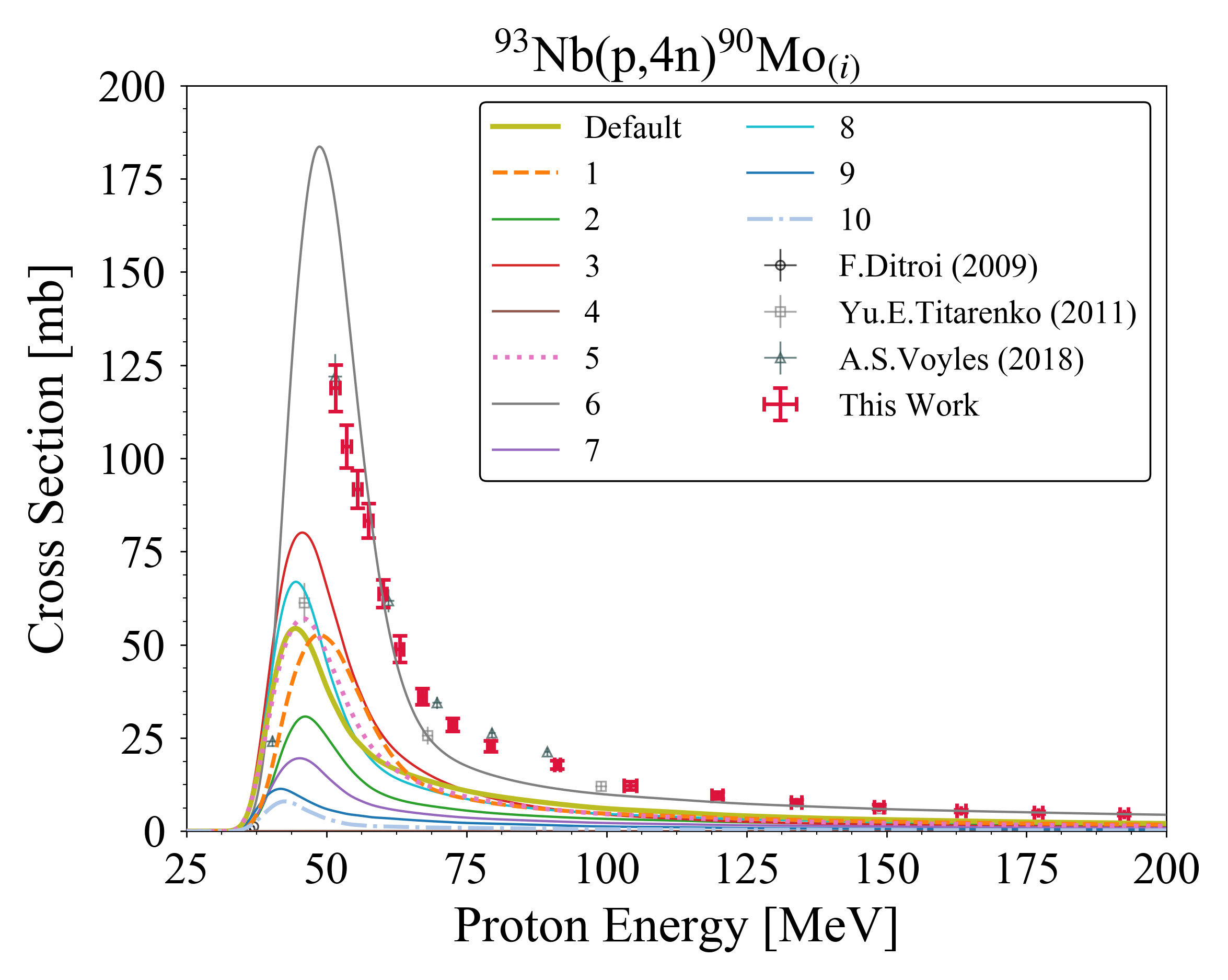}}
\vspace{-0.55cm}
\caption{Extension of model adjustments, optimized to singularly reproduce the (p,p3n) channel, to a neighbouring channel demonstrating poor fit behaviour, especially for the three dashed cases that previously performed best.}
\label{NbParametersNewChannel}
\end{figure}

Still, it could be argued that one set of changes is quantitatively the best to model this channel. A $\chi^2$-test using the experimental data demonstrates that models 1, 5, and 10 give the largest improvements over default. These models are indicated with dashed/dotted lines in Figure \ref{NbTALYSModels} and the $\chi^2$ result of each parameter set is listed as well in Appendix \ref{10Models}. Given these best fits, it consequently seems logical to search for meaning in the altered parameters and attribute their need to lacking physics in this charged-particle problem. However, simply applying these best fit models to surrounding reaction channels proves that these sets of parameter changes in fact do not improve the model's predictive capabilities. For example, in the $^{93}$Nb(p,4n)$^{90}$Mo channel, which also makes up a large share of the reaction cross section, models 1, 5, and 10 from Figure \ref{NbTALYSModels} perform extremely poorly, as shown in Figure \ref{NbParametersNewChannel}.

Instead, a more useful and realistic modeling approach should involve many prominent cross section channels and sensitivity studies. The inclusion of more experimental data and increased detail in the analysis process will yield a more unique and global solution along with the capability to justify the set of adjusted parameters while providing physics context for the predictions.

As outlined in Figure \ref{ProcedureFlowChart}, this suggested improved fitting procedure for spherical nuclei begins by identifying and having accurate experimental data for numerous prominent residual product channels. This approach is anchored in examining the most probable outcomes where it is possible to best isolate the impact of model changes. Experimental data for weaker production channels are still involved and relevant but are weighted less heavily due to their high sensitivity to the behaviour of the dominant reactions.

Once the largest reaction channels have been identified, the following step is to select a level density model for all the nuclei involved in the interaction being studied such that there is a concrete foundation, based on the well-established compound nucleus model, to build model adjustments upon and put their effects in context. TALYS-1.9 provides six level density models, three that are microscopic calculations, which are preferred in this procedure for their better care of the physics involved and use in predictive scenarios versus the remaining three phenomenological models \cite{TalysManual}. At this point, the proposed fitting approach reaches the key step of an exploration of the exciton model parameter space. Notably, the pre-equilibrium dynamics are adjusted the most in this suggested method. Both the OMP and exciton model parameterizations are based on very large global studies. However, deviations from the optical model default values represent a much greater change to the physics of the situation than tuning for the exciton model \cite{Koning2004:GlobalPEMexcitonOMP,TalysManual,KD2003:OMP,Carlson2010:OMP2}. The optical model fundamentally affects the nature of the particle-nucleus reaction while changing the exciton model parameters maintains the same pre-equilibrium physics basis but shifts evolution and emission rates within the model, which are not known precisely at the outset. In this manner, this fitting mechanism is specifically suited to isolate and gain insight into pre-equilibrium modeling for high-energy proton-induced reactions.

\begin{figure}[t]
{\includegraphics[width=1.0\columnwidth]{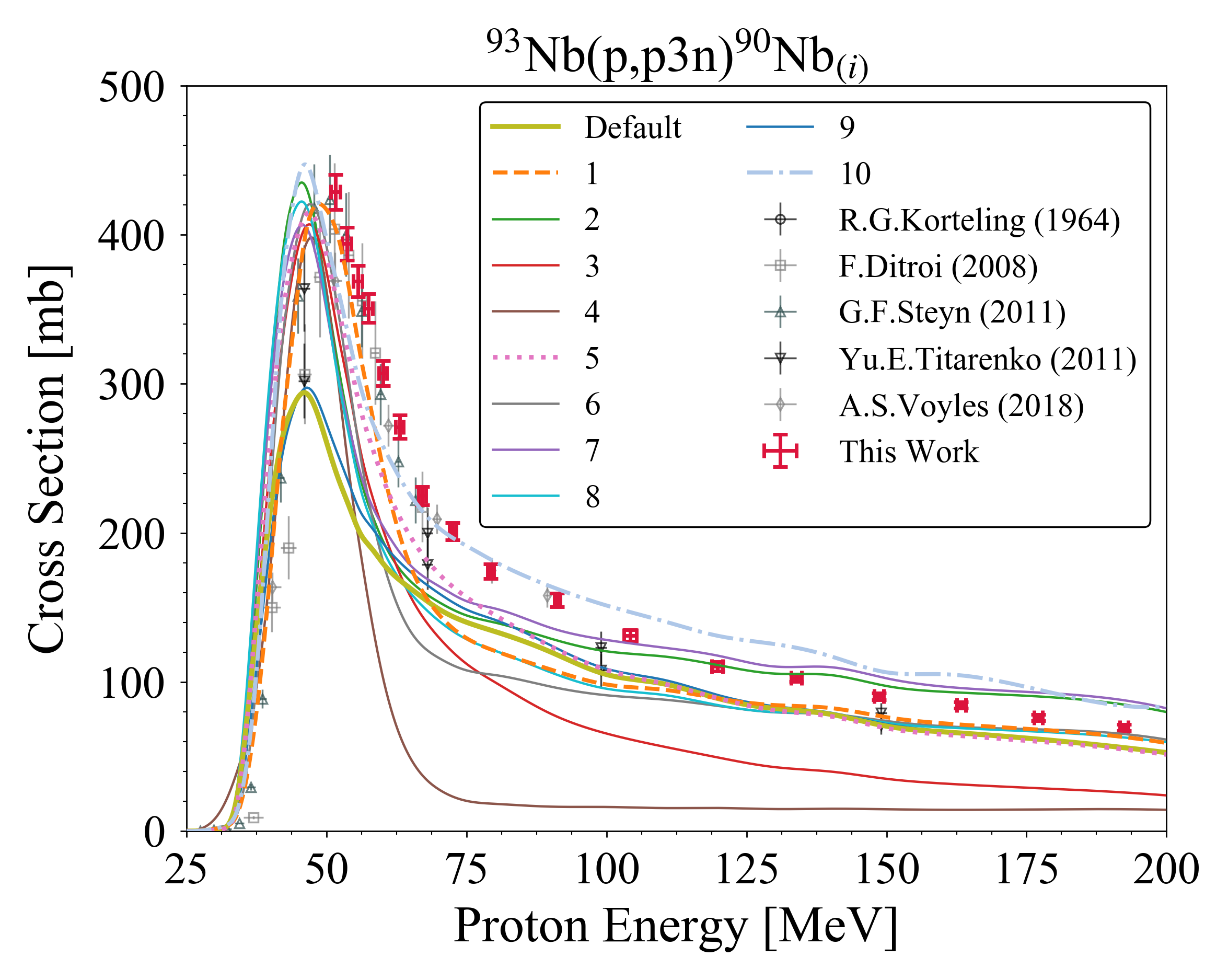}}
\vspace{-0.55cm}
\caption{Evidence for non-unique modeling solution when only considering one reaction channel. Ten sets of different parameter changes are shown to reproduce similar improvement over the default prediction, with the three dashed cases performing best as assessed by a statistical test.}
\label{NbTALYSModels}
\end{figure}

The most significant of the available exciton model free parameters within TALYS are \texttt{M2constant}, \texttt{M2limit}, and \texttt{M2shift}, which adjust $C_1$, $C_2$, and $C_3$, respectively, in Equation (\ref{M2eqn}). \texttt{M2constant}, \texttt{M2limit}, and \texttt{M2shift} are set to 1.0 as default in TALYS \cite{TalysManual}. A decrease in \texttt{M2constant} reduces the transition rate to more complex exciton states, thereby increasing pre-equilibrium emission in the initial interaction stages and creating an overall harder emission spectrum with an increased high-energy tail. The opposite effect applies for an increase in \texttt{M2constant}. The \texttt{M2limit} controls the asymptotic behaviour of $M^2$ and its increase leads to scattering to more complex states at high energies, thereby preventing an overestimation of the high-energy tail, which pulls reaction cross section from the evaporation peak \cite{Koning2004:GlobalPEMexcitonOMP}. The \texttt{M2shift} affects the total system energy and can shift the exciton model strength along the projectile energy axis. Other parameters that alter the pre-equilibrium effects to a lesser degree also exist such as \texttt{Rgamma, Cstrip, Rnupi, preeqspin, gpadjust} etc., which are all described in the TALYS-1.9 manual and should be considered as well \cite{TalysManual}.

Once the components of the exciton model are set according to the behaviour of the largest reaction channels, there is an opportunity to perform some studies of OMP and level density parameters. These aspects can help optimize the fit founded on the exciton model changes for smaller residual production channels or localized outstanding discrepancies between theory and experiment. The OMP and level density adjustments here are minor corrective factors to the broader deduced pre-equilibrium modeling. These adjustments may require some iterations to reach convergence \cite{Alhassan2019:ProtonsBestFitChiSquared}.

Lastly, a validation step is an important conclusion to this procedure. If the exciton, OMP, and level density adjustments set by the breadth of reaction channels considered are unique and correct, their application to channels not included in the initial sensitivity studies should yield appropriate fits. Cumulative excitation functions are good examples of unused data, where they may have large cross sections but the ambiguity from contributions of a chain of multiple nuclei and emission channels is not ideal for the initial sensitivity study. This is a test of the predictive capability of this procedure. Finally, a descriptive metric, such as a global $\chi^2$-test, can be applied to compare the adjusted fit in all utilized channels from this procedure to the default calculation \cite{Alhassan2019:ProtonsBestFitChiSquared,Alhassan2020:IterativeBayesianEval,Koning2015:BayesianEval}. Ideally, the metric is properly weighted to reflect the emphasis on the most prominent reaction channels. Formulae for these weights are discussed in Section \ref{FitNb}.

\subsection{\label{FitNb}Fitting Procedure Applied to $^{93}$Nb(p,x)}
This work demonstrates the procedure outlined in Figure \ref{ProcedureFlowChart} for high-energy proton reactions on niobium. At present, this sensitivity study work is performed manually to better gauge the physical effects of different parameters and to mimic typical cross section parameter adjustment work. Nine reaction channels are considered: $^{93}$Nb(p,x)$^{\textnormal{93m,90}}$Mo, $^{\textnormal{92m,90}}$Nb, $^{88,87,86}$Zr, $^{88,86}$Y, with $^{90}$Nb, $^{90}$Mo, and $^{88}$Zr production as the most prominent.

In the base level density model choice step, the microscopic models were indeed found to have greater predictive power than the phenomenological models. The $^{93}$Nb(p,4n)$^{90}$Mo reaction was found to be most sensitive to the level density model. Only the microscopic calculations from Goriely's tables using the Skyrme effective interaction (\texttt{ldmodel} 4) could produce a fit magnitude in the vicinity of the experimental data while maintaining adequate predictive power in the other considered channels \cite{TalysManual}. The apparent sensitivity of $^{90}$Mo production to angular momentum distributions in nuclei closer to the target $^{93}$Nb therefore made it the constraint for a level density choice.

Once the level density model was chosen, the adjustment of pre-equilibrium could take place. The sensitivity study of the exciton model parameters showed that reducing \texttt{M2constant} from its default 1.0 value could best benefit high-energy tail behaviour across the prominent residual product cross sections. The tail-shape improvement came at the cost of unwanted reduced compound peak magnitudes, which could be compensated by an increase in \texttt{M2limit} and a decrease in \texttt{M2shift}. Marginal variations of the three \texttt{M2} parameters relative to each other given these constraints demonstrated a best fit for the largest available channels when \texttt{M2constant}=0.875, \texttt{M2limit}=4.5, and \texttt{M2shift}=0.6. Furthermore, this pre-equilibrium correction for the larger channels introduced a cascade effect that improved the compound peak behaviour of smaller cross section channels, giving confidence that these adjustments were globally beneficial. The numerous other additional scaling factors and modeling choices for pre-equilibrium available in TALYS were also explored but were shown to be insensitive relative to the \texttt{M2} parameters or physically inconsistent across the nine considered reactions here.

However, while compound peak improvement was seen in the weaker far-from-target channels, issues arose with their higher-energy cross section predictions deviating from the experimental data. This applies to nuclei such as $^{87,86}$Zr and $^{86}$Y, which exist on the other side of the $N=50$ shell gap relative to the target $^{93}$Nb. The base level density model choice, which served calculations for the niobium and molybdenum excitation functions well, proved to be a root cause for these unpredictable emission issues further from the target nucleus. The level densities of all nuclei involved in this charged-particle interaction are not perfectly modelled by the base choice and may require specific variations, as outlined in Figure \ref{ProcedureFlowChart}. Adjusting the level density model for niobium and molybdenum nuclei relevant to emissions for these far-from-target residual products from \texttt{ldmodel} 4 to the Hilaire combinatorial calculation using the Skyrme force (\texttt{ldmodel} 5) was tested. This change produced a sufficient compensating effect to quell the incorrect high-energy behaviour in the majority of the far-from-target channels \cite{TalysManual,Koning2013:TalysBestParams}. Note that $^{93}$Mo and $^{92}$Mo needed to remain modelled by \texttt{ldmodel} 4 as these were key nuclei in the $^{90}$Mo angular momentum constraint discovered earlier in the base level density choice study.

Minor deviations to the optical model could then be considered to address outstanding discrepancies between prediction and experimental data. The key discrepancies remaining at this point in the analysis included a slight under-prediction of the $^{90}$Nb production compound peak and falling edge versus a slight over-prediction of the same aspects in $^{90}$Mo, as well as an incorrect competition between $^{86}$Zr and $^{86}$Y production, where the former was overestimated and pulled reaction flux from the latter. The zirconium and yttrium channels are inherently difficult to predict accurately as they are weaker reactions (with peak cross sections nearly an order of magnitude lower than the dominant channels comprising the initial tuning set) susceptible to large variations from compounding effects in the modeling. The larger $^{90}$Nb and $^{90}$Mo reactions were therefore the primary constraints for OMP parameter adjustments. Exploring the real and imaginary volume components of the OMP is the most physically sensible course for correcting the fit versus experimental data magnitude discrepancies, as these parameters directly affect particle flux loss and emission. The sensitivity study of the TALYS OMP volume terms revealed a significant reliance on only \texttt{rvadjust p/n/a} (multipliers to energy-independent radial factors of volume potentials)  and \texttt{w1adjust p} (direct multiplier to proton imaginary volume potential well depth) in this charged-particle reaction setting \cite{TalysManual,KD2003:OMP}. The other volume potential parameters may be relevant in a different context but are difficult to assess without double differential scattering information. Marginal changes to \texttt{rvadjust p/n/a} and \texttt{w1adjust p} demonstrated that only \texttt{w1adjust p} was needed to best improve the $^{90}$Nb peak magnitude and falling edge. \texttt{w1adjust p} affects the overall proton reactivity and emission. An increase to \texttt{w1adjust p} from its 1.0 default to a value of 2.2 increased the cross section reasonably of all channels but most noticeably for $^{90}$Nb production, especially relative to the $^{93}$Nb(p,4n)$^{90}$Mo reaction.

A slight errant local competition between $^{90}$Nb and $^{90}$Mo still existed that could be improved by manually adjusting level densities using the \texttt{ctable} and \texttt{ptable} TALYS commands. This level density table adjustment can be applied to an individual nuclide and when adjusted by reasonable amounts only has sensitivity for the selected nuclide and its neighbours, thereby maintaining the good global behaviour set by all the previous parameter changes. $^{90}$Mo required a \texttt{ctable} decrease to bring its production down while increasing the competing $^{90}$Nb channel, allowing both predictions to align well with experimental data. The zirconium and yttrium competition issues also required \texttt{ctable} decreases to be resolved and even prompted a slight $^{87}$Zr level density decrease as well. Adjusting the level densities in this manner for far-from-target nuclei holds a less clear physical meaning as the changes are potentially brought on by more complex reaction aspects, hidden from this sensitivity study work, that are lumped into this compensating correction. This is a part of the procedure described in Figure \ref{ProcedureFlowChart} but it should be emphasized that the most clear application of this approach is for dominant reaction channels.

\begin{figure}[H]
		{\includegraphics[width=1.0\columnwidth]{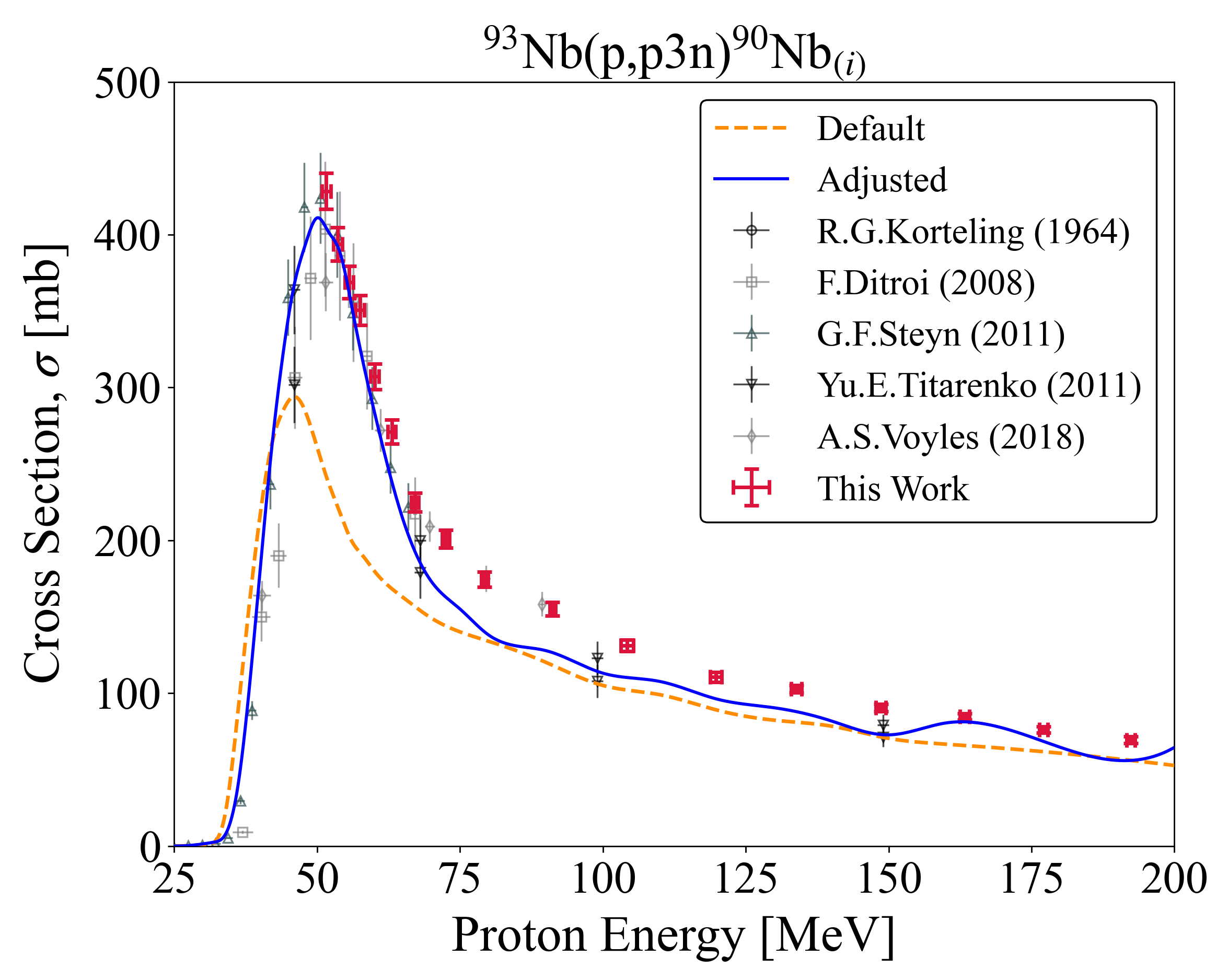}}
	\vspace{-0.65cm}
	\caption{TALYS default and adjusted calculation for $^{90}$Nb.}\label{talys45_90Nb}
\end{figure}

\begin{figure}[H]
		{\includegraphics[width=1.0\columnwidth]{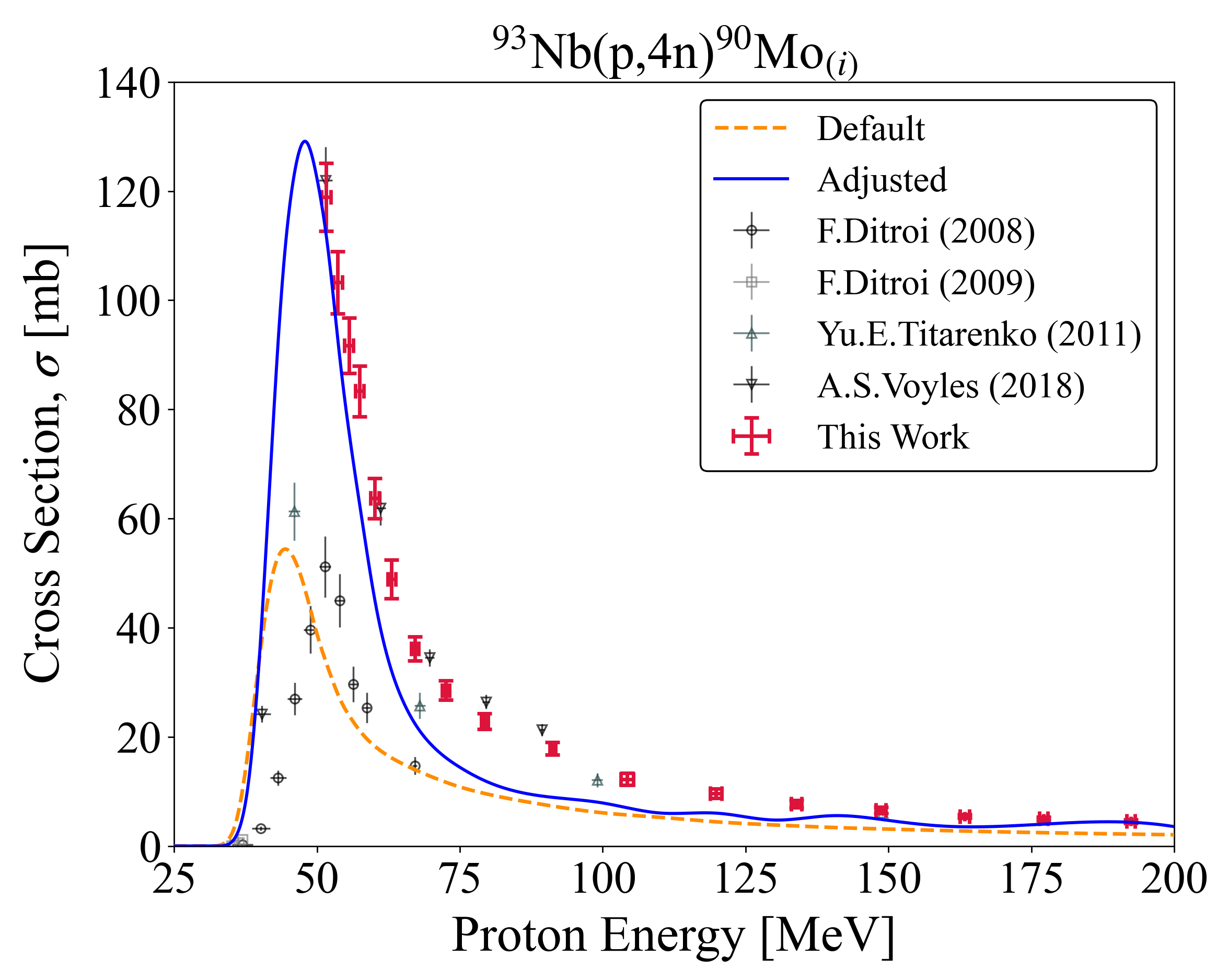}}
	\vspace{-0.65cm}
	\caption{TALYS default and adjusted calculation for $^{90}$Mo.}\label{talys45_90Mo}
		{\includegraphics[width=1.0\columnwidth]{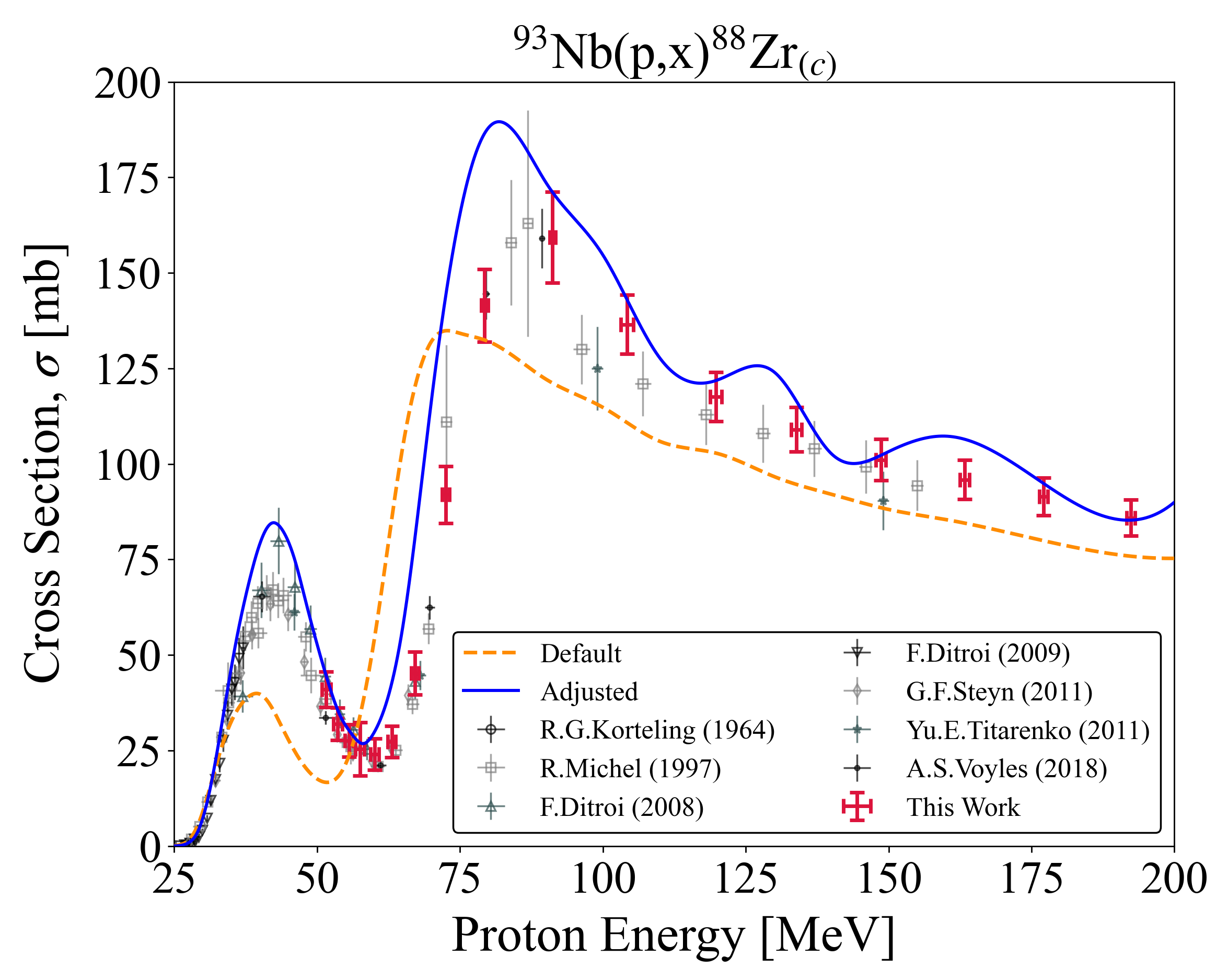}}
	\vspace{-0.65cm}
	\caption{TALYS default and adjusted calculation for $^{88}$Zr.}\label{talys45_88Zr}
		{\includegraphics[width=1.0\columnwidth]{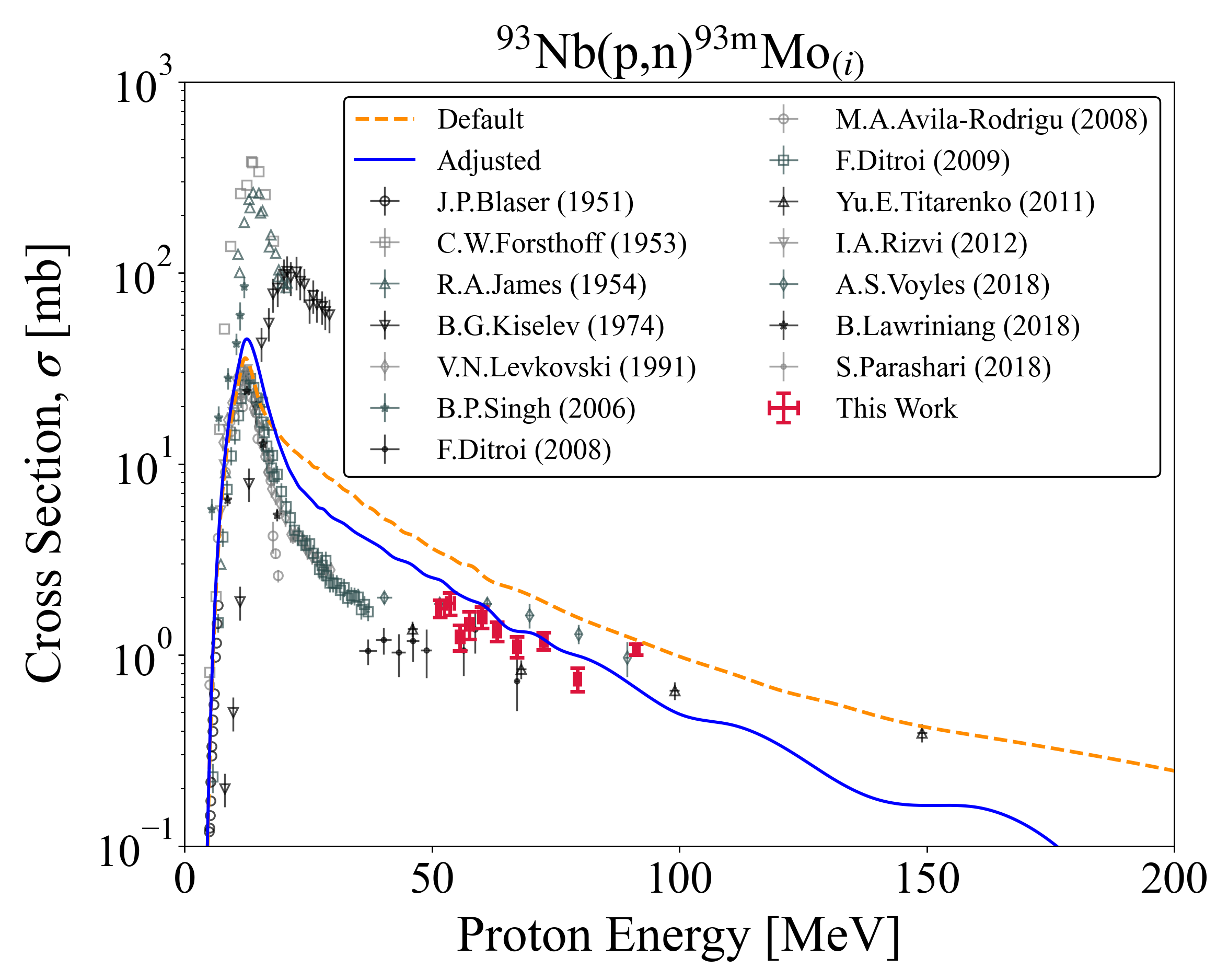}}
	\vspace{-0.65cm}
	\caption{TALYS default and adjusted calculation for $^{93\mathrm{m}}$Mo.}\label{talys45_93mMo}
\end{figure}

\begin{figure}[H]
		{\includegraphics[width=1.0\columnwidth]{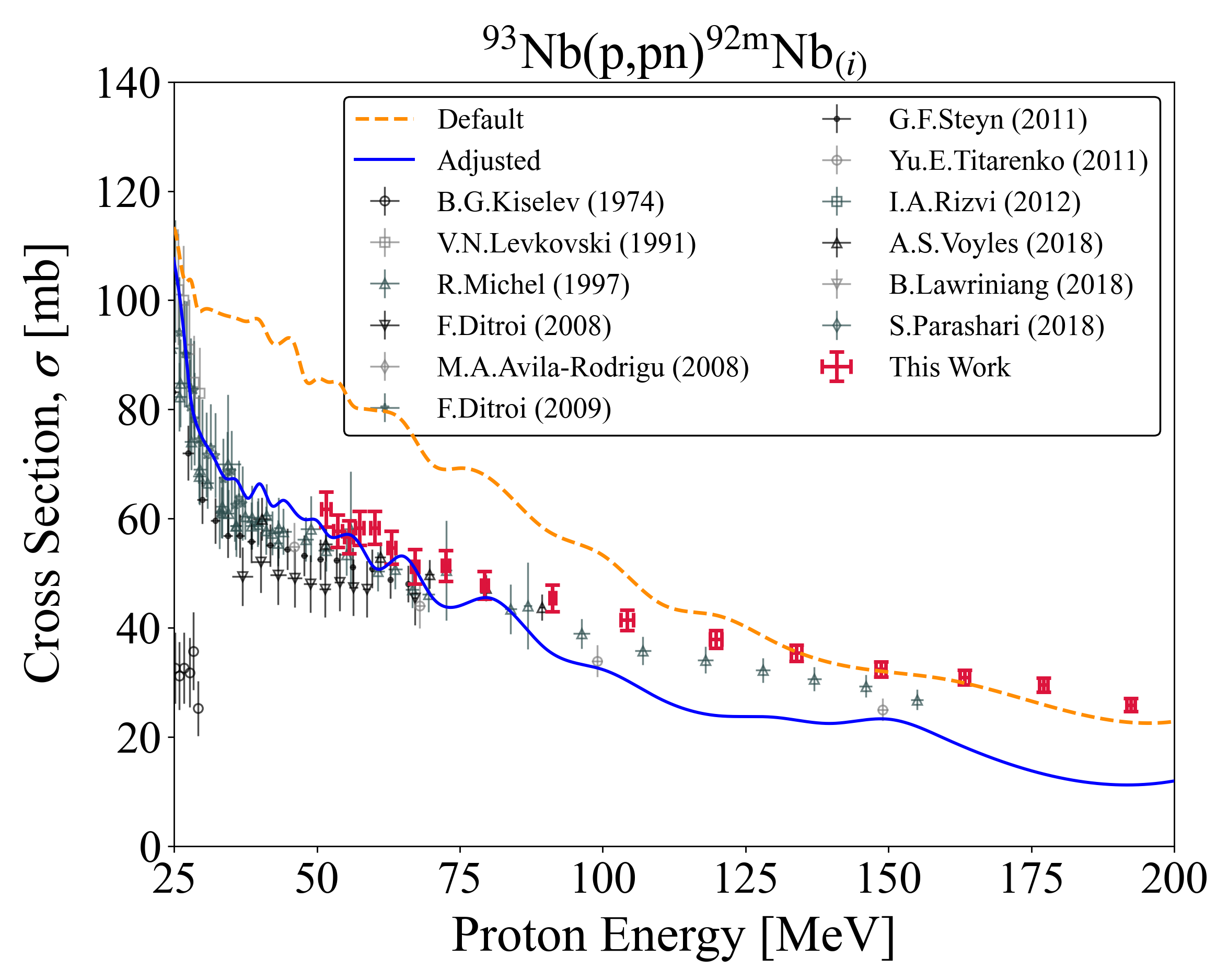}}
	\vspace{-0.65cm}
	\caption{TALYS default and adjusted calculation for $^{92\mathrm{m}}$Nb.}\label{talys45_92mNb}
		{\includegraphics[width=1.0\columnwidth]{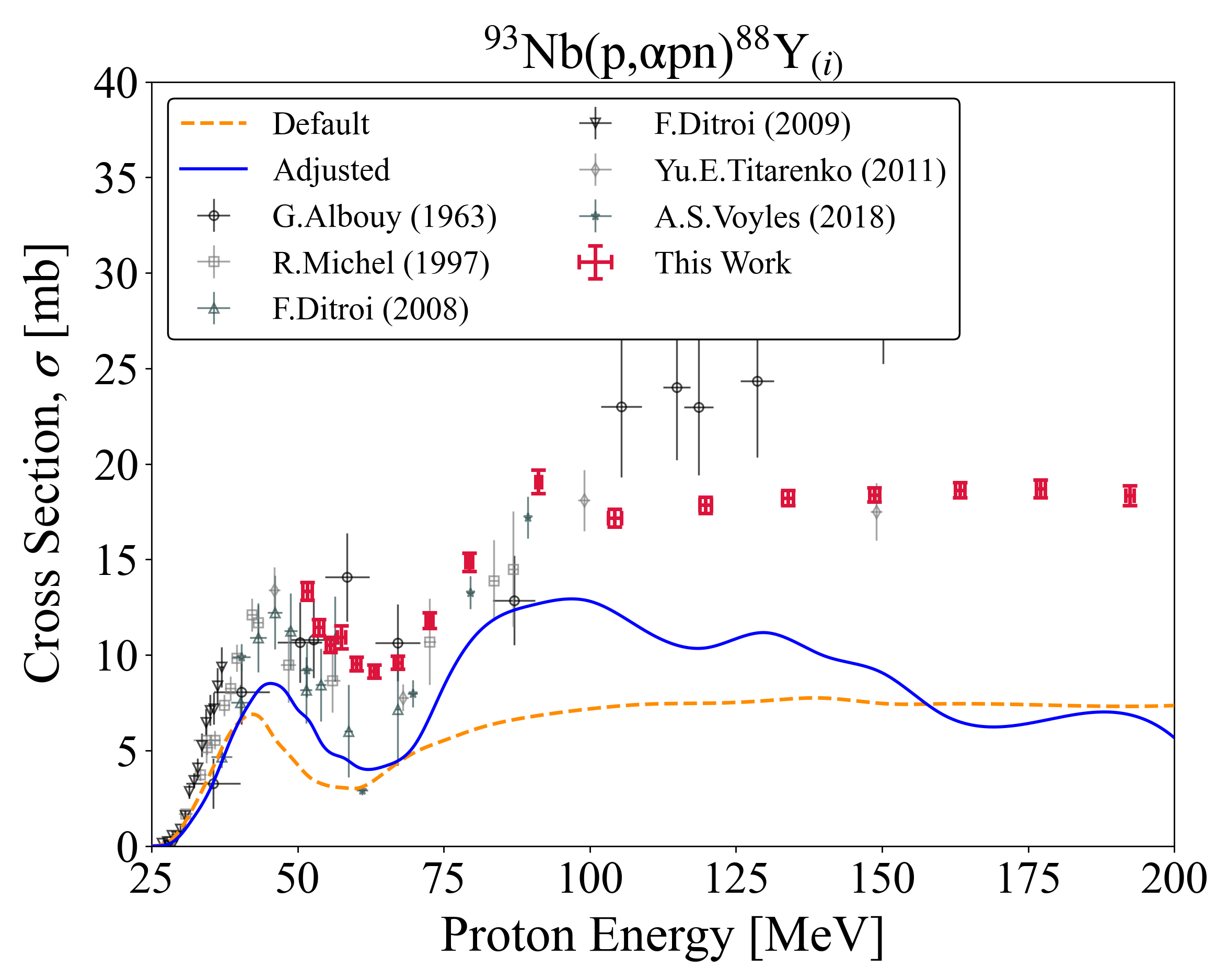}}
	\vspace{-0.65cm}
	\caption{TALYS default and adjusted calculation for $^{88}$Y.}\label{talys45_88Y}
		{\includegraphics[width=1.0\columnwidth]{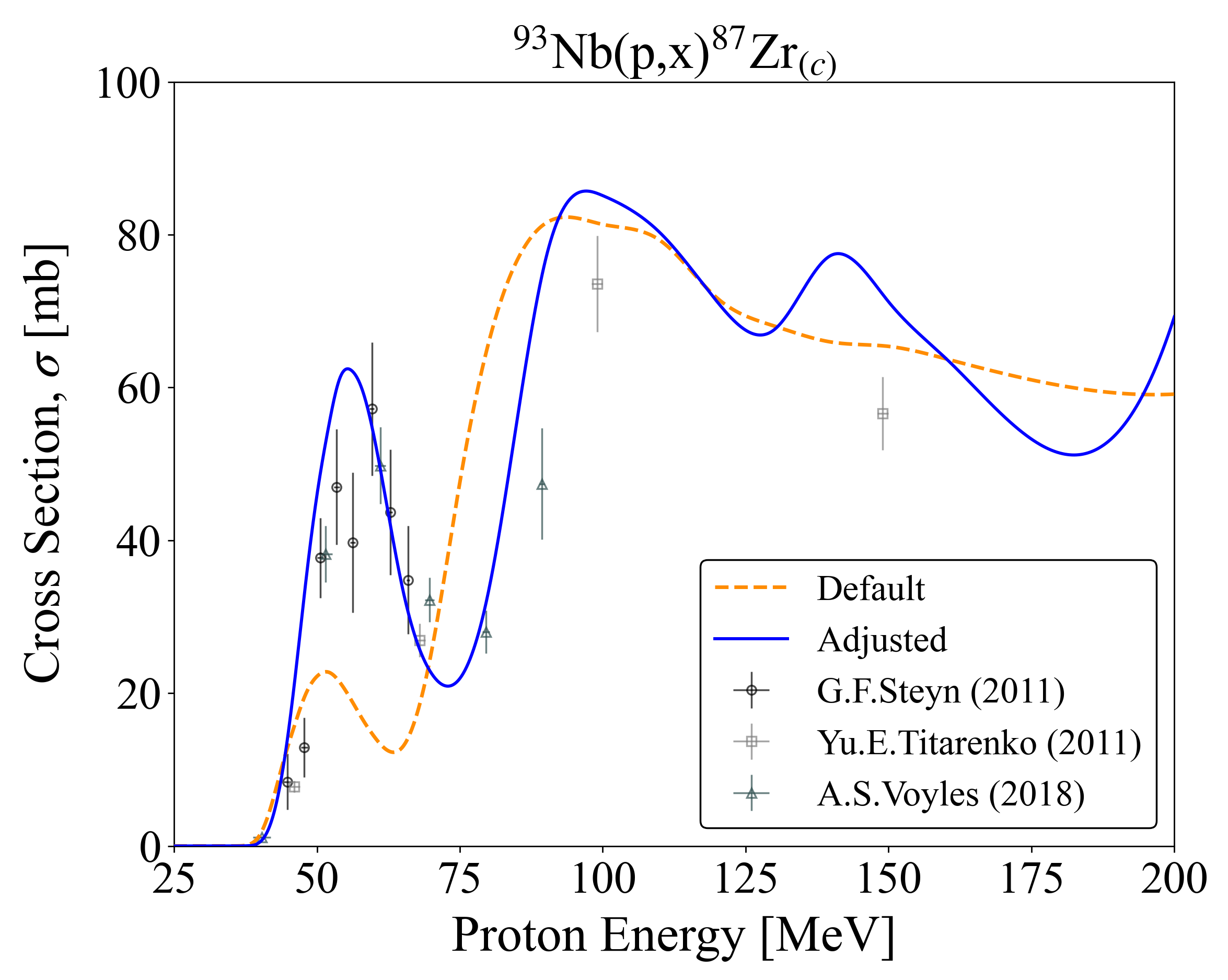}}
	\vspace{-0.65cm}
	\caption{TALYS default and adjusted calculation for $^{87}$Zr.}\label{talys45_87Zr}
\end{figure}

\begin{figure}[H]
		{\includegraphics[width=1.0\columnwidth]{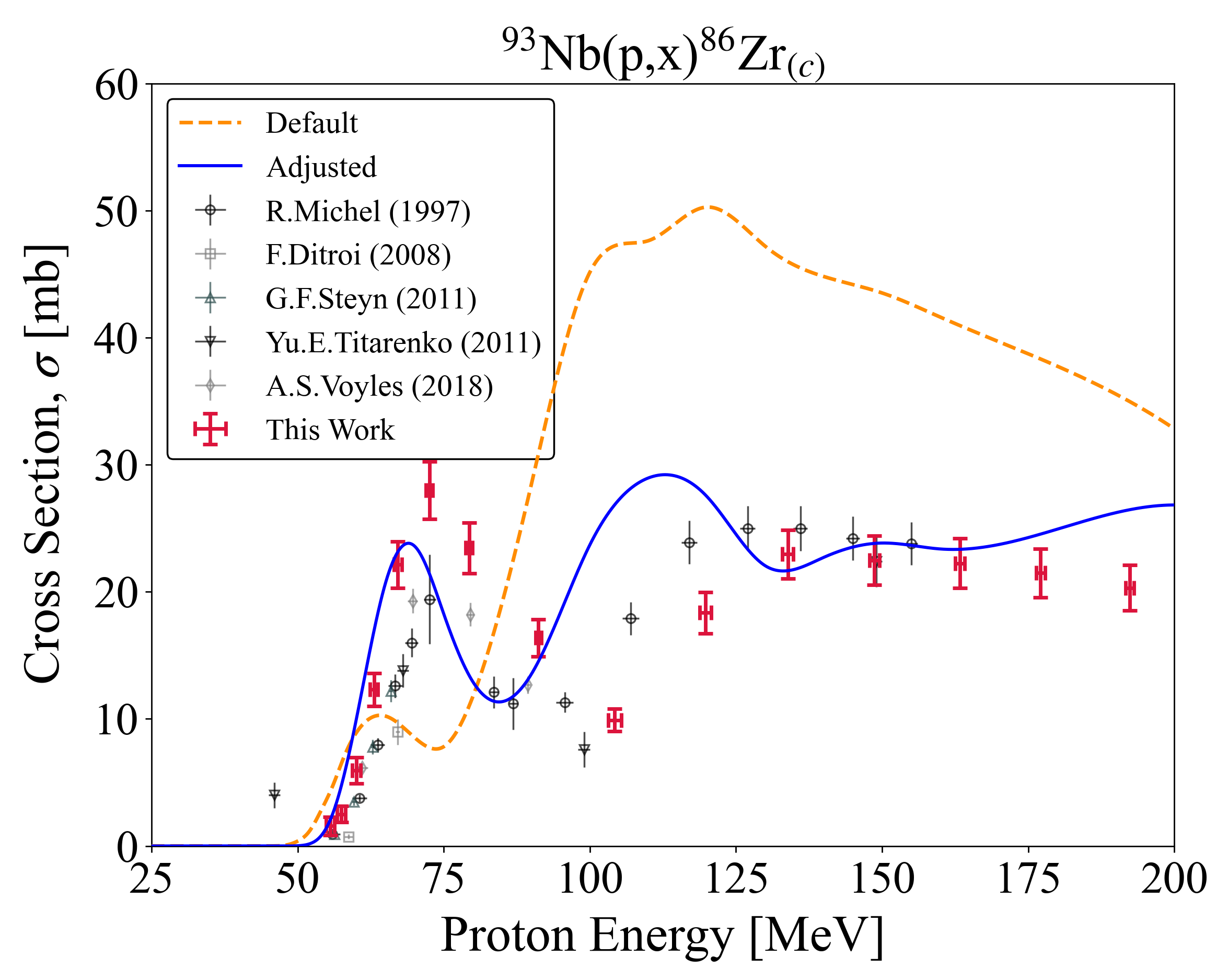}}
	\vspace{-0.65cm}
	\caption{TALYS default and adjusted calculation for $^{86}$Zr.}\label{talys45_86Zr}
		{\includegraphics[width=1.0\columnwidth]{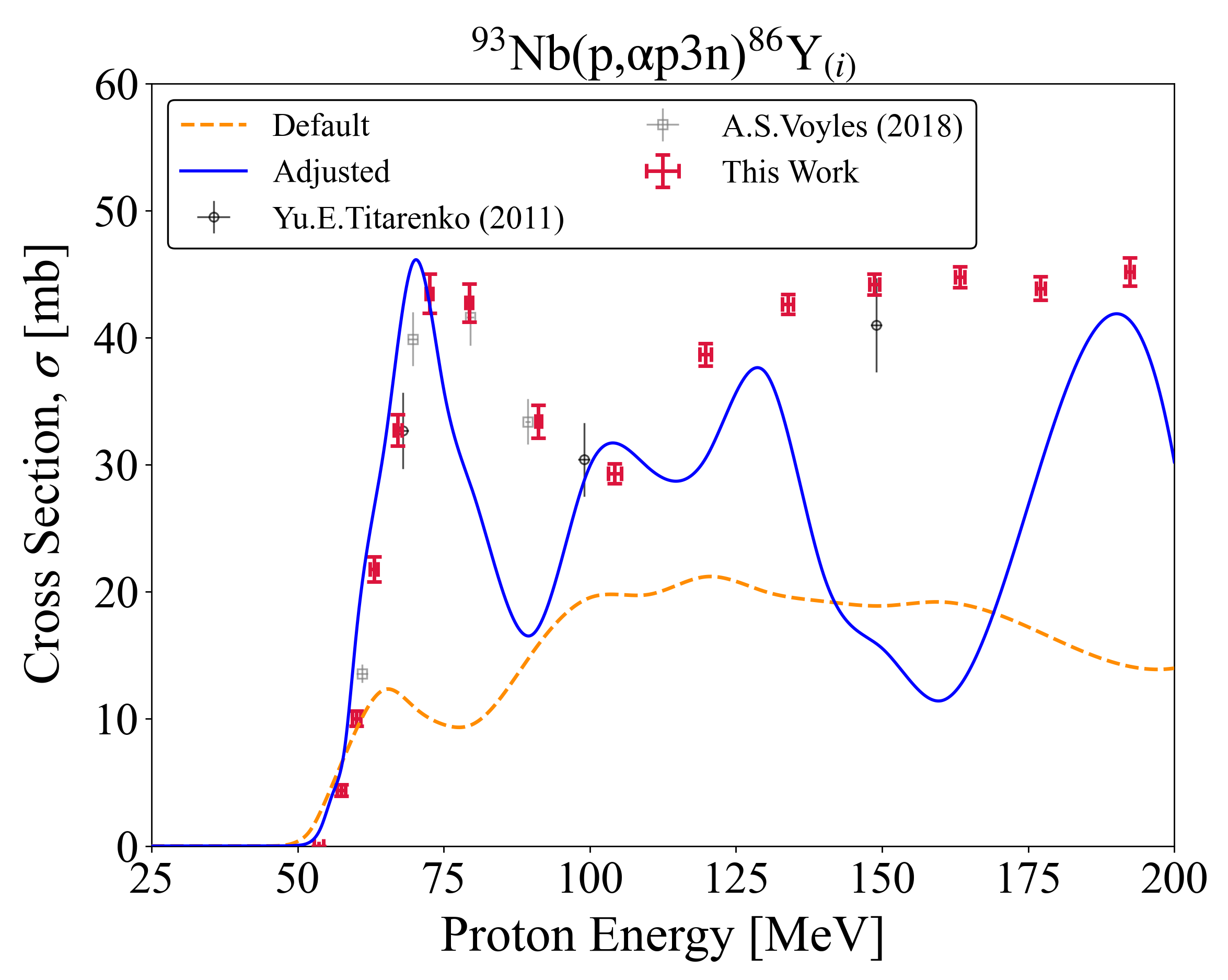}}
	\vspace{-0.65cm}
	\caption{TALYS default and adjusted calculation for $^{86}$Y.}\label{talys45_86Y}
\end{figure}

All of the final derived parameter changes for $^{93}$Nb(p,x) are listed in Appendix \ref{ParamsAdjustments} (Table \ref{NbAdjustedParams}). The adjusted fits accompanying this more detailed parameter study are shown compared to the default TALYS calculation for the nine considered reaction channels in Figures \ref{talys45_90Nb}--\ref{talys45_86Y}. The fits shown apply from 0 to 200 MeV.

\newpage
\newpage
\newpage

\subsubsection{\label{NbValidation}Parameter Adjustment Validation}
\begin{figure}[H]
		{\includegraphics[width=1.0\columnwidth]{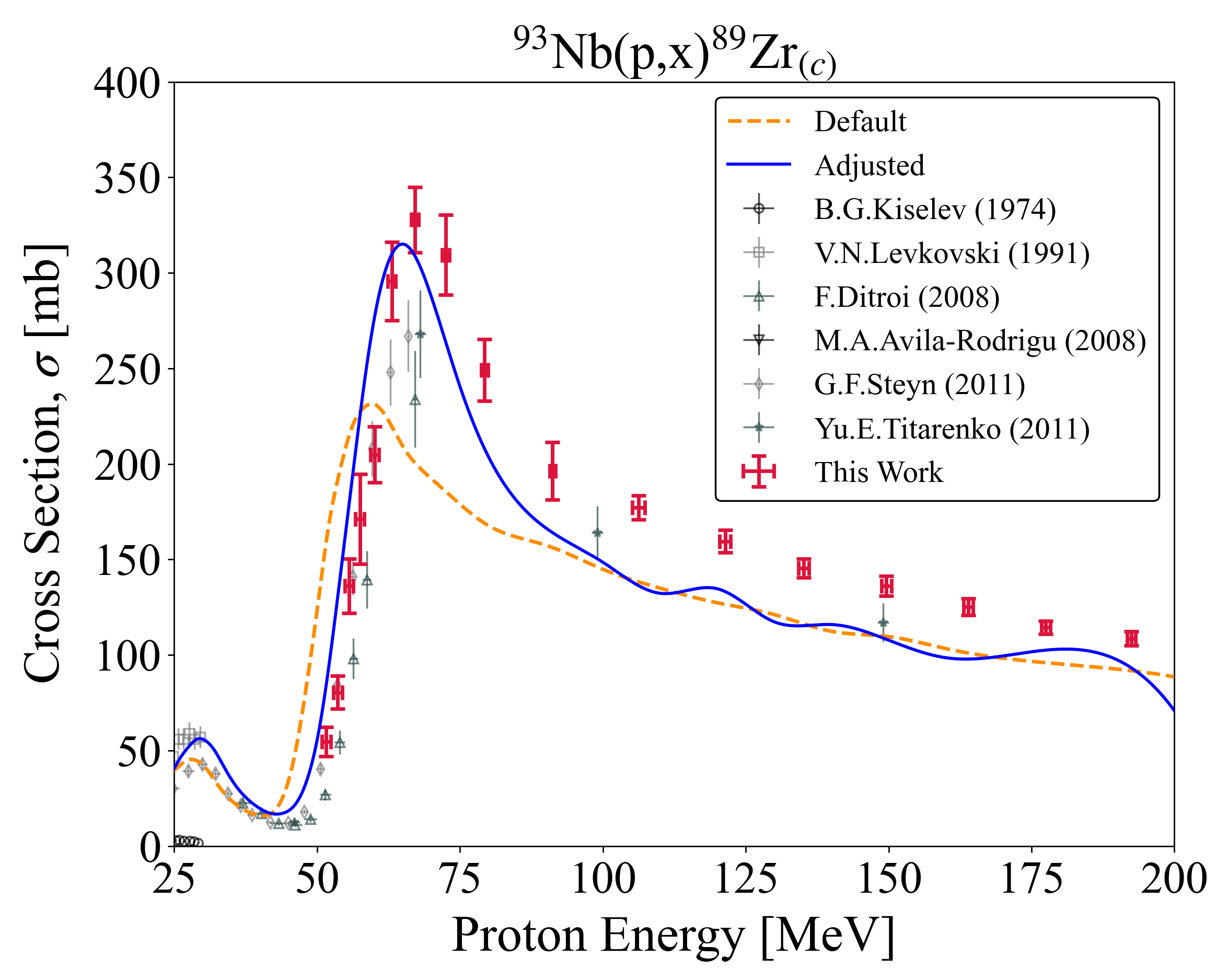}}
	\vspace{-0.65cm}
	\caption{TALYS default and adjusted extended to $^{89}$Zr.}\label{talys45_89Zr}
		{\includegraphics[width=1.0\columnwidth]{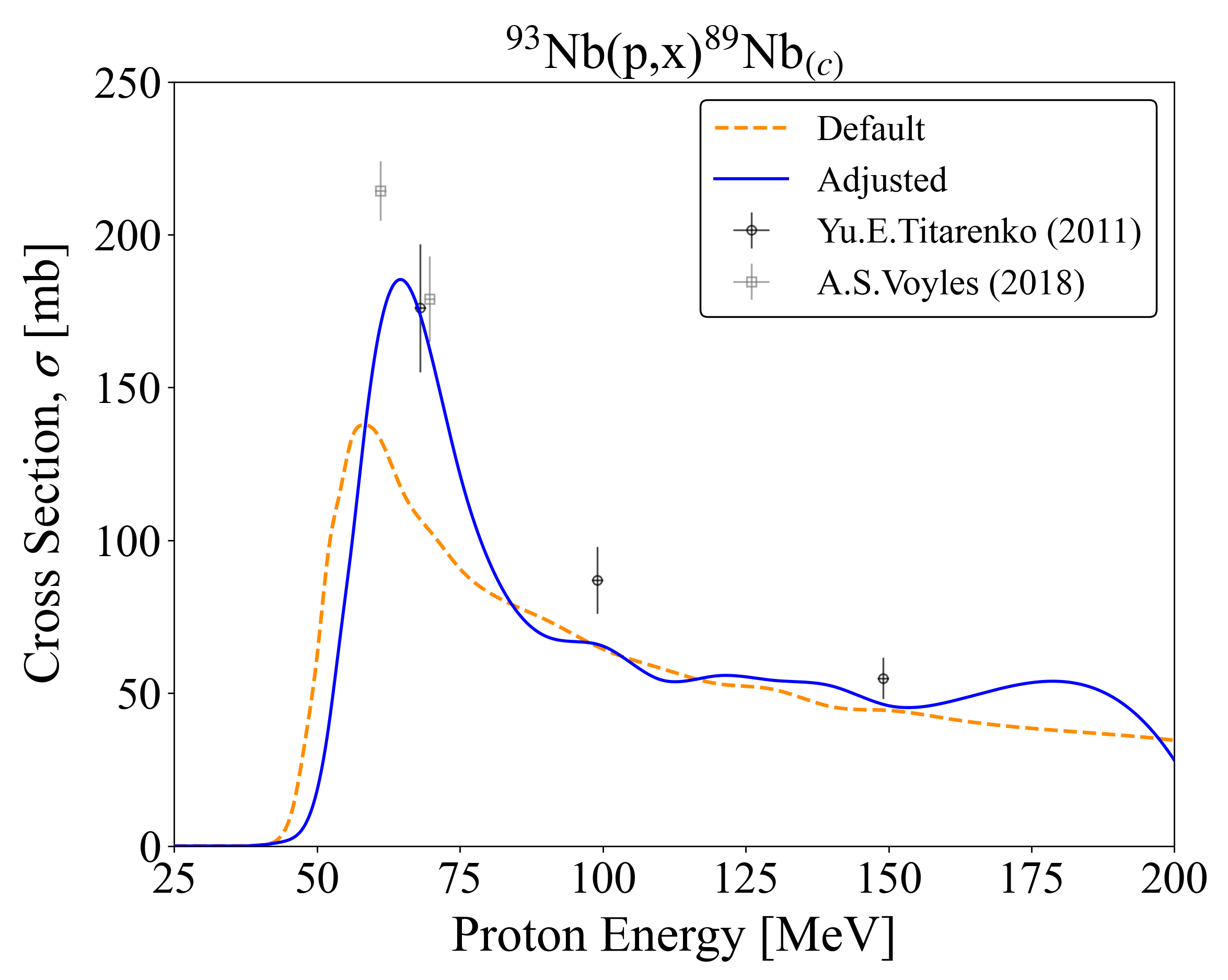}}
	\vspace{-0.65cm}
	\caption{TALYS default and adjusted extended to $^{89}$Nb.}\label{talys45_89Nb}
		{\includegraphics[width=1.0\columnwidth]{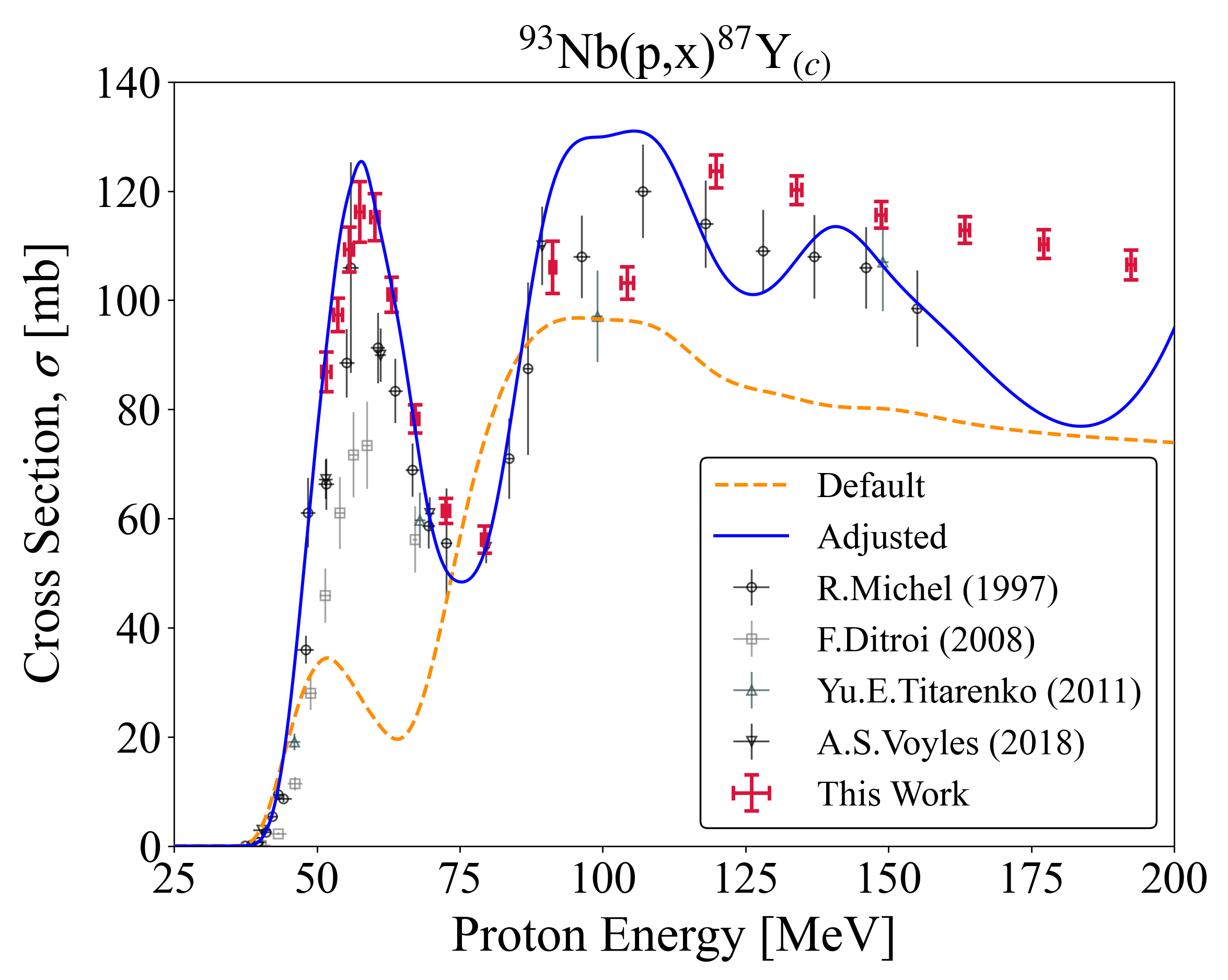}}
	\vspace{-0.65cm}
	\caption{TALYS default and adjusted extended to $^{87}$Y.}\label{talys45_87Y}
\end{figure}

\begin{figure}[H]
		{\includegraphics[width=1.0\columnwidth]{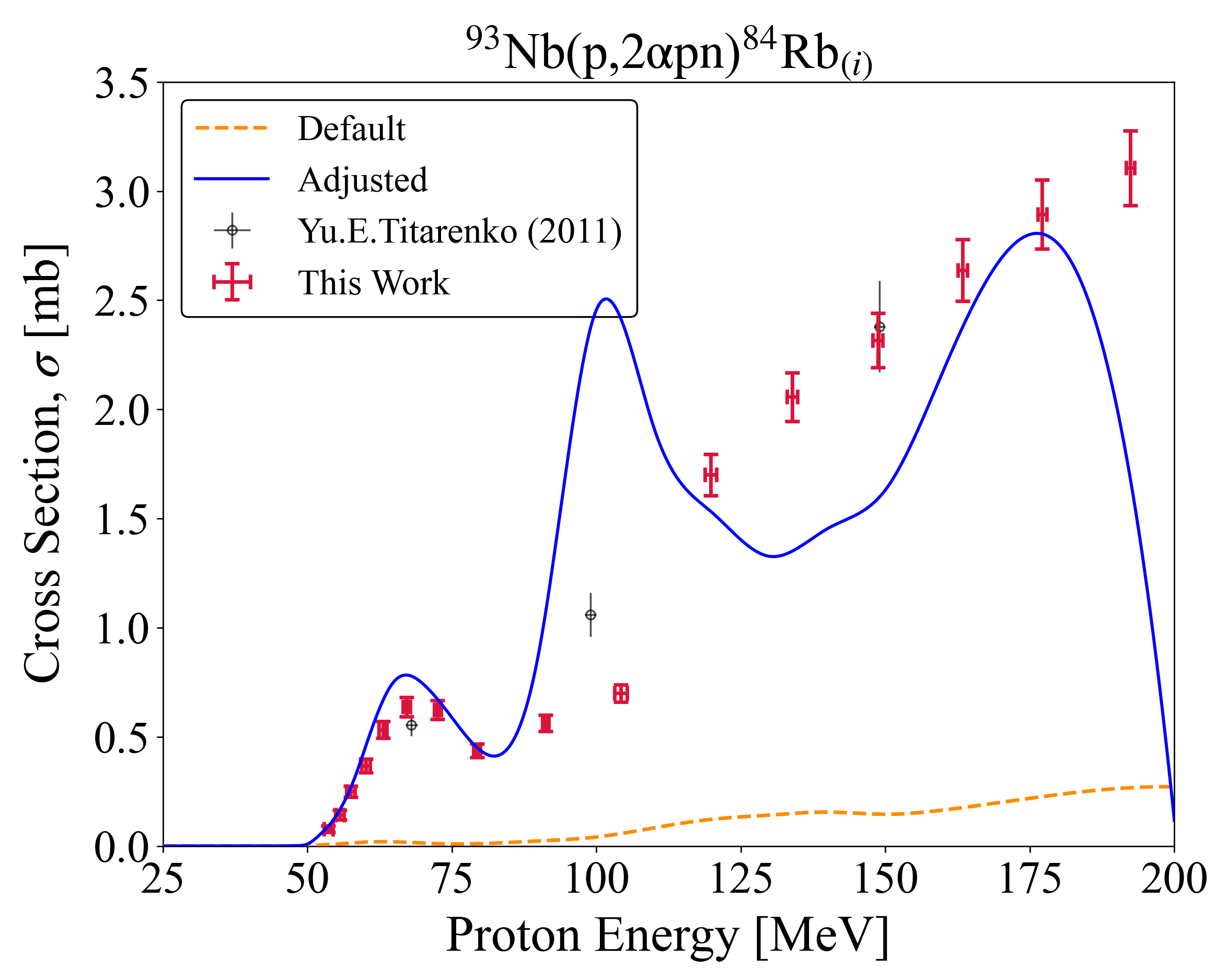}}
	\vspace{-0.65cm}
	\caption{TALYS default and adjusted extended to $^{84}$Rb.}\label{talys45_84Rb}
\end{figure}

A crucial aspect in this suggested approach is validation of the derived parameters to ensure that it is justified to attribute physical meaning to their values. The $^{93}$Nb(p,x)$^{89}$Zr,$^{89}$Nb,$^{87}$Y,$^{84}$Rb reaction channels, with all but $^{84}$Rb being cumulative data, were used for this purpose. The adjusted fit shown in Figures \ref{talys45_89Zr}--\ref{talys45_84Rb} continues to show improved behaviour over the default in these cases, especially in the compound peak regions.

\newpage

The total chi-squared, $\chi^2_{tot}$, used to compare the default and adjusted TALYS fit across all utilized and validation channels is given by:
\begin{gather}
\chi^2_{tot}=\frac{1}{N_c}\sum^{N_c}_{c=1}\chi^2_c w_c,
\end{gather}
where $N_c$ is the number of reaction channels considered, $\chi^2_c$ is the chi-squared value per channel, and $w_c$ is the weighting per channel \cite{Alhassan2019:ProtonsBestFitChiSquared,Koning2015:BayesianEval}. Each $\chi^2_c$ is defined by:
\begin{gather}
\chi^2_{c}=\frac{1}{N_p}\sum^{N_p}_{i=1}\left(\frac{\sigma_T^i-\sigma_E^i}{\Delta\sigma_E^i}\right)^2,
\end{gather}
where $N_p$ is the number of data points from all experimental datasets in a given channel, $\sigma_E^i$ are the experimental cross sections with $\Delta\sigma_E^i$ uncertainty, and $\sigma_T^i$ is the TALYS cross section calculation \cite{Alhassan2019:ProtonsBestFitChiSquared,Koning2015:BayesianEval}. No exclusions or preference was given to the quality of data beyond weighting by uncertainty, which is in opposition to techniques typically used in an evaluation \cite{Alhassan2019:ProtonsBestFitChiSquared,Alhassan2020:BayesianMCEval}. Two weighting calculations were considered in this application, both of which tried to emphasize the importance of fits to the most prominent channels. One weighting methodology is to use the cumulative cross section of the TALYS calculation in a given channel relative to the sum of all channels' cumulative cross sections:
\begin{gather}
w_c=\frac{\sum_{i=1}^{N_p}\sigma^{ci}_T(E)}{\sum_{c=1}^{N_c}\sum_{i=1}^{N_p}\sigma^{ci}_T(E)}.
\end{gather}
The above ``Cumulative $\sigma$" weighting potentially poses a risk of washing out the importance of large compound peaks that were significant to parameter adjustment studies but fall off at high energies such as the case with $^{90}$Mo production. This issue could be resolved with an alternative ``Maximum $\sigma$" weighting that considers the maximum production cross section reached in each channel relative to the sum of all channels' maximums:
\begin{gather}
w_c=\frac{\sigma^{c}_{T,max}}{\sum_{c=1}^{N_c}\sigma^{c}_{T,max}}.
\end{gather}
The $\chi^2_{tot}$ results based on both weighting methods are given in Table \ref{chi2Nb}. In this case both weighting techniques yield similar results, which clearly show that the adjusted parameters fit performs much better for high-energy proton-induced reactions on niobium than the default prediction. Ultimately, this more realistic analysis method, even as a manual search, has produced a fit with a better performance than the default calculations with a justifiable limited set of parameter changes built from measured experimental data. This analysis is therefore an improved standard over the one-channel adjustment norm and can be a reasonable expectation for future parameter optimization data work.

\vspace{-0.2cm}
\begin{table}[H]
\caption{Global $\chi^2$ metric describing goodness-of-fit for the default and adjusted TALYS calculations of $^{93}$Nb(p,x).}
\label{chi2Nb}
\begin{ruledtabular}
\begin{tabular}{lcc}
Weighting Method&Default $\chi^2_{tot}$ &Adjusted $\chi^2_{tot}$ \\[0.1cm]
\hline\\[-0.25cm]
Cumulative $\sigma$     &15.6 & 3.37\\[0.1cm]

Maximum $\sigma$        &16.0 & 3.28\\
\end{tabular}
\end{ruledtabular}
\end{table}
\vspace{-0.3cm}

\subsection{\label{LaFit}Fitting Procedure Applied to $^{139}$La(p,x)}
The same fitting approach detailed for niobium was also applied to high-energy proton-induced reactions on lanthanum. Eight reaction channels were used in the study: $^{139}$La(p,x)$^{137\textnormal{m},137\textnormal{g},135,134,133\textnormal{m},132}$Ce, $^{135}$La, $^{133\textnormal{m}}$Ba, with $^{135}$Ce, $^{134}$Ce, $^{137\textnormal{m}}$Ce, and $^{135}$La production as the most prominent.

The cross section data for $^{139}$La(p,x) are more limited than what was available in the niobium case. These eight channels only contain the three datasets of \textcite{Tarkanyi2017:La}, \textcite{Becker2020:protonsLa}, and \textcite{Morrell2020}, with the latter two characterizations utilizing stacked-target activation at LANL and LBNL, respectively, consistent with the work performed here.

In addition to a sparser body of data, there is a limited diversity of reaction products, where only the $^{135}$La production gives insight into proton emission behaviour and only the $^{133\textnormal{m}}$Ba production gives insight into alpha emission behaviour. The measured cerium channels, comprising the bulk of the available data, are solely (p,xn) reactions. That being said, the restricted dataset makes $^{139}$La(p,x) a valuable application of the suggested fitting procedure as it can show the amount of predictive power that can be gained even from reactions that are being partially measured for the first time.

Note that the default TALYS calculations for lanthanum were significantly better than for niobium, whose dominant channels were predicted with extremely discrepant shapes, magnitudes, and positioning from the experimental data. As a result, the amount of parameter adjustments, fine tuning, and iteration needed to properly model the niobium can be considered higher than typical.

Firstly, the application of microscopic level densities over phenomenological ones in the lanthanum calculations provided immediate benefit, matching the observed rising edges and shapes of the dominant $^{135}$Ce and $^{134}$Ce compound peaks quite well. Similarly to the niobium, \texttt{ldmodel} 4 performed best and was chosen, though there was no apparent constraining residual product in this case and \texttt{ldmodel} 5 was a close next best choice.

The pre-equilibrium portion of the procedure revealed a need for adjustments of \texttt{M2constant}=0.85, \texttt{M2limit}=2.5, and \texttt{M2shift}=0.9 to the exciton model matrix parameterization. It should be noted that these parameters are all shifted in the same directions as in the niobium case, simply to a lesser extent, which emphasizes the better initial default guess here. A last additional pre-equilibrium change also included \texttt{Cstrip\,a}=2.0, where \texttt{Cstrip a} affects the transfer reaction contribution of $\mathrm{(p,\alpha)}$ to the overall pre-equilibrium cross section. This helps to increase $^{133\textnormal{m}}$Ba production without much noticeable effect to the other considered channels.

For OMP fine tuning, the $^{135}$La and $^{133\textnormal{m}}$Ba channels necessarily played important roles due to their particle emission diversity. The prevailing discrepancies in these two channels at this point included a slight overprediction of $^{135}$La production and a minor underprediction of the $^{133\textnormal{m}}$Ba compound peak falling edge. A testing of the available TALYS OMP parameters demonstrated that \texttt{rvadjust p} and \texttt{rvadjust a} held the most sensitivity. The most accurate behaviour was extracted solely using \texttt{rvadjust p}=0.96. Finally, there was a small local competition error between $^{135}$Ce and $^{134}$Ce that could be corrected by a \texttt{ctable} increase to $^{135}$Ce. There were far fewer confounding level density changes for the lanthanum relative to the niobium.

\begin{figure}[H]
		{\includegraphics[width=1.0\columnwidth]{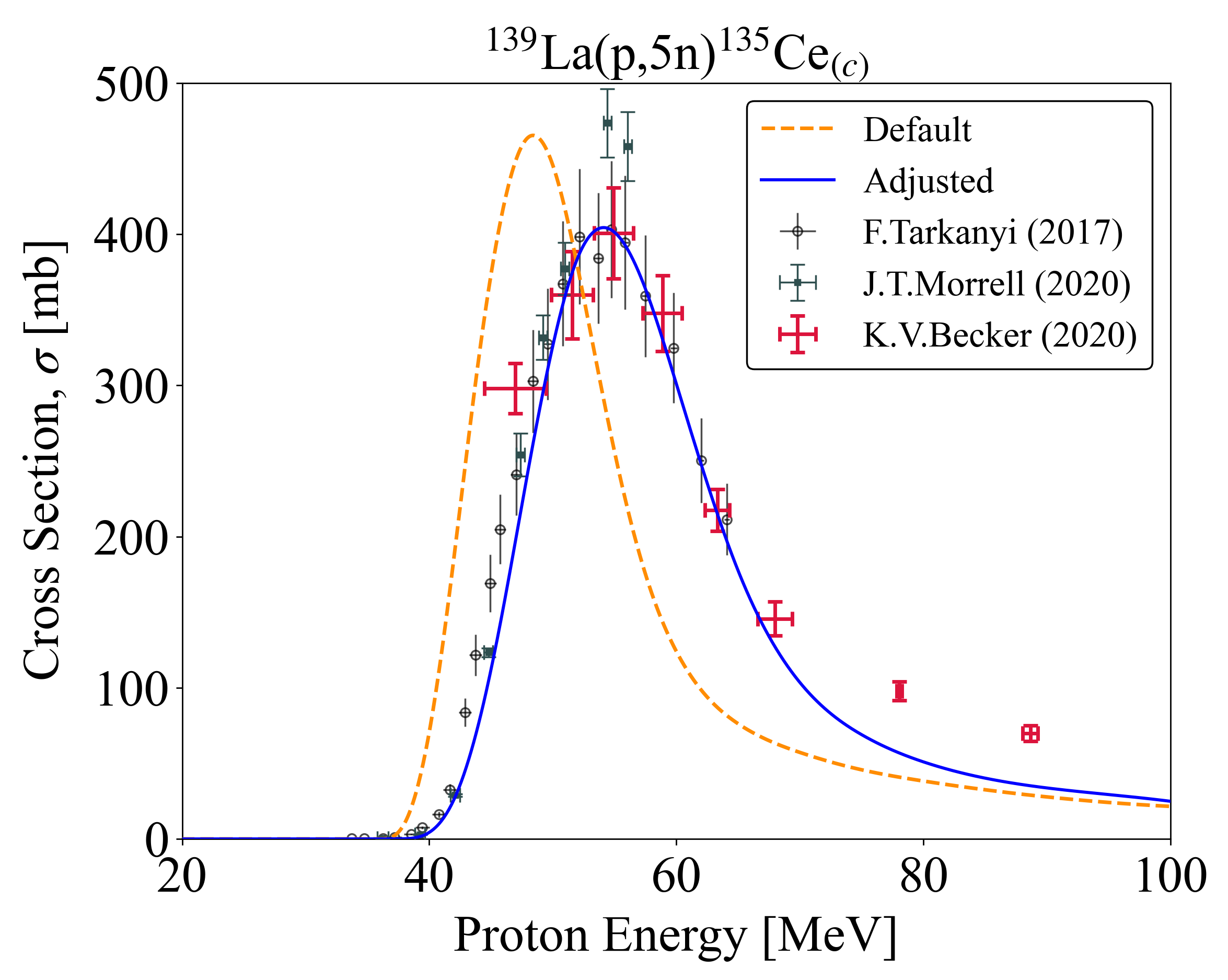}}
	\vspace{-0.65cm}
	\caption{TALYS default and adjusted calculation for $^{135}$Ce.}\label{135Ce}
		{\includegraphics[width=1.0\columnwidth]{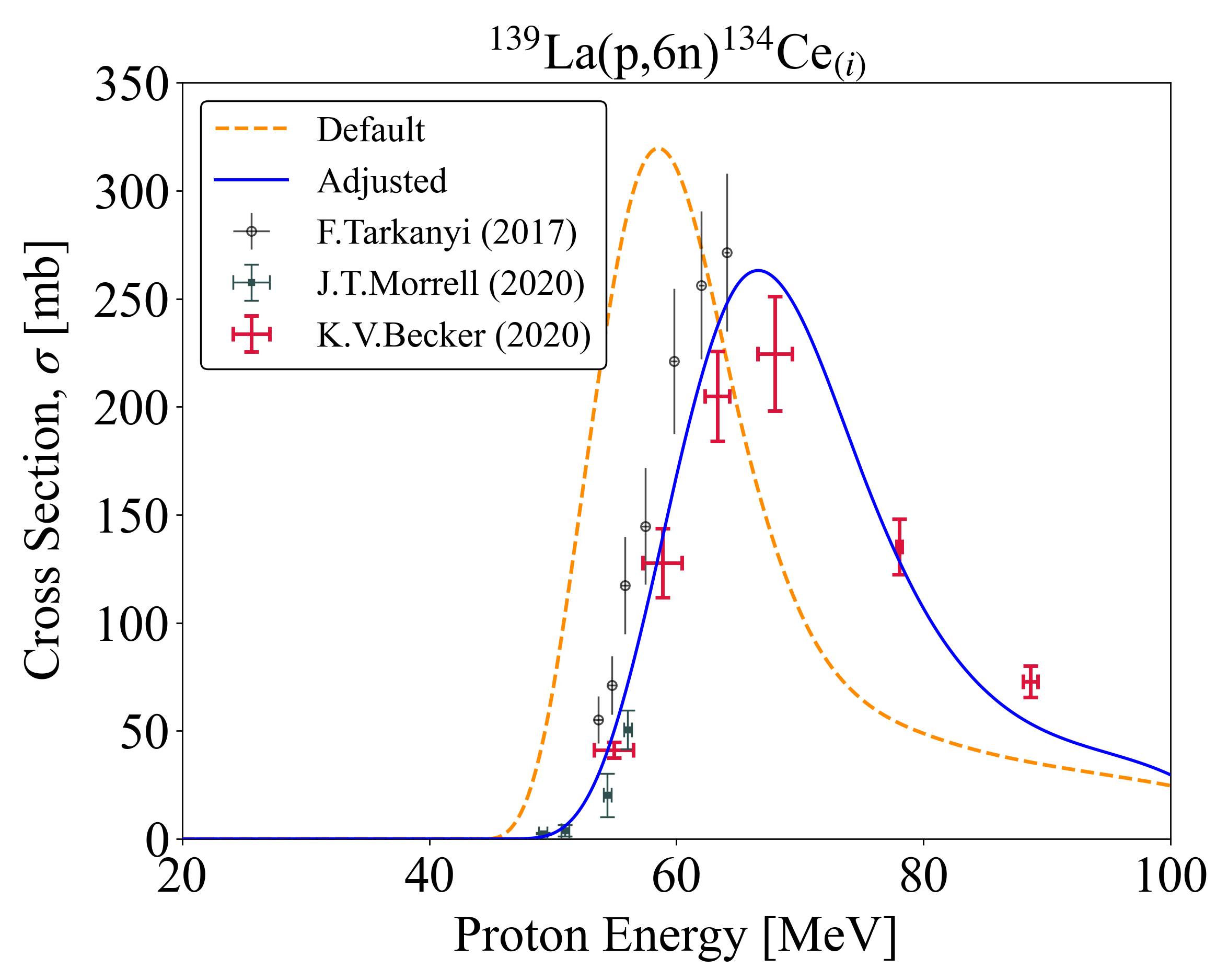}}
	\vspace{-0.65cm}
	\caption{TALYS default and adjusted calculation for $^{134}$Ce.}\label{134Ce}
		{\includegraphics[width=1.0\columnwidth]{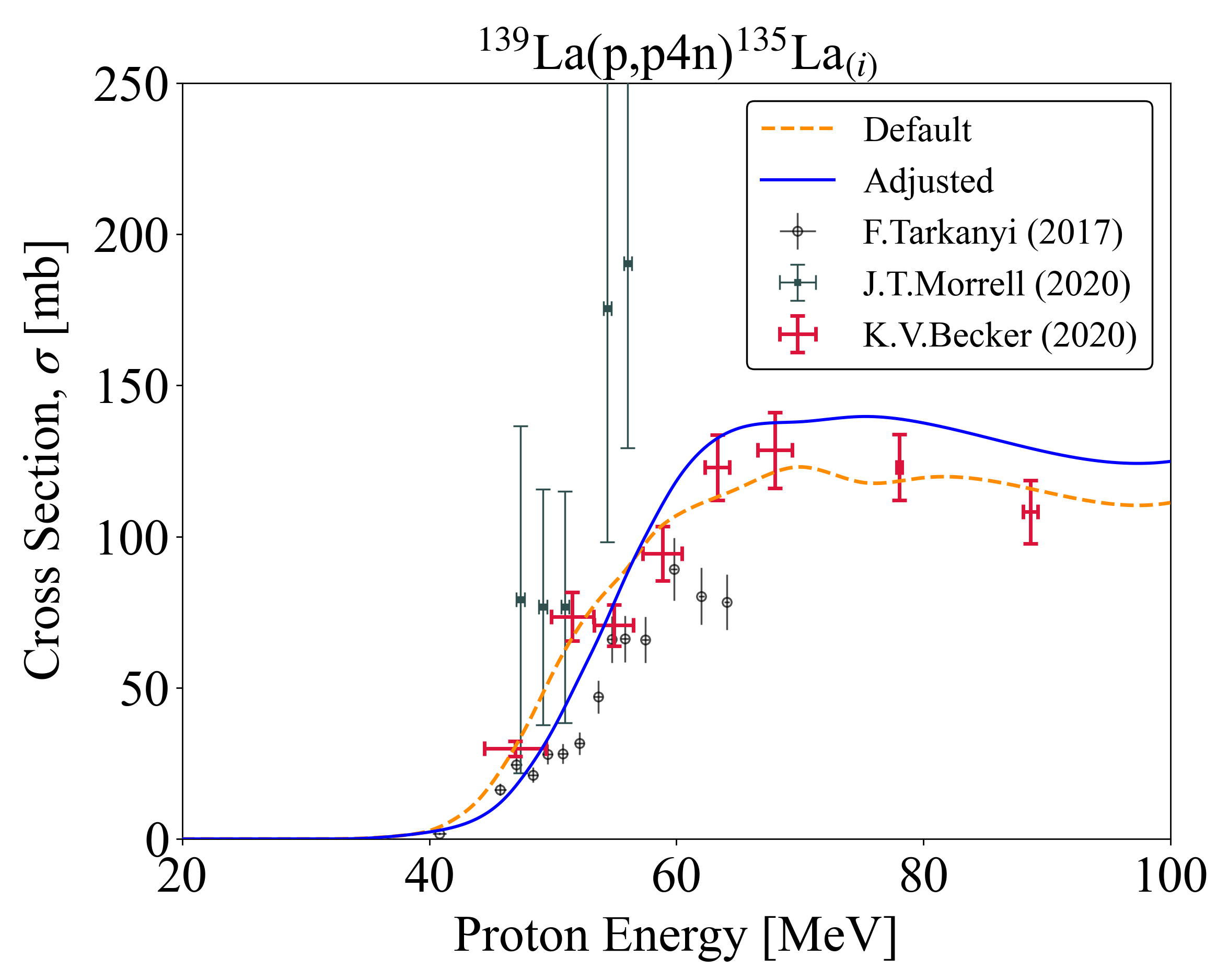}}
	\vspace{-0.65cm}
	\caption{TALYS default and adjusted calculation for $^{135}$La.}\label{135La}
\end{figure}

\begin{figure}[H]
		{\includegraphics[width=1.0\columnwidth]{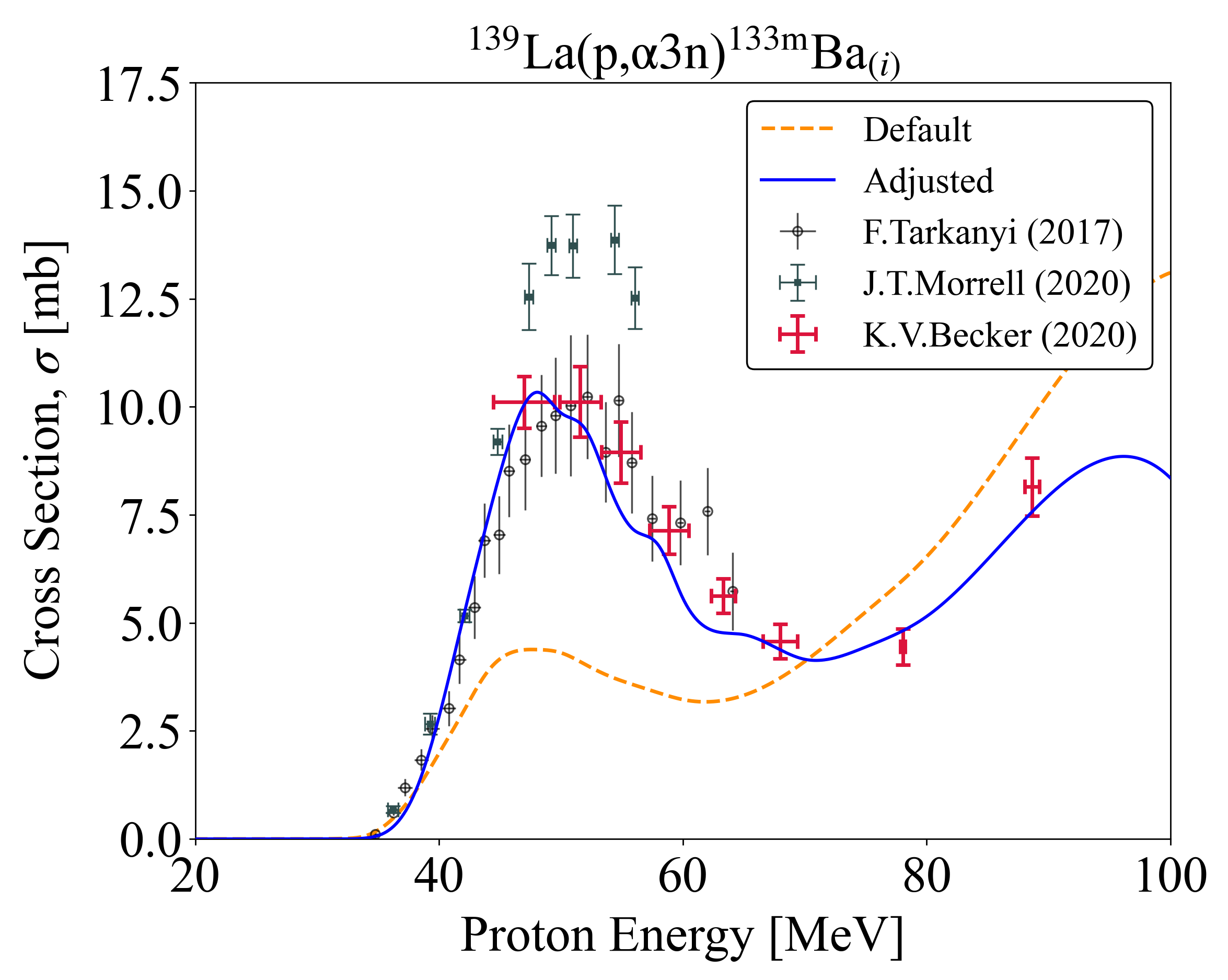}}
	\vspace{-0.65cm}
	\caption{TALYS default and adjusted calculation for $^{133\mathrm{m}}$Ba.}\label{133mBa}
		{\includegraphics[width=1.0\columnwidth]{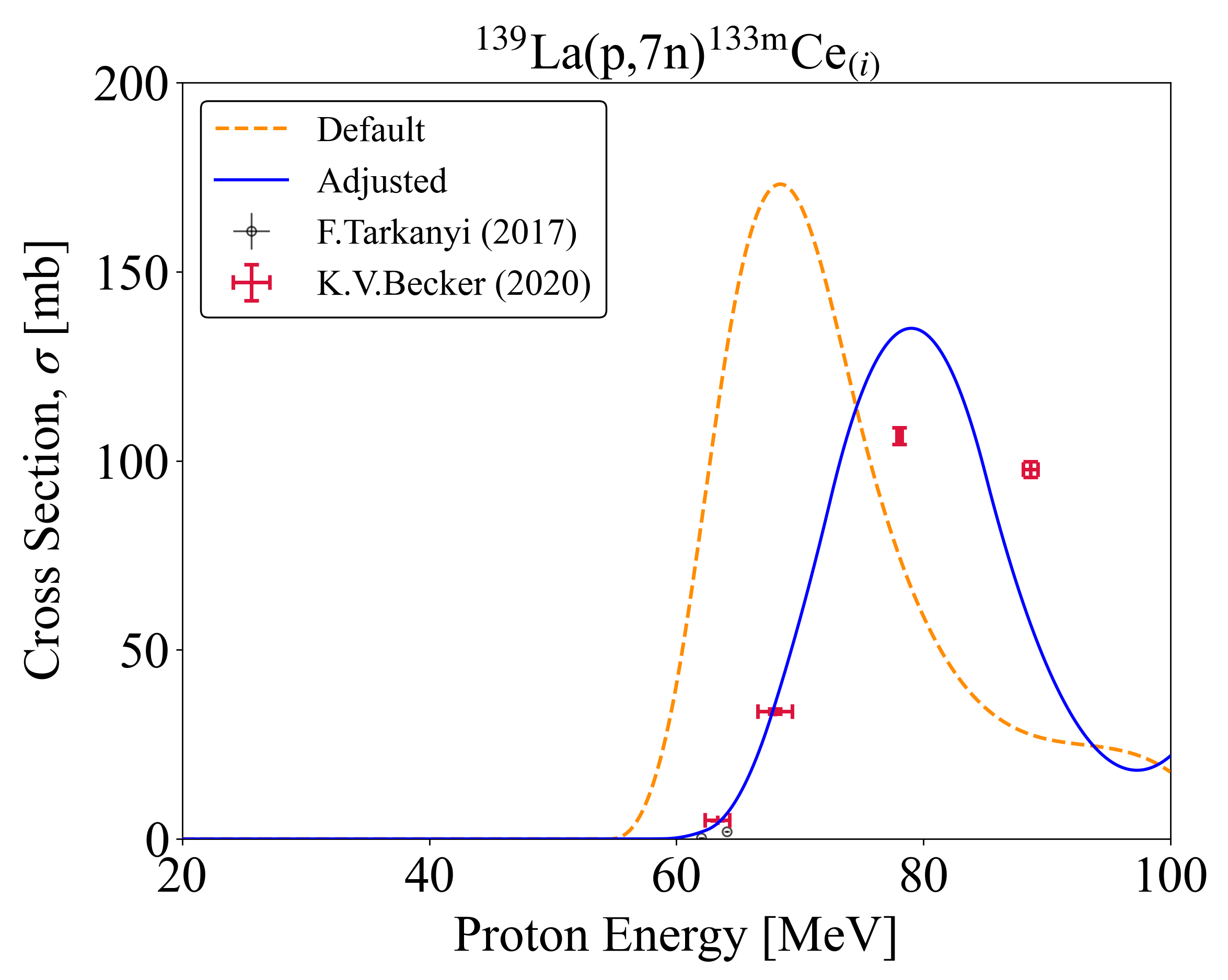}}
	\vspace{-0.65cm}
	\caption{TALYS default and adjusted calculation for $^{133\mathrm{m}}$Ce.}\label{133mCe}
		{\includegraphics[width=1.0\columnwidth]{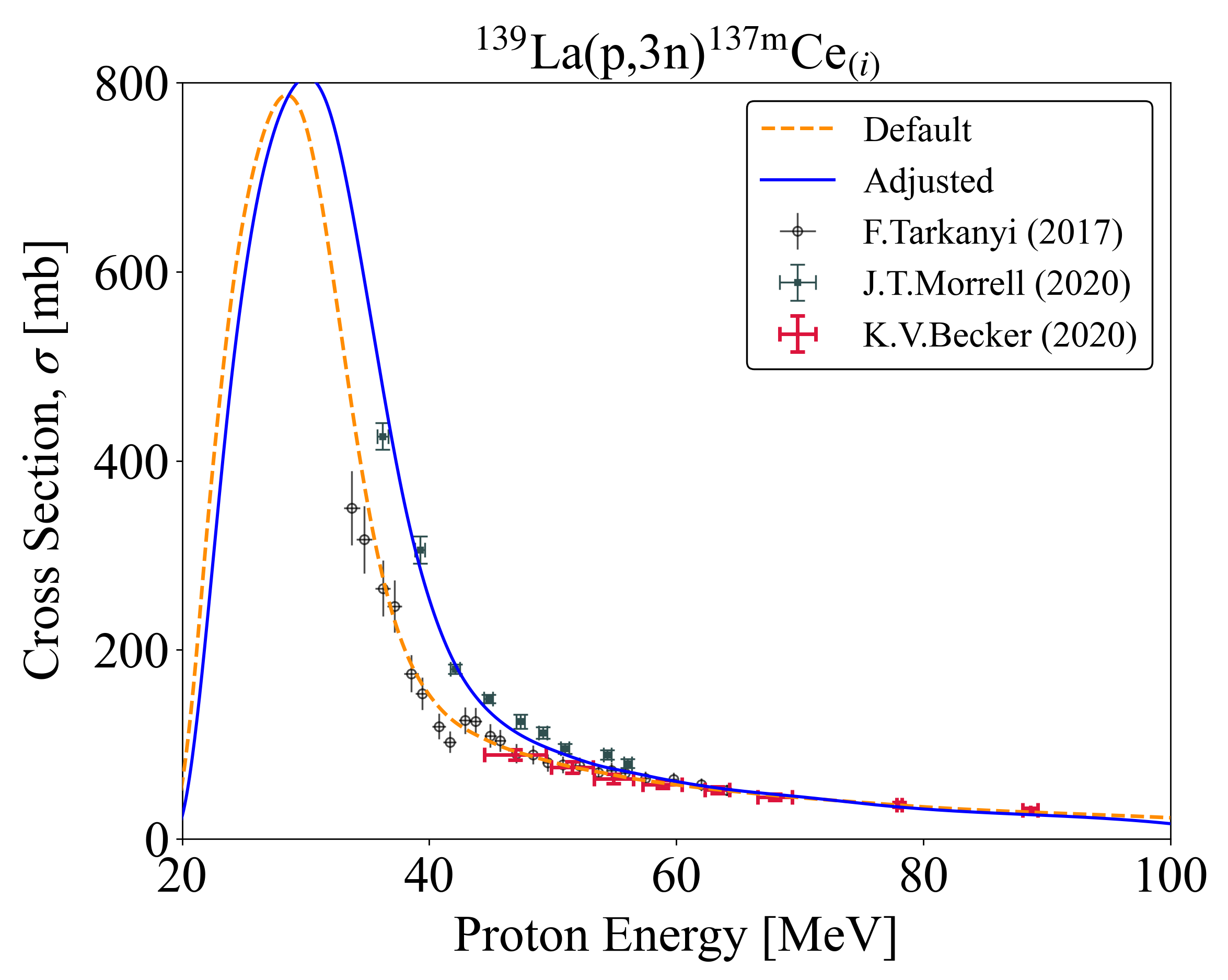}}
	\vspace{-0.65cm}
	\caption{TALYS default and adjusted calculation for $^{137\mathrm{m}}$Ce.}\label{137mCe}
\end{figure}

\begin{figure}[H]
		{\includegraphics[width=1.0\columnwidth]{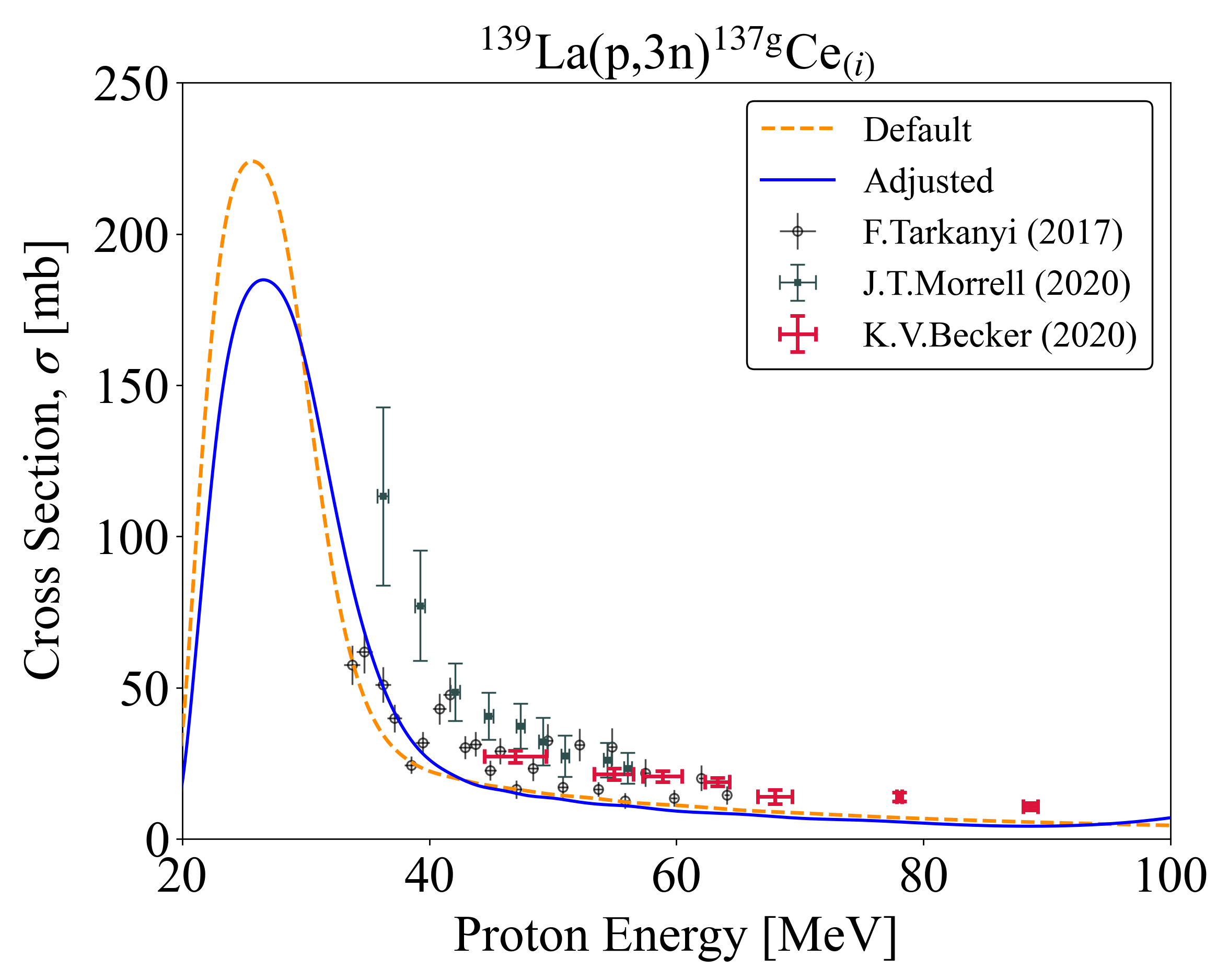}}
	\vspace{-0.65cm}
	\caption{TALYS default and adjusted calculation for $^{137\mathrm{g}}$Ce.}\label{137gCe}
		{\includegraphics[width=1.0\columnwidth]{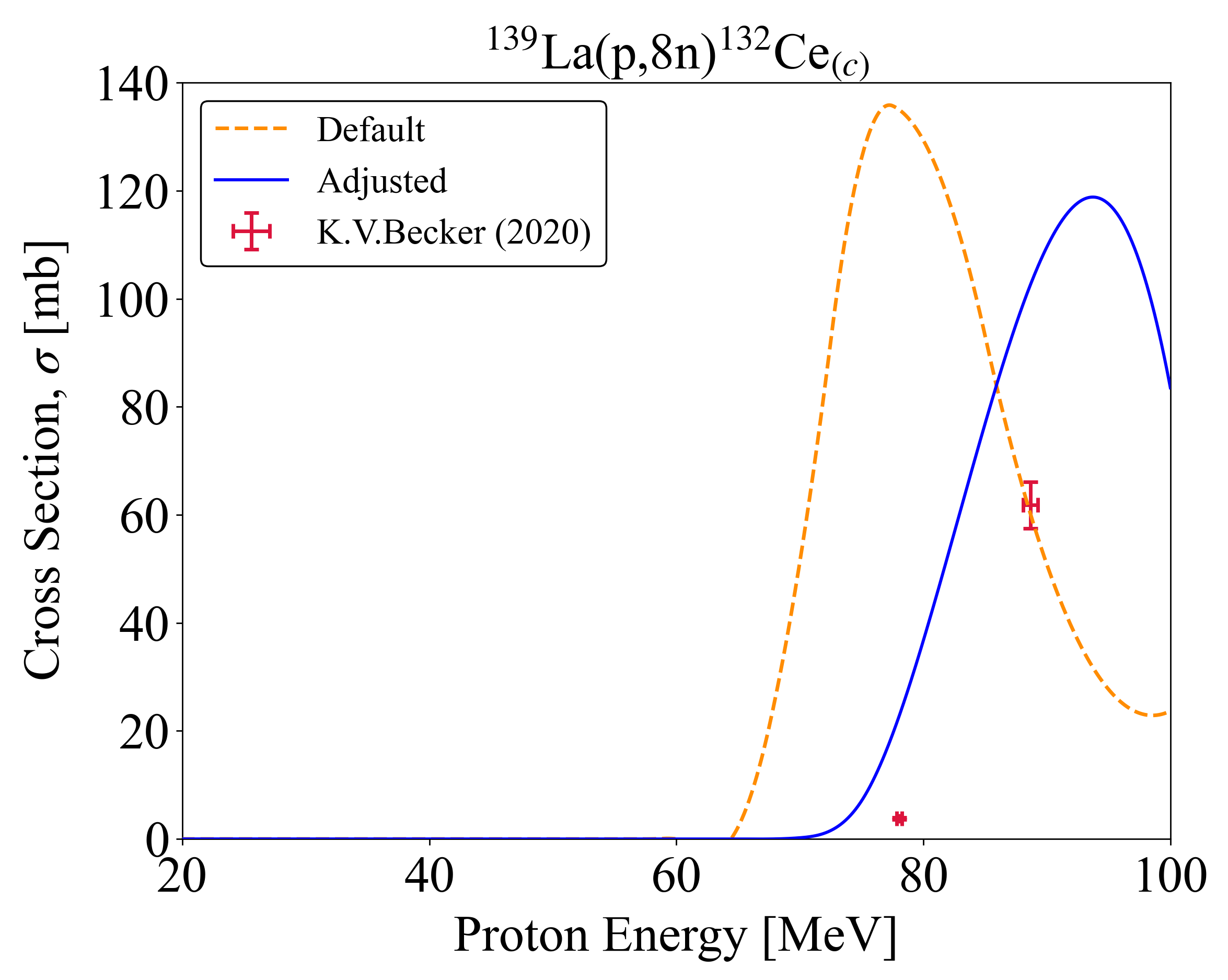}}
	\vspace{-0.65cm}
	\caption{TALYS default and adjusted calculation for $^{132}$Ce.}\label{132Ce}
\end{figure}

The total derived parameter changes for $^{139}$La(p,x) are listed in Appendix \ref{ParamsAdjustments} (Table \ref{LaAdjustedParams}). The adjusted TALYS fits from this procedure are given in Figures \ref{135Ce}--\ref{132Ce} compared to the default calculation and EXFOR data for the eight used reaction channels \cite{Tarkanyi2017:La,Becker2020:protonsLa,Morrell2020}. Given that the experimental data do not extend beyond 100 MeV, the fits are shown only up to this point.

\newpage
\newpage
\newpage

\subsubsection{Parameter Adjustment Validation}
\begin{figure}[H]
		{\includegraphics[width=1.0\columnwidth]{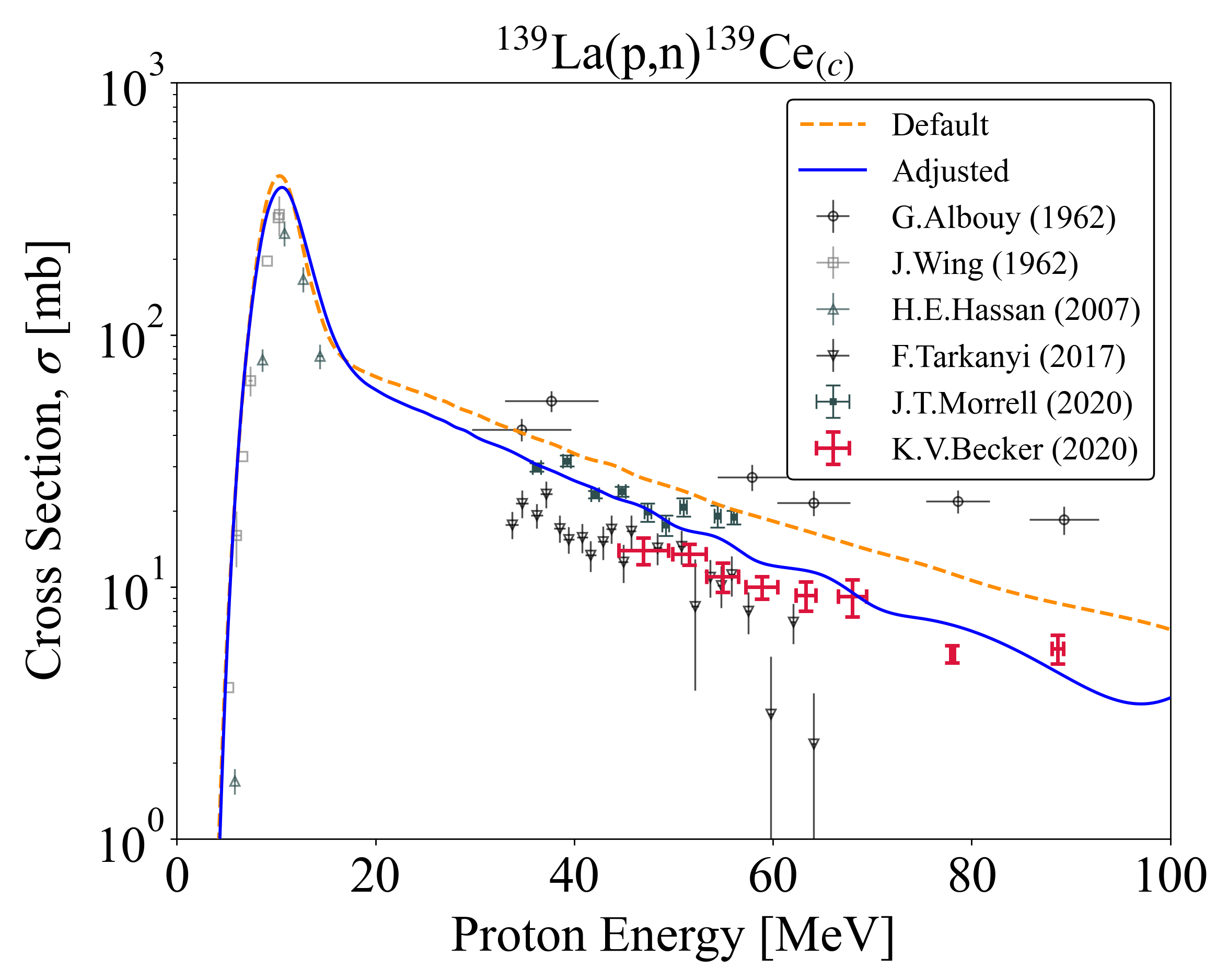}}
	\vspace{-0.65cm}
	\caption{TALYS default and adjusted extended to $^{139}$Ce.}\label{139Ce}
		{\includegraphics[width=1.0\columnwidth]{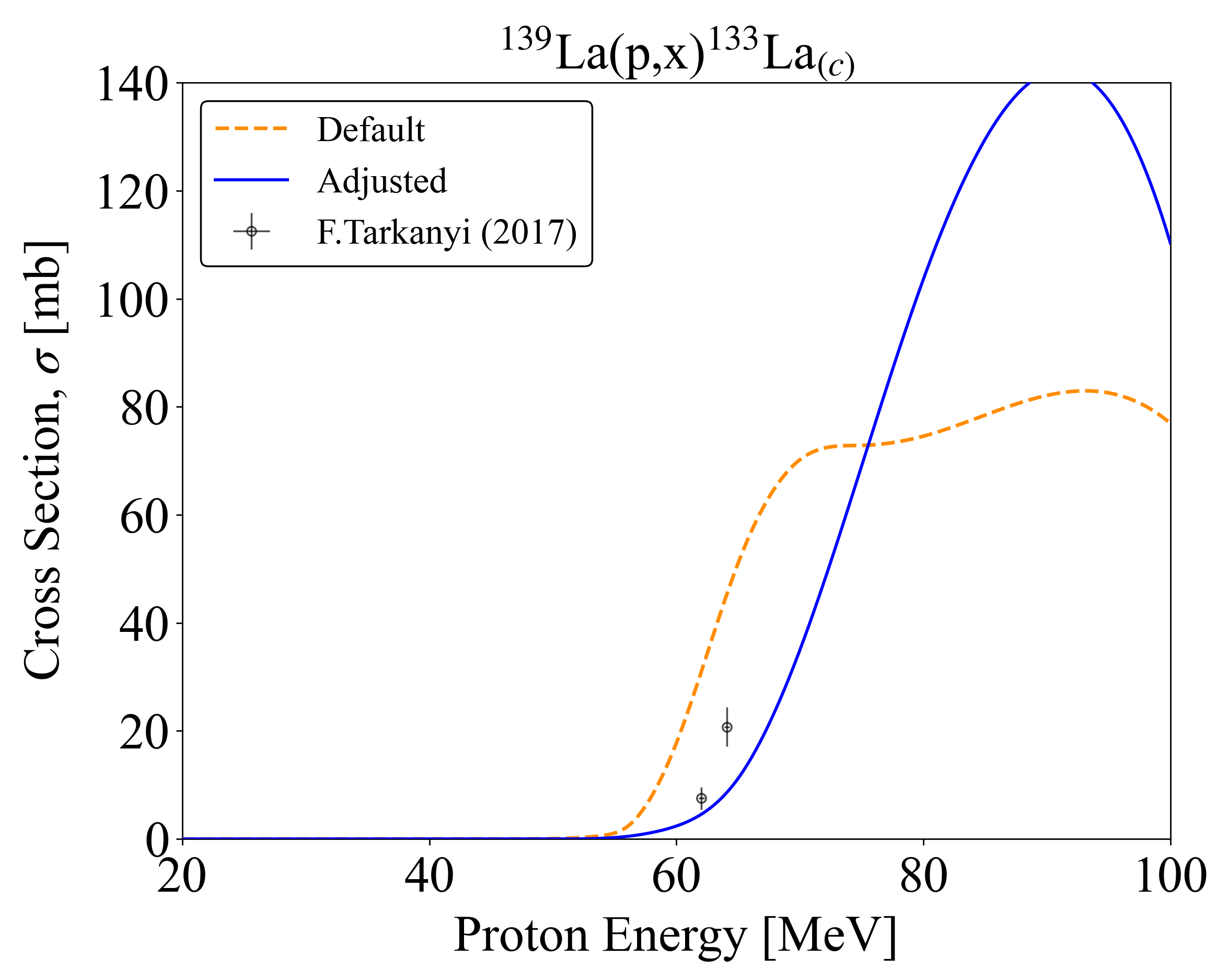}}
	\vspace{-0.65cm}
	\caption{TALYS default and adjusted extended to $^{133}$La.}\label{133La}
\end{figure}

\begin{figure}[H]
		{\includegraphics[width=1.0\columnwidth]{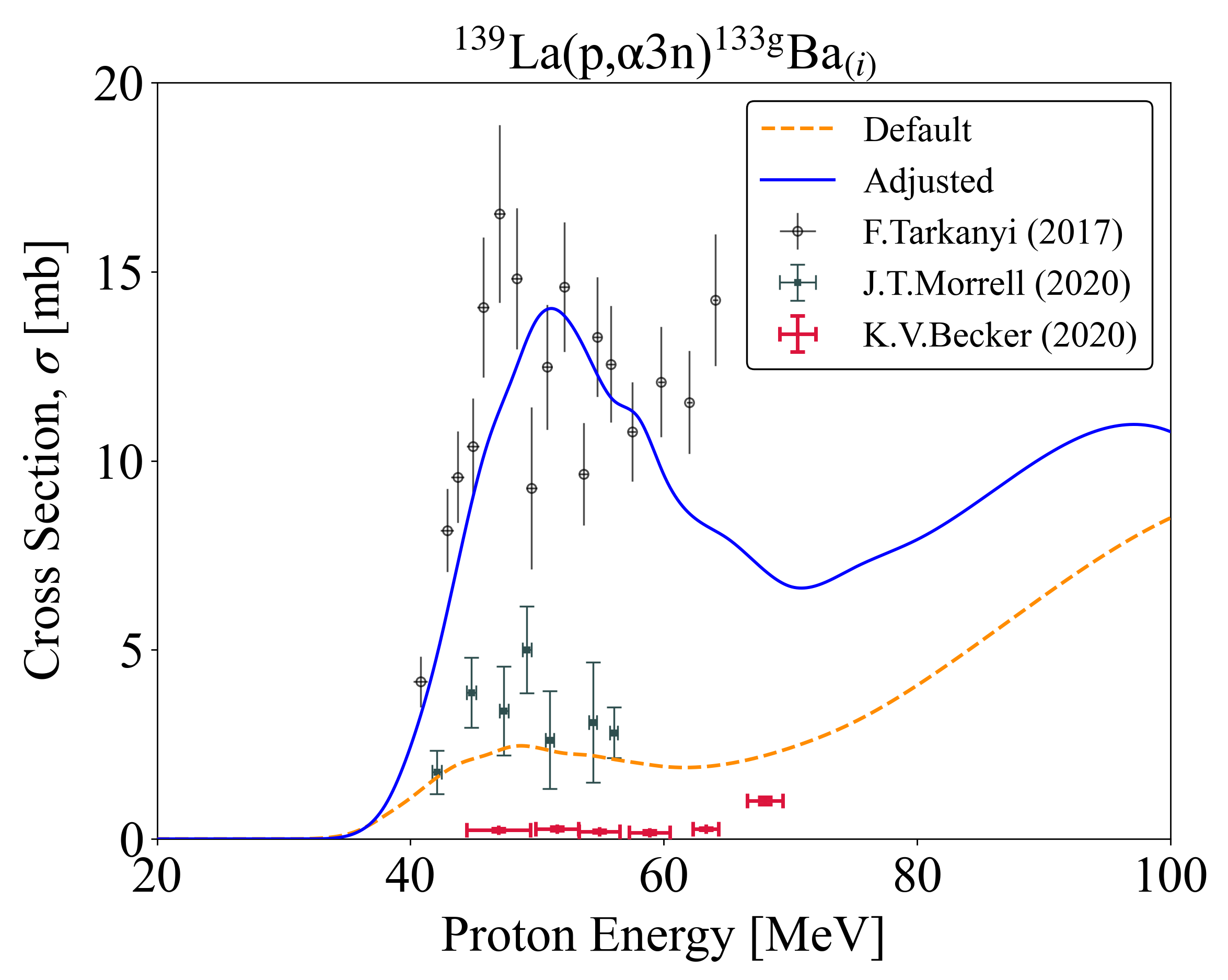}}
	\vspace{-0.65cm}
	\caption{TALYS default and adjusted extended to $^{133\mathrm{g}}$Ba.}\label{133gBa}
		{\includegraphics[width=1.0\columnwidth]{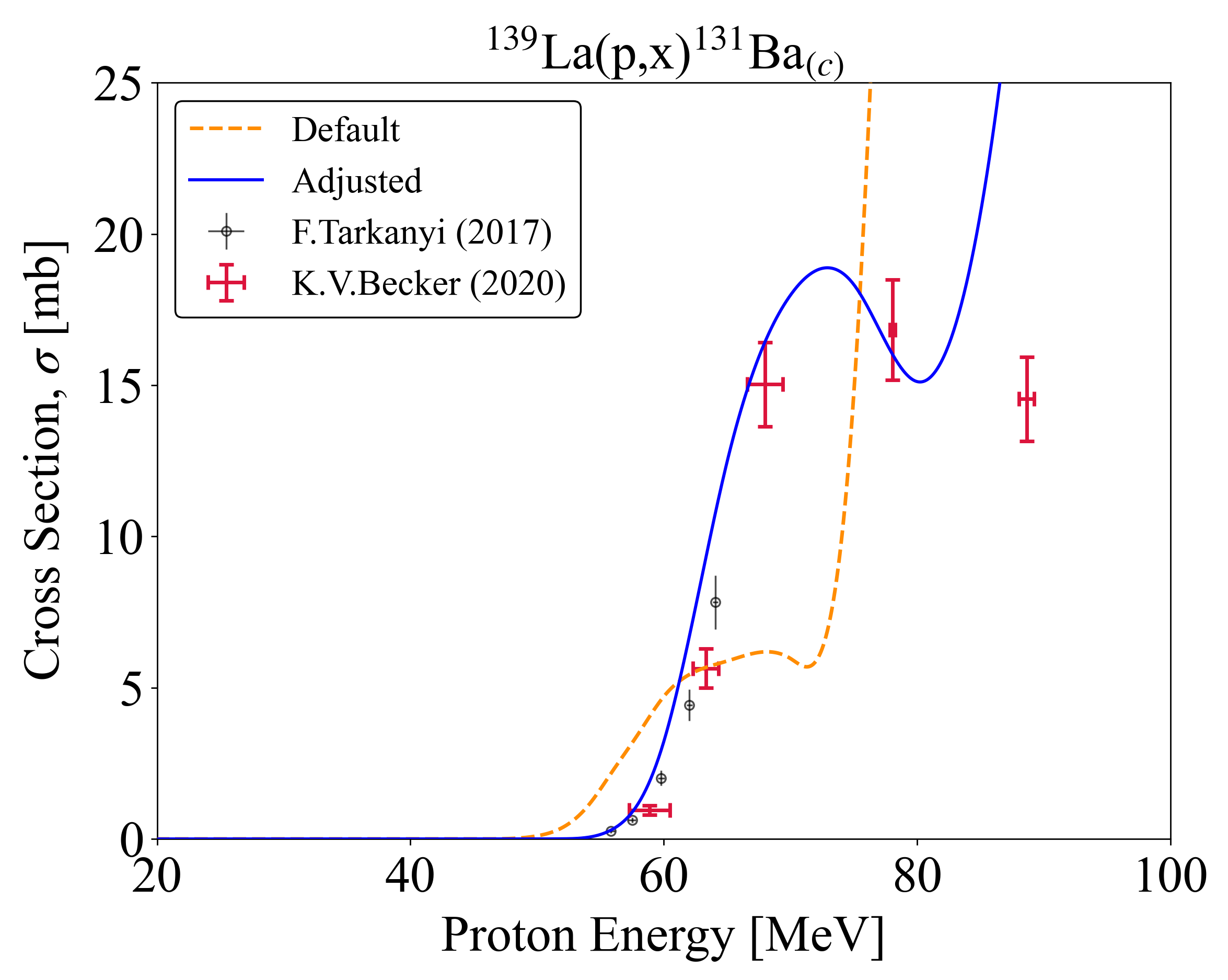}}
	\vspace{-0.65cm}
	\caption{TALYS default and adjusted extended to $^{131}$Ba.}\label{131Ba}
		{\includegraphics[width=1.0\columnwidth]{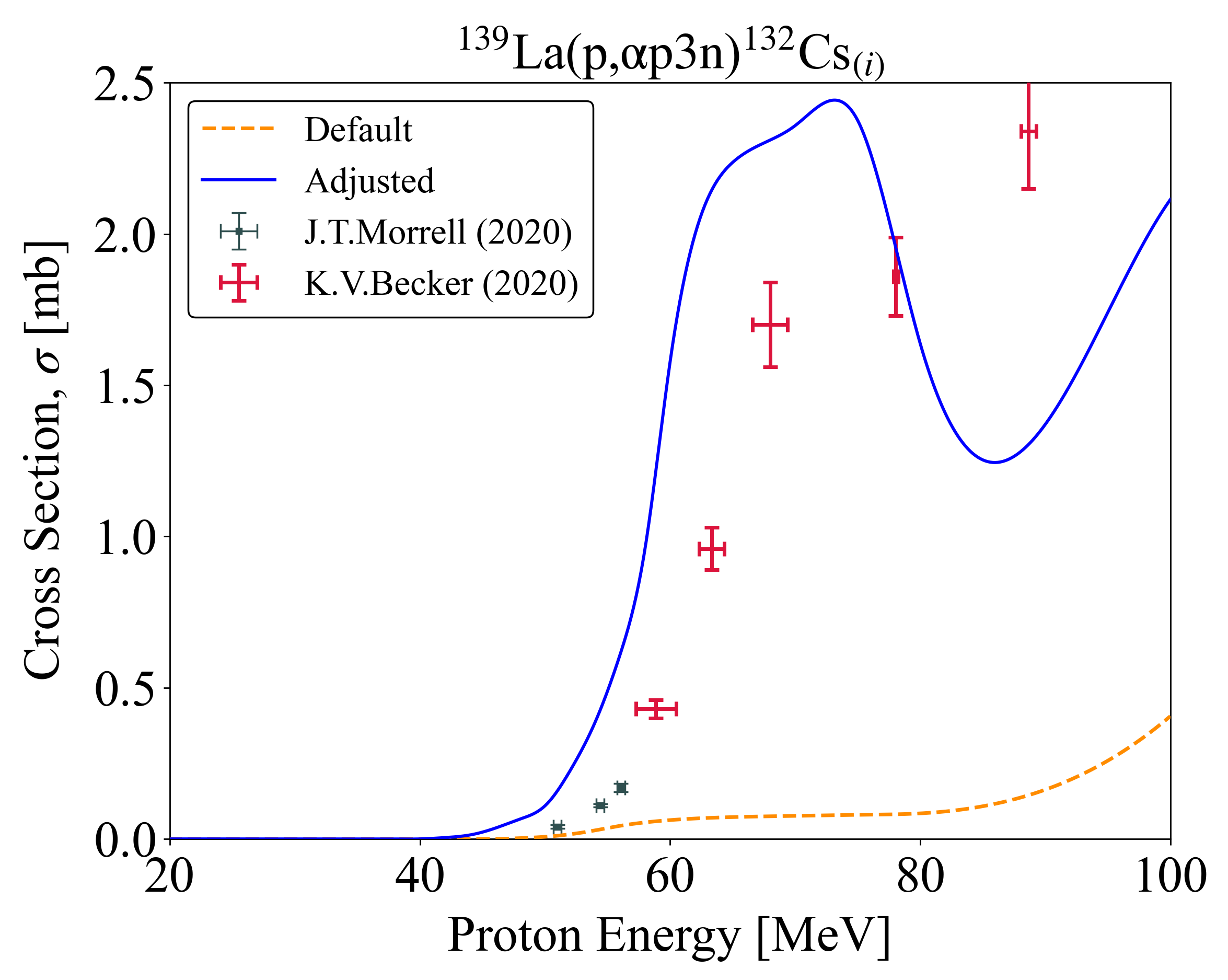}}
	\vspace{-0.65cm}
	\caption{TALYS default and adjusted extended to $^{132}$Cs.}\label{132Cs}
\end{figure}

Validation of this adjusted fit is performed via comparison to the $^{139}$La(p,x)$^{139}$Ce,$^{133}$La,$^{133\textnormal{g},131}$Ba,$^{132}$Cs channels, which were not used in the fitting approach due to their magnitudes or ambiguity/lack of data \cite{Albouy1962:La,Hassan2007:La,Wing1962:La}. However, even in these channels, the adjusted fit is shown in Figures \ref{139Ce}--\ref{132Cs} to have impressive predictive power versus the default. Specifically, the predictive success for the single-particle out $^{139}$La(p,n)$^{139}$Ce reaction, necessarily heavily influenced by pre-equilibrium, instills confidence in the adjusted parameters.

\newpage
\newpage

The $\chi^2_{tot}$ results comparing the adjusted and default fit globally based on both weighting methods described in Section \ref{NbValidation} are given in Table \ref{chi2La}. Again, both weighting methodologies yield similar results, and it is evident that the adjusted fit outperforms the default prediction. In both the niobium and lanthanum presented cases of this work, the suggested standardized fitting procedure has produced improved results over the TALYS default in a comprehensive and justifiable manner.

\vspace{-0.2cm}
\begin{table}[H]
\caption{Global $\chi^2$ metric describing goodness-of-fit for the default and adjusted TALYS calculations of $^{139}$La(p,x). The very large improvement in $\chi^2$ for the adjusted case may imply that the applied weights were too large, contributing to an inflated change versus the default.}
\label{chi2La}
\begin{ruledtabular}
\begin{tabular}{lcc}
Weighting Method&Default $\chi^2_{tot}$ &Adjusted $\chi^2_{tot}$ \\[0.1cm]
\hline\\[-0.25cm]
Cumulative $\sigma$     &87.8 & 1.89\\[0.1cm]

Maximum $\sigma$        &96.4 & 3.34\\
\end{tabular}
\end{ruledtabular}
\end{table}

\subsection{Interpretation of Parameter Adjustments}
\begin{figure*}[!t]
{\includegraphics[trim=0cm 0cm 0cm 0cm, clip=true,scale=0.475]{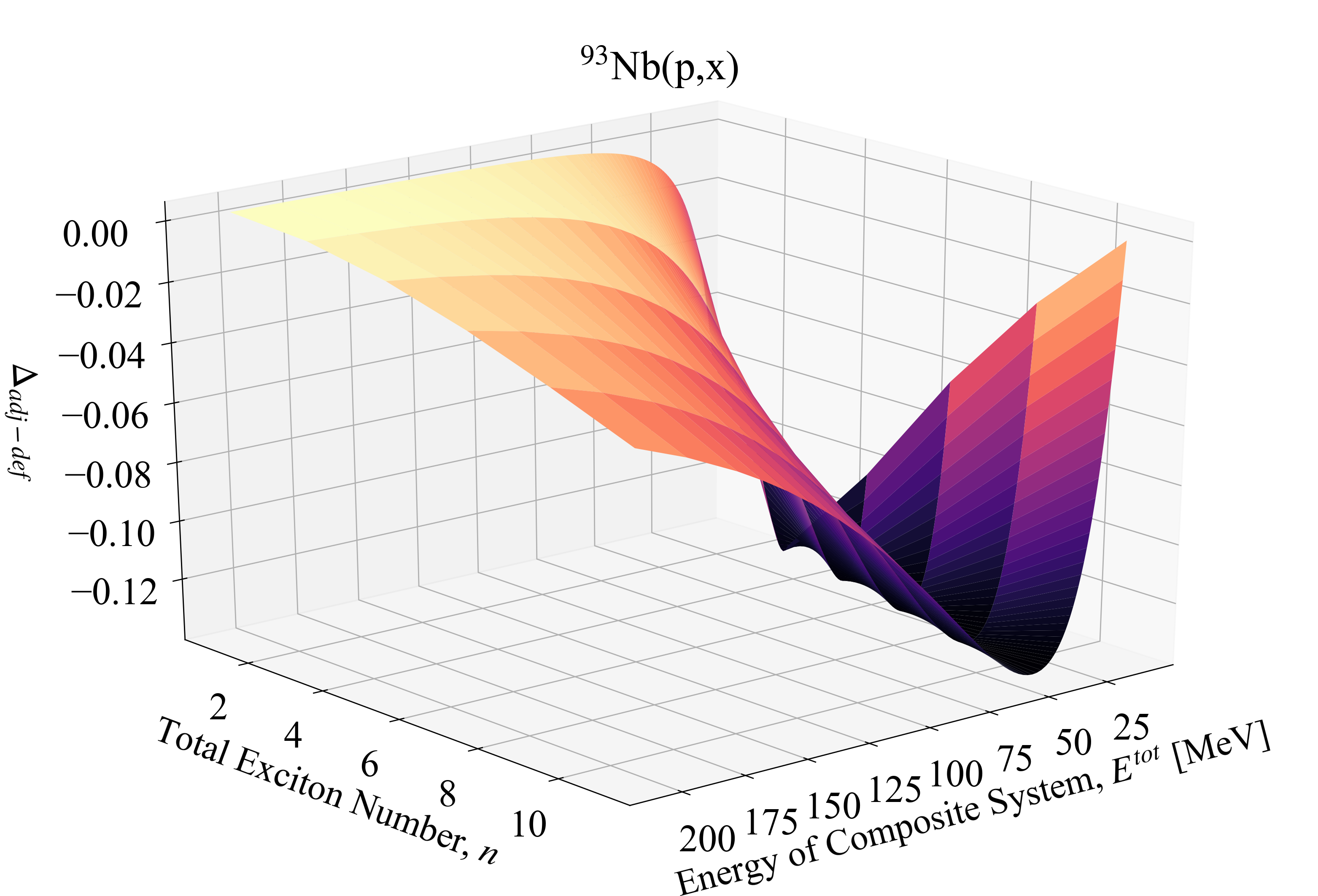}}
\vspace{-0.35cm}
{\includegraphics[trim=0cm 0cm 0cm 0cm, clip=true,scale=0.475]{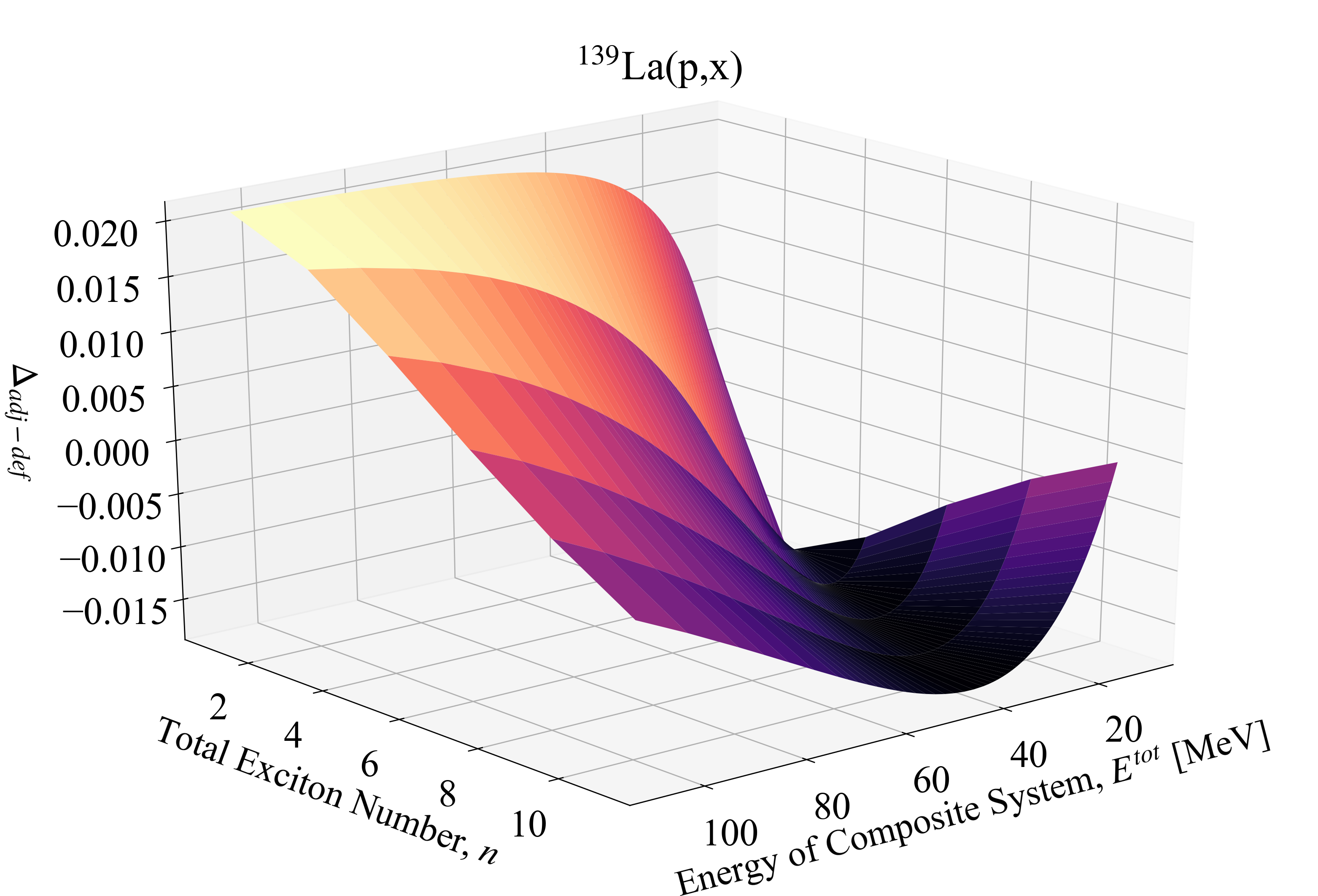}}
\vspace{0.25cm}
\caption{Visualization of impact from pre-equilibrium parameter adjustments across reaction phase space on the exciton model squared matrix element for the effective residual interaction. A consistent pattern is seen in the adjustments for the niobium and lanthanum cases, with more pronounced behaviour for the niobium. The colour scale is a mapping of the z-axis in each case.}
\label{RelativeM2}
\end{figure*}

The success of this fitting approach suggests that physical meaning could be inferred from the adjustments made to the exciton model parameters. Moreover, the consistent adjustments made to the \texttt{M2} exciton parameters in both the niobium and lanthanum cases appears to reveal a systematic trend in how residual product excitation functions for high-energy proton-induced reactions on spherical nuclei are miscalculated in the current exciton model scheme. Across the prominent reaction channels explored in this work, there was a consistent underprediction of both the high-energy pre-equilibrium tails and compound peak magnitudes. It was seen that enforcing \texttt{M2constant}$<$1.0 could improve lacking tail behaviour while \texttt{M2limit}$>$1.0 with \texttt{M2shift}$<$1.0 helped compensate for the increased tail by creating more production in the compound peak. It is possible to further visualize and quantify this trend by plotting the magnitude of the squared effective interaction matrix element within the ($E^{tot},n$) reaction phase space. Specifically, defining $\Delta_{adj-def}$ as the difference of normalized $M^2$ between the adjusted fit and the default calculation by:
\begin{gather}
\Delta_{adj-def}=\frac{M^2(E^{tot},n)_{adj}}{M^2(E^{tot},n)_{adj,max}}-\frac{M^2(E^{tot},n)_{def}}{M^2(E^{tot},n)_{def,max}},
\end{gather}
the relative strength of $M^2$ for the adjusted case can be compared to the relative strength of $M^2$ in the default case across all of the reaction phase space. The $\Delta_{adj-def}$ results for both the $^{93}$Nb(p,x) and $^{139}$La(p,x) modeling are plotted in Figure \ref{RelativeM2}. It is seen that the adjustments for both targets exhibit the same trend that better modeling fits were achieved when there was a relative decrease for internal transition rates at intermediate proton energies ($E_p=20-60$ MeV) in the exciton model as compared to default values. The relative decrease reduces the probability of formation of complex exciton states, and in turn the compound nucleus equilibration limit, in favour of pre-equilibrium emission. Furthermore, the location of the relative decrease in reaction phase space indicates that there is difficulty transitioning between the Hauser-Feshbach and exciton models for nuclear reactions. These exciton adjustments appear to act as a surrogate for better damping into the compound nucleus system.

The results of Figure \ref{RelativeM2} are additionally interesting because of the variation between the $\Delta_{adj-def}$ magnitudes for $^{93}$Nb(p,x) and $^{139}$La(p,x). The $\Delta_{adj-def}$ for $^{139}$La(p,x) are smaller as a function of the better initial default residual product calculations in TALYS compared to $^{93}$Nb(p,x). However, the root cause of this more pronounced default model failure in the niobium case is unknown, especially given that both niobium and lanthanum are structurally similar.

In total, the modeling adjustments in this work suggest the need to incorporate residual product excitation function data in some capacity into future exciton model parameterizations. Further, this trend applies for proton-induced reactions and perhaps implies a need to release the strict generality of having the same exciton model formulas for both incident protons and neutrons \cite{Koning2004:GlobalPEMexcitonOMP}.

\subsection{Future Considerations}
Residual product excitation functions were not used in the initial exciton model parameterization by \textcite{Koning2004:GlobalPEMexcitonOMP} because of the complexity and uncertainties brought in by the additional level density and transmission coefficient models. This study has included this complexity and tried to isolate for these competing issues and uncertainties through the order of the fitting procedure and the focus on fitting many of the prominent channels, though difficulties still remain with their incorporation.

\begin{figure}[!t]
	\subfloat[\label{talys45_tot}]
		{\includegraphics[width=1.0\columnwidth]{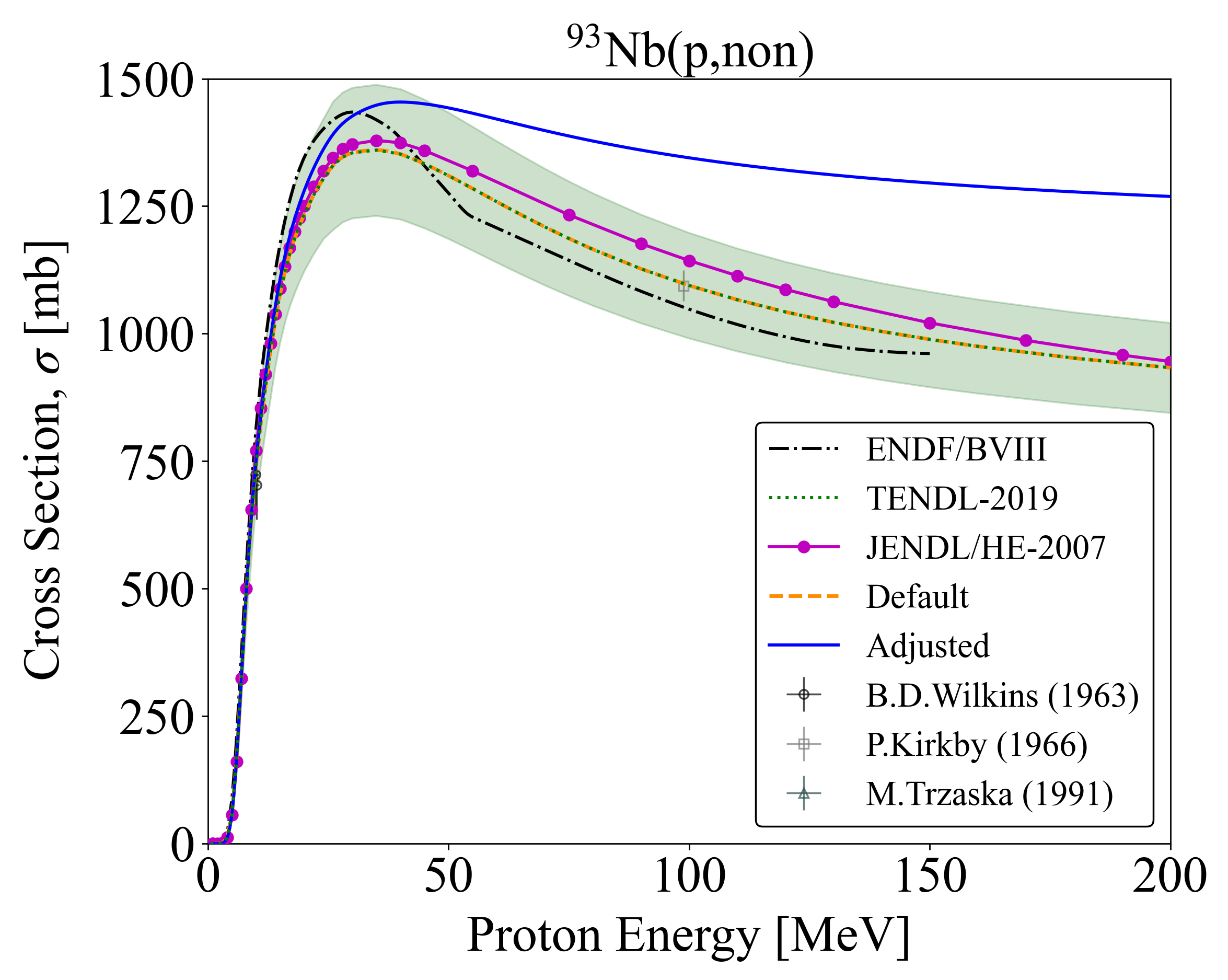}}
	\vspace{-0.25cm}
	\subfloat[\label{La_non}]
		{\includegraphics[width=1.0\columnwidth]{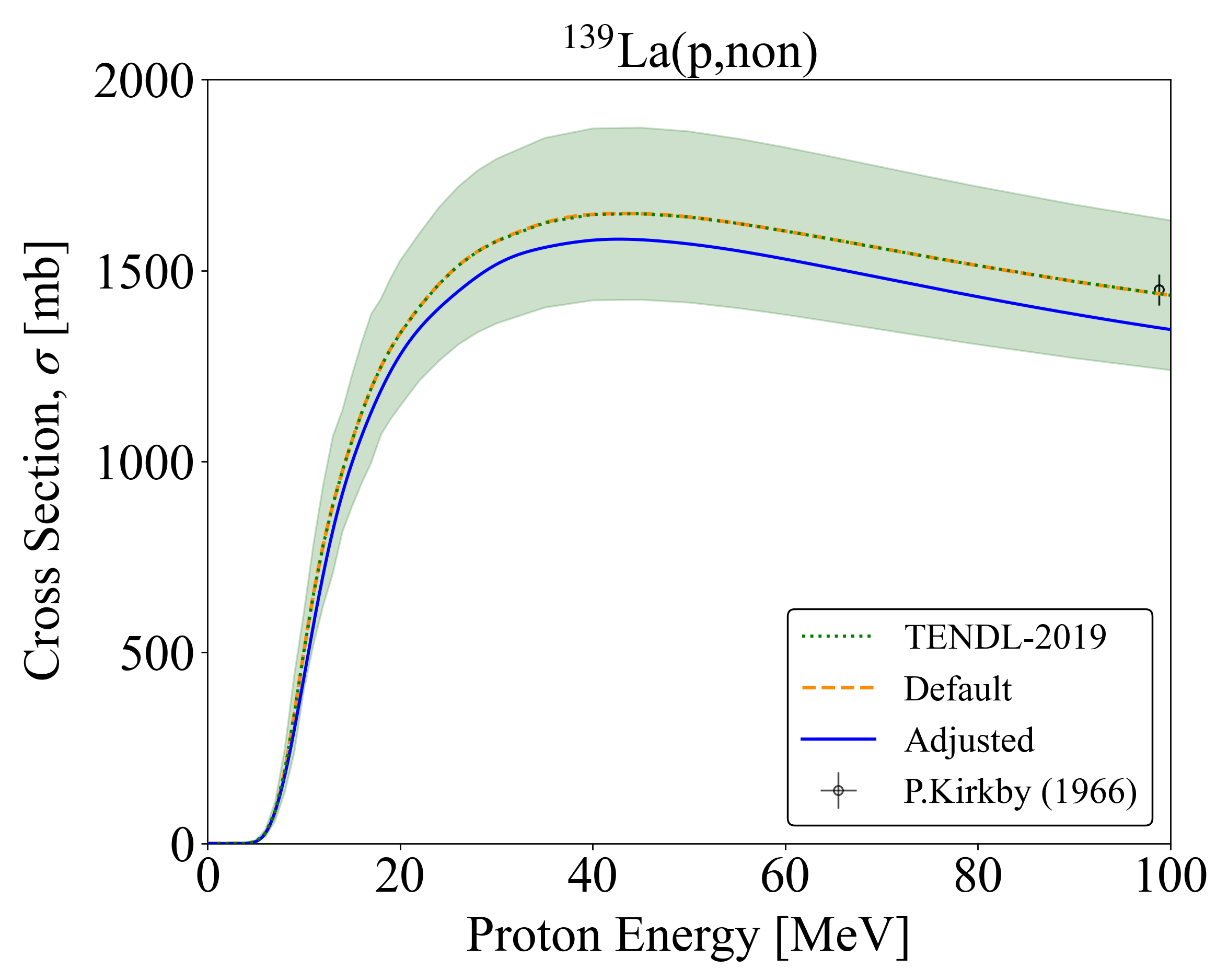}}
	\vspace{-0.15cm}
	\caption{Comparison of experimental, evaluated, and theoretical non-elastic cross sections. The filled error bands are associated with the TENDL data.}\label{nonelastic_compares}
\end{figure}

Furthermore, the adjusted parameters lead not only to changes in specific product reaction channels, but to the total non-elastic channel as well. Consider the difference in total non-elastic cross section for protons incident on niobium between the TALYS default, other evaluation databases, and the TALYS adjusted case, as given in Figure \ref{talys45_tot} \cite{Kirby1966,Trzaska1991,Wilkins1963,JENDL,ENDF}. The adjusted case argues for an increased high-energy cross section. While below 50 MeV, the adjusted calculation seems quite reasonable, above 50 MeV it is evident that there is a large discrepancy between it and the other predictions. However, it should be noted that the evaluations are all heavily constrained by a single high-energy data point, which may not fully represent reality. Nonetheless, they suggest that there should be less confidence in extension of the adjusted TALYS fit to far-from-target residual products such as Kr, Se, and As. It is possible that the poorer fit at high energies is also a reflection of the deterioration in the quality of level density predictions in general at such high excitations. It is likely that the employed microscopic models used in the fitting are less appropriate at such high energies than a more simple stochastic model such as a Fermi gas calculation, though this model too may break down near 200 MeV excitation energy \cite{Grimes1990:LevelDensityHighE}. This is a difficult consideration to experimentally check but might be a more realistic cause for error than the shell gap effects discussed in Section \ref{FitNb}.

A further neglected effect, which may be relevant to the code mispredictions seen at high energies for far-from-target products, is the incorporation of isospin conservation in the modeled reactions. The theoretical calculations of \textcite{Grimes1972:Isospin} and \textcite{Robson1973:Isospin} using a modified Hauser-Feshbach formalism including isospin effects and the experimental findings from works such as \textcite{Lu1971:Isospin} and \textcite{KalbachCline1974:Isospin} explored this factor. They demonstrate that isospin conservation yields cross sections and particle emission spectra different from the Bohr independence hypothesis of compound nuclear decay including only angular momentum and from the typical exciton model for pre-equilibrium decay. Particularly, \textcite{Grimes1972:Isospin} and \textcite{Lu1971:Isospin} show that isospin selection rules for proton-induced reactions result in enhanced proton emission. These publications explored proton bombardment energies in the $10-20$\,MeV range. Although the adjusted modeling fits in this work were appropriate at those incident energies, it is possible that the choice of level density parameters were an unknowing compensating factor for neglected isospin effects, which did not remain effectively compensating at higher energies. It is also possible that isospin effects are simply small for the target mass and energies under consideration here. We believe it would be a worthwhile experiment for the community to explore these isospin considerations through a study of particle emission spectra resulting from both p+$^{93}$Nb and $\alpha$+$^{90}$Zr irradiations. Specifically, these reactions populate the same $^{94}$Mo compound system with different isospins and the proximity of $^{94}$Mo to the $N=50$ shell gap may mean that pure isospin states exist that can be well-defined, making the compound system a suitable candidate for this type of structure investigation.

\clearpage

Unfortunately, it is not possible to derive any $^{93}$Nb(p,non) data points from summed residual product cross sections measured in this work for a more in-depth fit comparison. The presented cross section results are not exhaustive enough for this calculation since stable and very short-lived isotope production was not measured. This potential non-elastic cross section issue, or the possible high-energy theoretical shortcomings, do not discredit the procedure shown here but instead emphasize that the approach suggested in this work is not meant to be on par with complete reaction evaluations. In general, this approach is a holistic and realistic methodology, grounded on observables and experimental data, that experimenters can perform to benefit theory and support further predictive work. Although, it is clear that the niobium fitting is an extreme case and looking at the total non-elastic cross section for protons incident on lanthanum in Figure \ref{La_non} instills more confidence in this overall fitting process \cite{Kirby1966}.

A worthwhile different way of continuing study on the departure of equal matrix elements for neutron-induced or proton-induced reactions may be to systematically study one reaction channel, instead of all reaction channels simultaneously as in this work. Hence, one could investigate whether (p,n) reactions for different nuclides would show the same exciton adjustment trends discovered here.

In the future, this fitting procedure could expand to include emission spectra and double-differential data to try and improve the elastic versus non-elastic competition and potentially determine other corrective parameter adjustments that are simply not sensitive in the purely residual product data analysis \cite{KD2003:OMP}. Including the extra datasets can help clarify effects between level density models, the optical model, and pre-equilibrium parameterizations. Such a procedure could be an inspiration and act as a stepping stone to the development of a charged-particle evaluated data database \cite{WANDA2019}.

Although the sensitivity work performed in this paper was a manual search, it would be useful to incorporate automation, such as search techniques within a Bayesian framework, with the acquired exciton adjustment knowledge. This would help to more accurately determine a global minimum for parameter optimization and to better express the resolving power of different parameters and channels in a more quantitative fashion.
\newpage

\vspace{-0.45cm}
\section{Conclusions}
This work reports 23 sets of measured $^{93}$Nb(p,x) residual product cross sections between 50--200\,MeV as part of a Tri-lab collaboration between LBNL, LANL, and BNL. The reported cross sections greatly extend the datasets for numerous products and are of higher precision than a majority of previous measurements. The $^{93}$Nb(p,4n)$^{90}$Mo monitor reaction of particular interest for intermediate proton energy stacked-target activation experiments was characterized beyond 100 MeV for the first time.

Given the measured data, an in-depth investigation of reaction modeling and pre-equilibrium mechanisms was conducted. A standardized parameter adjustment fitting procedure to improve default code predictions in a physically justifiable manner was proposed and applied to $^{93}$Nb(p,x) and $^{139}$La(p,x) cross section data as tests. The fitting approach focused on the current parameterization of the squared matrix element in the pre-equilibrium two-component exciton model. A systematic trend for the exciton parameter adjustments to correct high-energy tails and compound peak magnitudes was seen that implied the current parameterization is not wholly correct. This result suggests the need to incorporate residual product excitation function data in some capacity into future exciton model parameterizations and potentially create different parameterizations altogether for incident protons and neutrons.

The focus of this work was on presenting and interpreting the results from (p,x) reactions on spherical target nuclei (Nb and La). Subsequent papers will discuss additional data results from the Tri-lab collaboration for $^{75}$As(p,x) reactions as well as the production and characterization of thin arsenic targets.

\section*{Data Availability Statement}
The gamma-ray spectra and all other raw data created during this research are openly available \cite{ZenodoTREND}. Upon publication, the experimentally determined cross sections will be uploaded to the EXFOR database.

\section*{Acknowledgements}
This research was supported by the Isotope Program within the U.S. Department of Energy's Office of Science, carried out under Lawrence Berkeley National Laboratory (Contract No. DE-AC02-05CH11231), Los Alamos National Laboratory (Contract No. 89233218CNA000001), and Brookhaven National Laboratory (Contract No. DEAC02-98CH10886). The authors would like to acknowledge the assistance and support of Brien Ninemire, Scott Small, Nick Brickner, Devin Thatcher, and all the rest of the operations, research, and facilities staff of the LBNL 88-Inch Cyclotron. We would also like to thank David Reass and Mike Connors at LANSCE-IPF, the LANL C-NR Countroom operators, and the LANSCE Accelerator Operations staff. The authors would like to acknowledge Deepak Raparia, head of the Pre-Injector Systems group at CAD-BNL, for LINAC beam tuning for the experiment, and all members of the BNL Medical Isotope Research and Production group for their assistance. We are grateful to Patrick Sullivan and John Aloi of the BNL Radiological Control Division for the Health Physics support. Sumanta Nayak is acknowledged for the engineering and Frank Naase for the IT support.
\newpage
\appendix
\section{\label{StackTables}Target Stack Designs}
Details of the stacked-targets irradiated in this work are given in Tables \ref{LANLStack} and \ref{BNLStack}.

\begin{longtable}{lP{1.3cm}P{1.8cm}P{1.8cm}}
\caption{Target stack design for irradiation at IPF. The proton beam initially hits the stainless steel plate (SS-SN1) after passing through the upstream Inconel beam entrance window, a water cooling channel, and the target box aluminum window. The thickness and areal density measurements are prior to any application of the variance minimization techniques described in this work.}\label{LANLStack}\\ 

\hline\hline\\[-0.27cm]
\multirow{4}{*}{Target Layer}&\multirow{4}{=}{\centering Thickness [$\mmicro$m]}&\multirow{4}{=}{\centering Areal Density [mg/cm$^2$]}&\multirow{4}{=}{\centering Areal Density Uncertainty [\%]}\\ \\ \\ \\[0.1cm]
\hline\\[-0.22cm]
\endfirsthead

{{\tablename\ \thetable{} -- cont.}} \\
\hline\hline\\[-0.27cm]
\multirow{4}{*}{Target Layer}&\multirow{4}{=}{\centering Thickness [$\mmicro$m]}&\multirow{4}{=}{\centering Areal Density [mg/cm$^2$]}&\multirow{4}{=}{\centering Areal Density Uncertainty [\%]}\\ \\ \\ \\[0.1cm]
\hline\\[-0.22cm]
\endhead

\hline Continued on next page&&&
\endfoot

\hline\hline
\endlastfoot

SS-SN1 Profile Monitor    &130.0 &100.12&0.07 \\[0.1cm]

Al-SN1                             &27.33 &7.51  &0.21  \\[0.1cm]

Nb-SN1                           &25.75  &23.08 &0.12  \\[0.1cm]

As-SN1                            &4.27 &2.45 &8.2 \\[0.1cm]

Ti-SN1                               &25.00 &11.265 &1.0 \\[0.1cm]

Cu-SN1                             &24.33 &19.04 &0.13 \\[0.1cm]

Al Degrader 01                 &6307.0 &1702.89 &0.001 \\[0.1cm]

Al-SN2                              &26.67 &7.58  &0.32  \\[0.1cm]

Nb-SN2                            &24.75  &22.67 &0.08  \\[0.1cm]

As-SN2                            &4.30 &2.46 &8.3 \\[0.1cm]

Ti-SN2                             &25.00 &11.265 &1.0 \\[0.1cm]

Cu-SN2                           &24.00 &18.90 &0.36 \\[0.1cm]

Al Degrader 02                 &3185.5 &860.09 &0.02 \\[0.1cm]

Al-SN3                             &26.67 &7.38  &0.22  \\[0.1cm]

Nb-SN3                          &24.50  &22.83 &0.03  \\[0.1cm]

As-SN3                          &3.62 &2.07 &9.0 \\[0.1cm]

Ti-SN3                            &25.00 &11.265 &1.0 \\[0.1cm]

Cu-SN3                           &23.33 &19.38 &0.11 \\[0.1cm]

Al Degrader 03                 &2304.5  &622.22 &0.06 \\[0.1cm]

Al-SN4                            &28.00 &7.34  &0.18  \\[0.1cm]

Nb-SN4                           &25.50  &22.57 &0.16  \\[0.1cm]

As-SN4                           &3.54 &2.03 &9.2 \\[0.1cm]

Ti-SN4                             &25.00 &11.265 &1.0 \\[0.1cm]

Cu-SN4                             &24.67 &19.24 &0.11 \\[0.1cm]

Al Degrader 04                  &1581.3  &426.94&0.04 \\[0.1cm]

Al-SN5                               &27.00 &7.48  &0.44  \\[0.1cm]

Nb-SN5                              &24.75  &22.78 &0.12  \\[0.1cm]

As-SN5                               &3.90 &2.23 &8.7 \\[0.1cm]

Ti-SN5                               &25.00 &11.265 &1.0 \\[0.1cm]

Cu-SN5                             &25.00 &19.09 &0.17 \\[0.1cm]

Al Degrader 05                  &1033.8 &279.11 &0.06 \\[0.1cm]

Al-SN6                               &28.67 &7.44  &0.25  \\[0.1cm]

Nb-SN6                              &25.25  &22.80 &0.08  \\[0.1cm]

As-SN6                               &3.11 &1.78 &10 \\[0.1cm]

Ti-SN6                                 &25.00 &11.265 &1.0 \\[0.1cm]

Cu-SN6                               &24.33 &19.50 &0.16 \\[0.1cm]

Al Degrader 06                    &834.8  &225.38 &0.22 \\[0.1cm]

Al-SN7                                  &28.33 &7.56  &0.15  \\[0.1cm]

Nb-SN7                              &25.50  &22.62 &0.06  \\[0.1cm]

As-SN7                               &2.79 &1.59 &9.2 \\[0.1cm]

Ti-SN7                                 &25.00 &11.265 &1.0 \\[0.1cm]

Cu-SN7                               &23.67 &18.79 &0.04 \\[0.1cm]

Al Degrader 07                     &513.5  &138.65 &0.10 \\[0.1cm]

Al-SN8                                  &27.67 &7.56  &0.10  \\[0.1cm]

Nb-SN8                               &25.50  &22.95 &0.45  \\[0.1cm]

As-SN8                               &2.20 &1.26 &9.0 \\[0.1cm]

Ti-SN8                                 &25.00 &11.265 &1.0 \\[0.1cm]

Cu-SN8                               &24.00 &19.06 &0.23 \\[0.1cm]
                 
Al Degrader 08                   &517.3  &139.66 &0.43 \\[0.1cm]

Al-SN9                                &27.00 &7.47  &0.36  \\[0.1cm]

Nb-SN9                               &25.00  &22.53 &0.24  \\[0.1cm]

As-SN9                               &2.57 &1.47 &9.9 \\[0.1cm]

Ti-SN9                                  &25.00 &11.265 &1.0 \\[0.1cm]

Cu-SN9                                 &26.33 &19.19 &0.12 \\[0.1cm]

Al Degrader 09                     &517.8  &139.79 &0.09 \\[0.1cm]

Al-SN10                                &28.00 &7.41  &0.17  \\[0.1cm]

Nb-SN10                               &24.75  &22.82 &0.02  \\[0.1cm]

As-SN10                               &1.94 &1.11 &10 \\[0.1cm]

Ti-SN10                                  &25.00 &11.265 &1.0 \\[0.1cm]

Cu-SN10                                 &25.67 &18.87 &0.18 \\[0.1cm]

SS-SN10 Profile Monitor          &130.0 &100.12&0.07 \\[0.1cm]
\end{longtable}
\newpage

\newpage
\begin{table}[!t]
\caption{Target stack design for irradiation at BLIP. The proton beam initially hits the stainless steel plate after passing through the upstream beam windows, water cooling channels, and target box aluminum window. The thickness and areal density measurements are prior to any application of the variance minimization techniques described in this work.}
\label{BNLStack}
\begin{ruledtabular}
\begin{tabular}{lP{1.3cm}P{1.8cm}P{1.8cm}}
\multirow{4}{*}{Target Layer}&\multirow{4}{=}{\centering Thickness [$\mmicro$m]}&\multirow{4}{=}{\centering Areal Density [mg/cm$^2$]}&\multirow{4}{=}{\centering Areal Density Uncertainty [\%]}\\ \\ \\ \\[0.1cm]
\hline\\[-0.22cm]
SS Profile Monitor               &120.2 &95.16&0.58 \\[0.1cm]

Cu-SN1                              &26.00 &22.34  &0.10  \\[0.1cm]

Nb-SN1                             &25.75  &22.75 &0.25  \\[0.1cm]

As-SN1                             &1.89 &1.08 &9.9 \\[0.1cm]

Ti-SN1                                &25.00 &11.265 &1.0 \\[0.1cm]

Cu Degrader 01               &5261.1 &4708.07 &0.02 \\[0.1cm]

Cu-SN2                           &26.75 &22.41  &0.11  \\[0.1cm]

Nb-SN2                          &24.75  &22.91 &0.19  \\[0.1cm]

As-SN2                            &2.94 &1.68 &9.0 \\[0.1cm]

Ti-SN2                            &25.00 &11.265 &1.0 \\[0.1cm]

Cu Degrader 02              &4490.7 &4018.99 &0.04 \\[0.1cm]

Cu-SN3                            &26.50 &22.26  &0.05  \\[0.1cm]

Nb-SN3                           &24.00  &22.67 &0.31  \\[0.1cm]

As-SN3                           &3.06 &1.75 &10 \\[0.1cm]

Ti-SN3                            &25.00 &11.265 &1.0 \\[0.1cm]

Cu Degrader 03             &4501.8  &4028.84 &0.03 \\[0.1cm]

Cu-SN4                          &26.00 &22.29  &0.15  \\[0.1cm]

Nb-SN4                          &24.75  &22.70 &0.23  \\[0.1cm]

As-SN4                           &4.85 &2.78 &9.9 \\[0.1cm]

Ti-SN4                              &25.00 &11.265 &1.0 \\[0.1cm]

Cu Degrader 04               &4243.9  &3797.96&0.03 \\[0.1cm]

Cu-SN5                            &25.50 &22.35  &0.04  \\[0.1cm]

Nb-SN5                            &25.00  &22.54 &0.12  \\[0.1cm]

As-SN5                            &7.26 &4.15 &12 \\[0.1cm]

Ti-SN5                             &25.00 &11.265 &1.0 \\[0.1cm]

Cu Degrader 05               &3733.8 &3341.56 &0.03 \\[0.1cm]

Cu-SN6                           &26.25 &22.34  &0.08  \\[0.1cm]

Nb-SN6                           &25.00  &22.36 &0.24  \\[0.1cm]

As-SN6                           &4.93 &2.82 &9.0 \\[0.1cm]

Ti-SN6                             &25.00 &11.265 &1.0 \\[0.1cm]

Cu Degrader 06               &3783.0  &3385.41 &0.04 \\[0.1cm]

Cu-SN7                            &25.75 &22.26  &0.09  \\[0.1cm]

Nb-SN7                            &25.75  &22.62 &0.10  \\[0.1cm]

As-SN7                            &12.62 &7.22&9.3 \\[0.1cm]

Ti-SN7                            &25.00 &11.265 &1.0 \\[0.1cm]
\end{tabular}
\end{ruledtabular}
\end{table}
\clearpage

\section{\label{Appendix_Plots}Measured Excitation Functions}
Plots of extracted cross sections in this work are given with reference to existing literature data, TENDL-2019, and reaction modeling codes TALYS-1.9, EMPIRE-3.2.3, CoH-3.5.3, and ALICE-20 using default parameters \cite{Michel1997:ProtonsTiCuNb,Titarenko2011,Voyles2018:Nb,Ditroi2008,Steyn2011,Ditroi2009,Albouy1963,Korteling1964,AvilaRodriguez2008,Levkovskii1991,Kiselev1974,Rizvi2012,Lawriniang2018,Parashari2018,Singh2006:ProtonAlphaModelExcitationFunc,James1954,Forsthoff1953,Blaser1951}. Subscripts $(i)$ and $(c)$ in figure titles indicate independent and cumulative cross sections, respectively.

\begin{figure}[H]
	{\includegraphics[width=1.0\columnwidth]{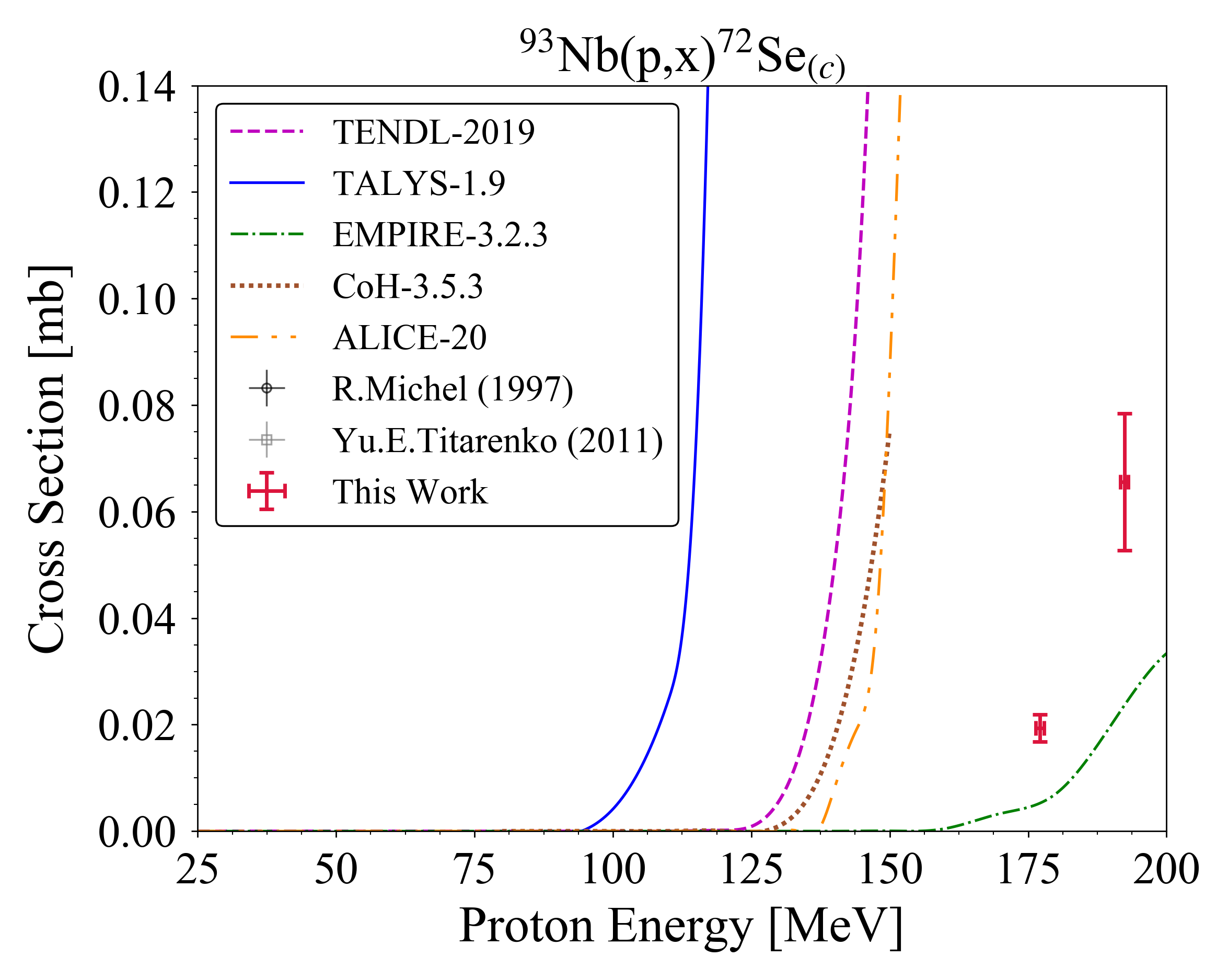}}
\vspace{-0.65cm}
\caption{Experimental and theoretical cross sections for $^{72}$Se production.}\label{Nb_72Se}
	{\includegraphics[width=1.0\columnwidth]{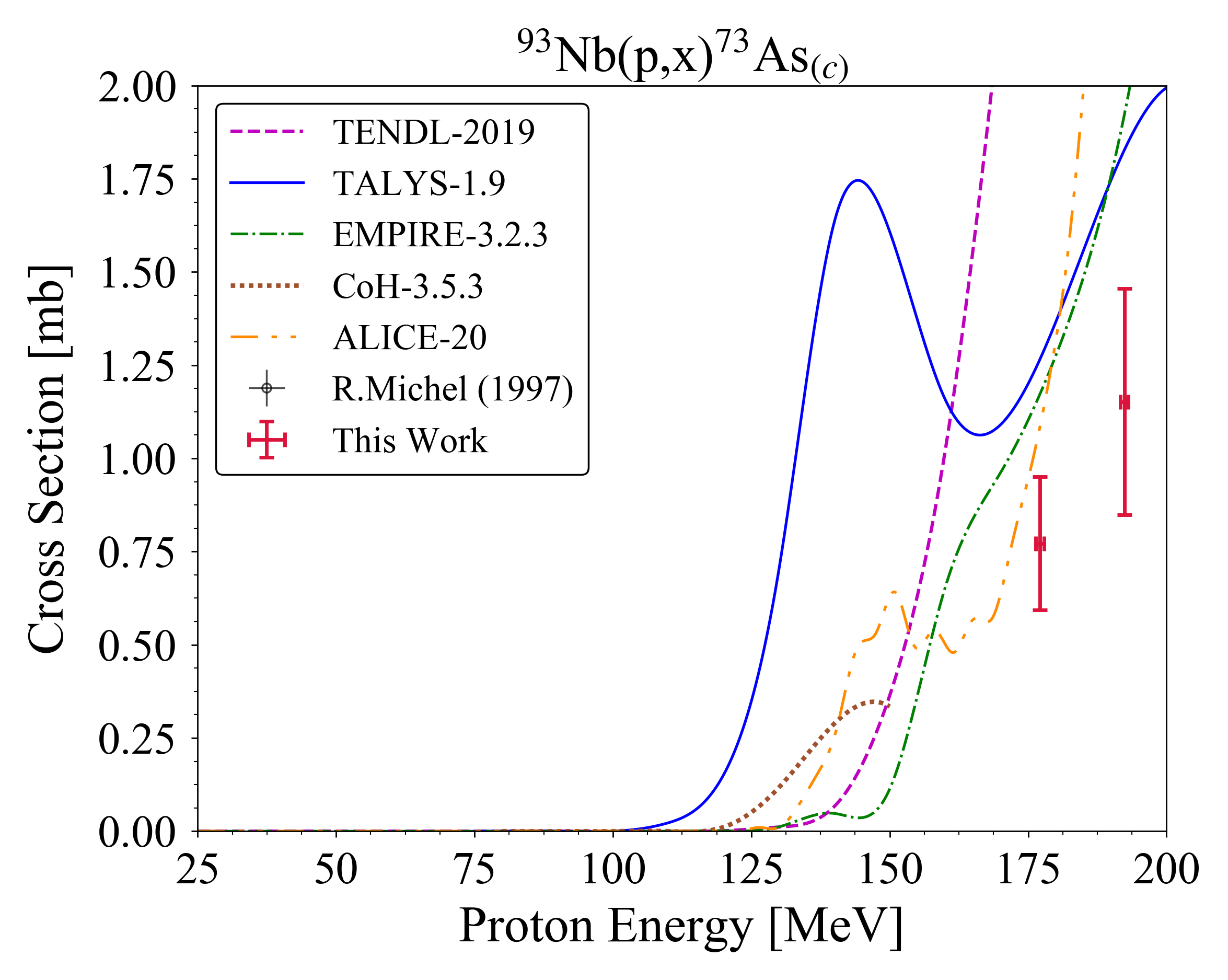}}
\vspace{-0.65cm}
\caption{Experimental and theoretical cross sections for $^{73}$As production.}\label{Nb_73As}
\end{figure}

\begin{figure}[H]
	{\includegraphics[width=1.0\columnwidth]{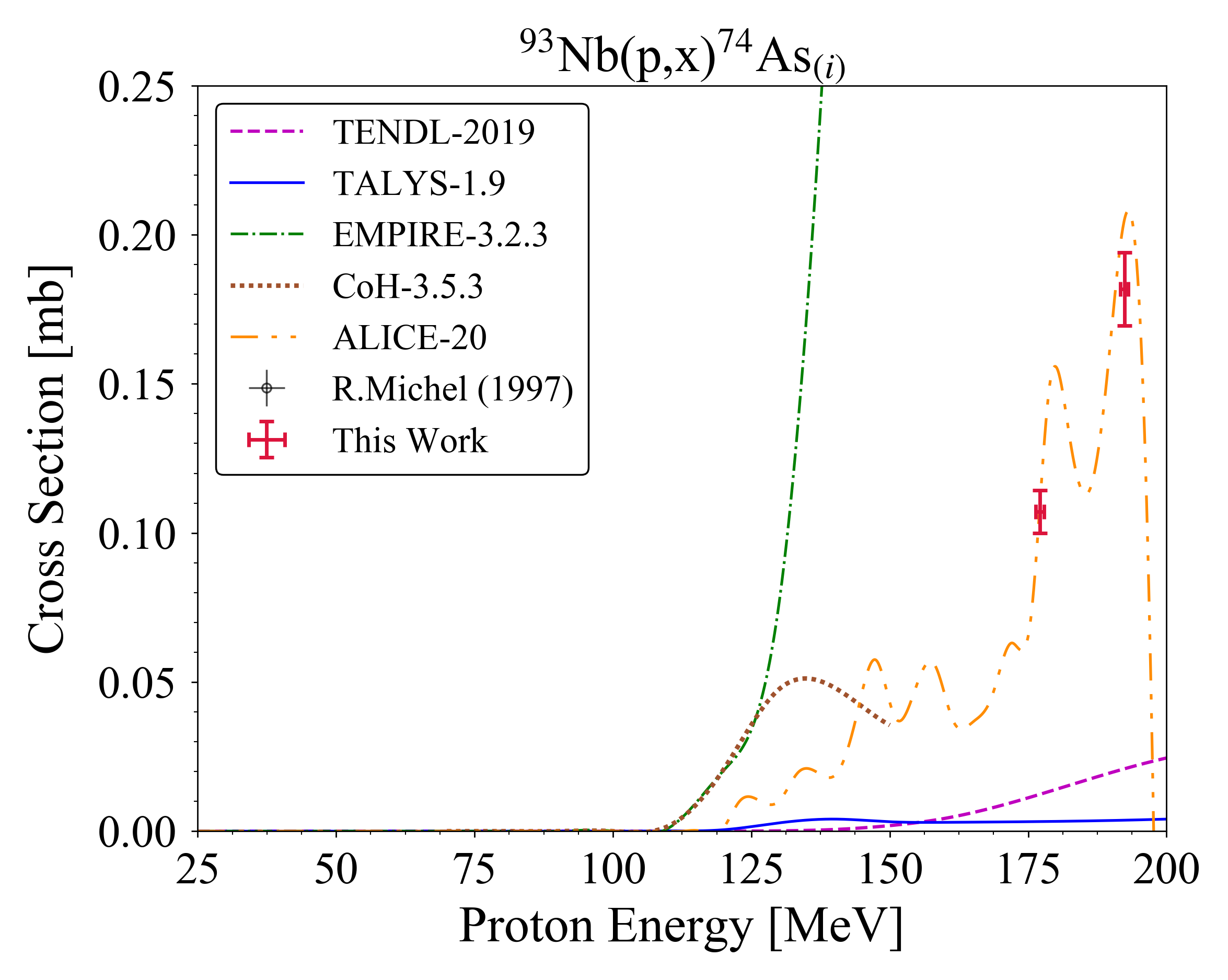}}
\vspace{-0.65cm}
\caption{Experimental and theoretical cross sections for $^{74}$As production.}\label{Nb_74As}
	{\includegraphics[width=1.0\columnwidth]{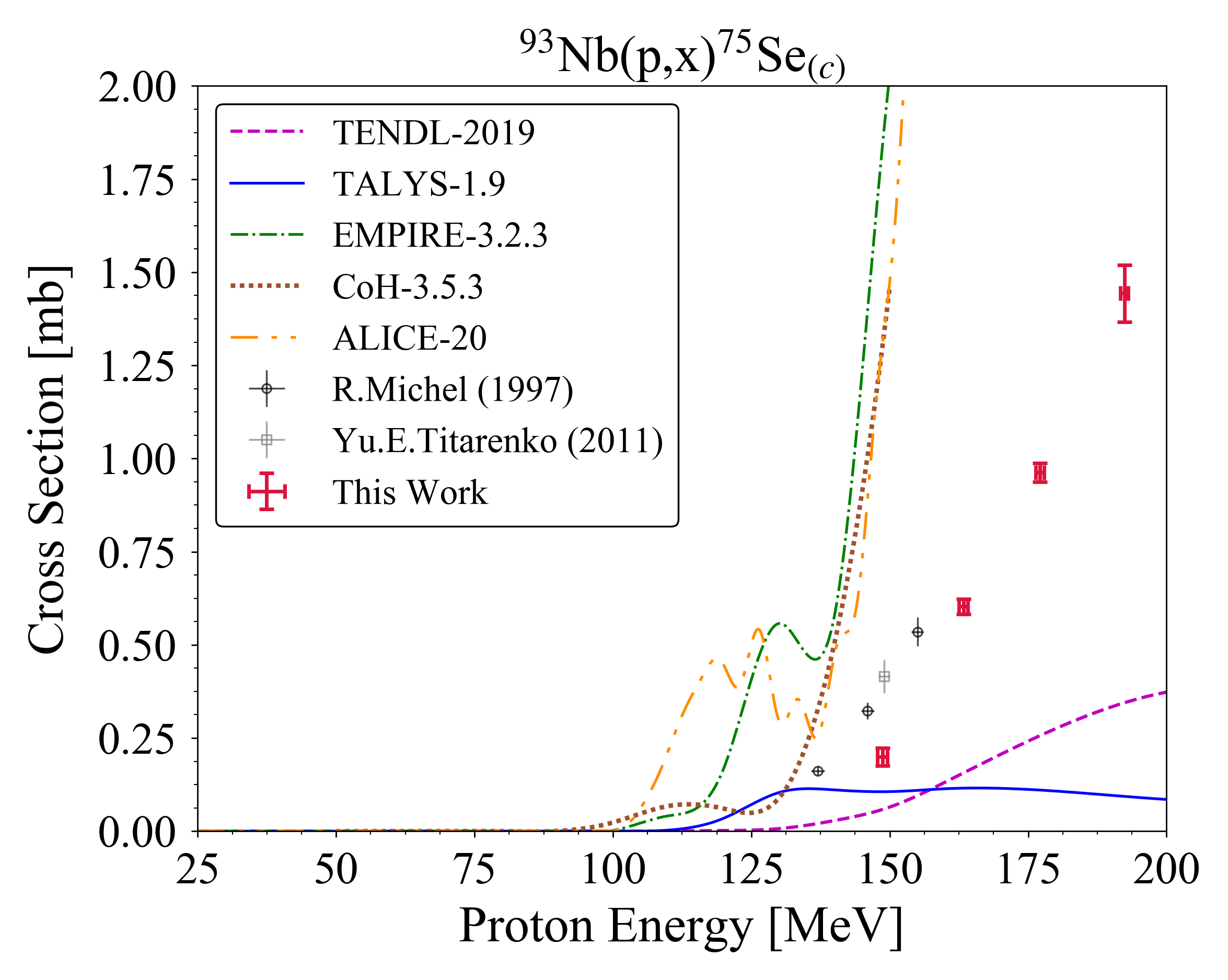}}
\vspace{-0.65cm}
\caption{Experimental and theoretical cross sections for $^{75}$Se production.}\label{Nb_75Se}
	{\includegraphics[width=1.0\columnwidth]{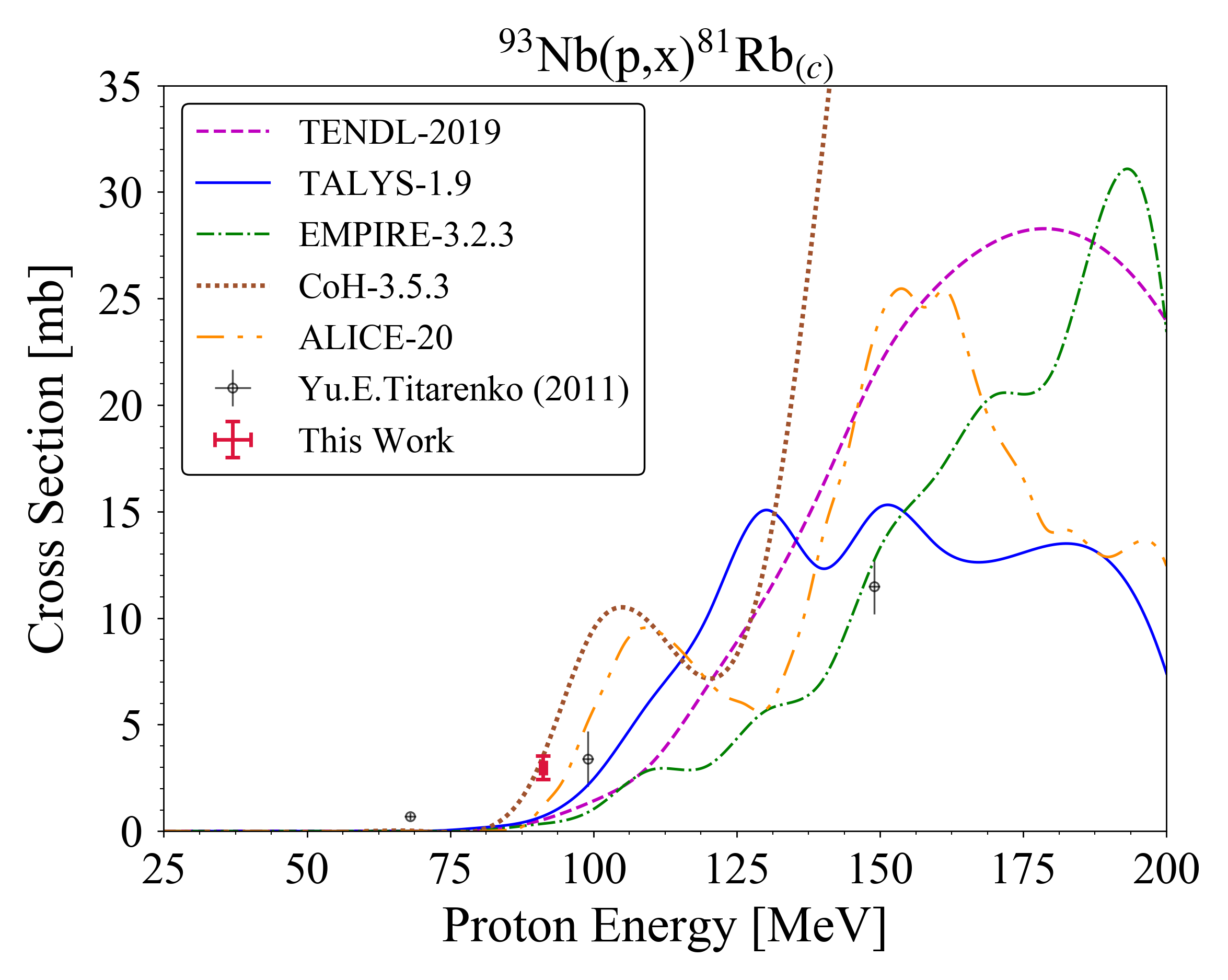}}
\vspace{-0.65cm}
\caption{Experimental and theoretical cross sections for $^{81}$Rb production.}\label{Nb_81Rb}
\end{figure}

\begin{figure}[H]
	{\includegraphics[width=1.0\columnwidth]{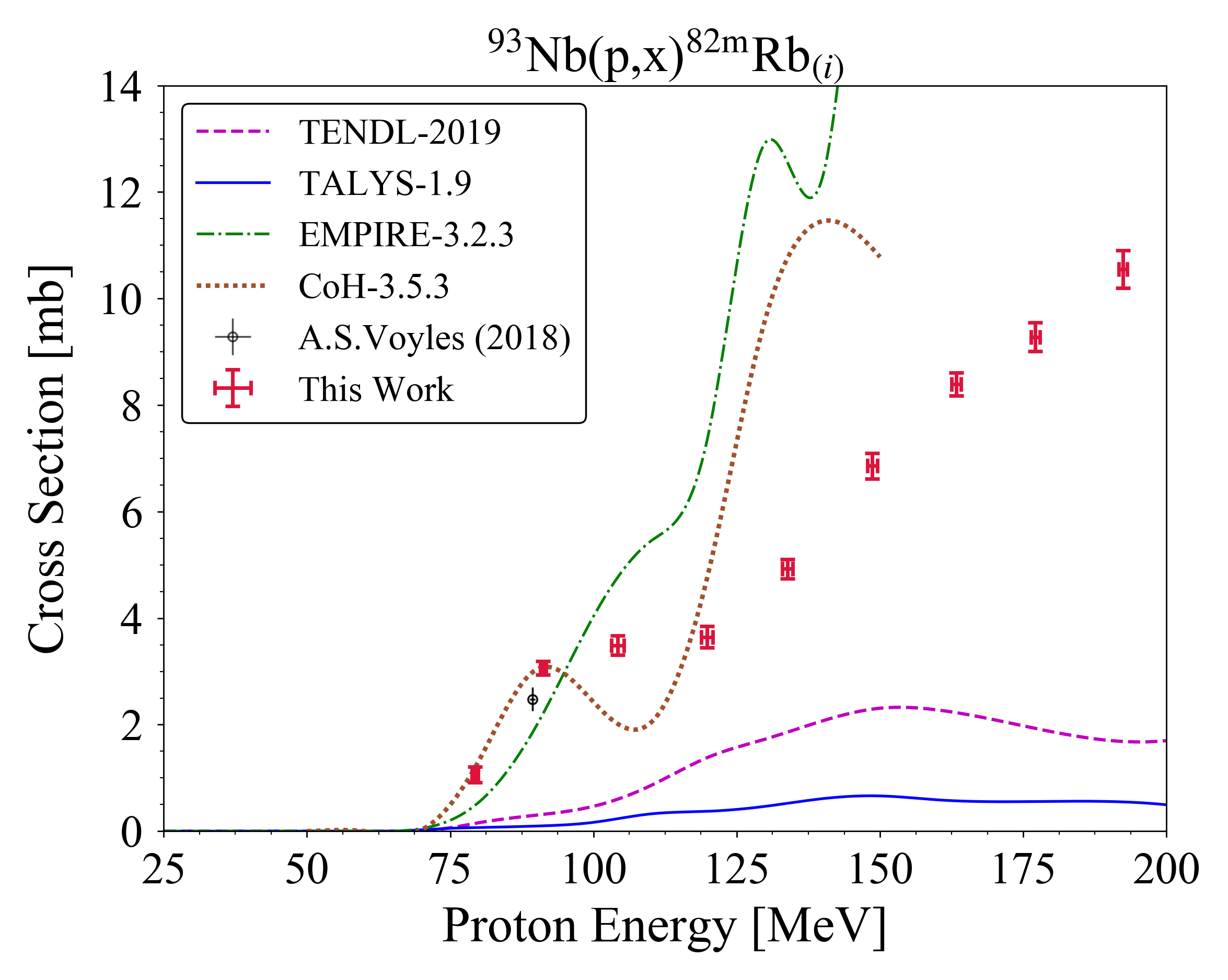}}
\vspace{-0.65cm}
\caption{Experimental and theoretical cross sections for $^{\textnormal{82m}}$Rb production.}\label{Nb_82mRb}
	{\includegraphics[width=1.0\columnwidth]{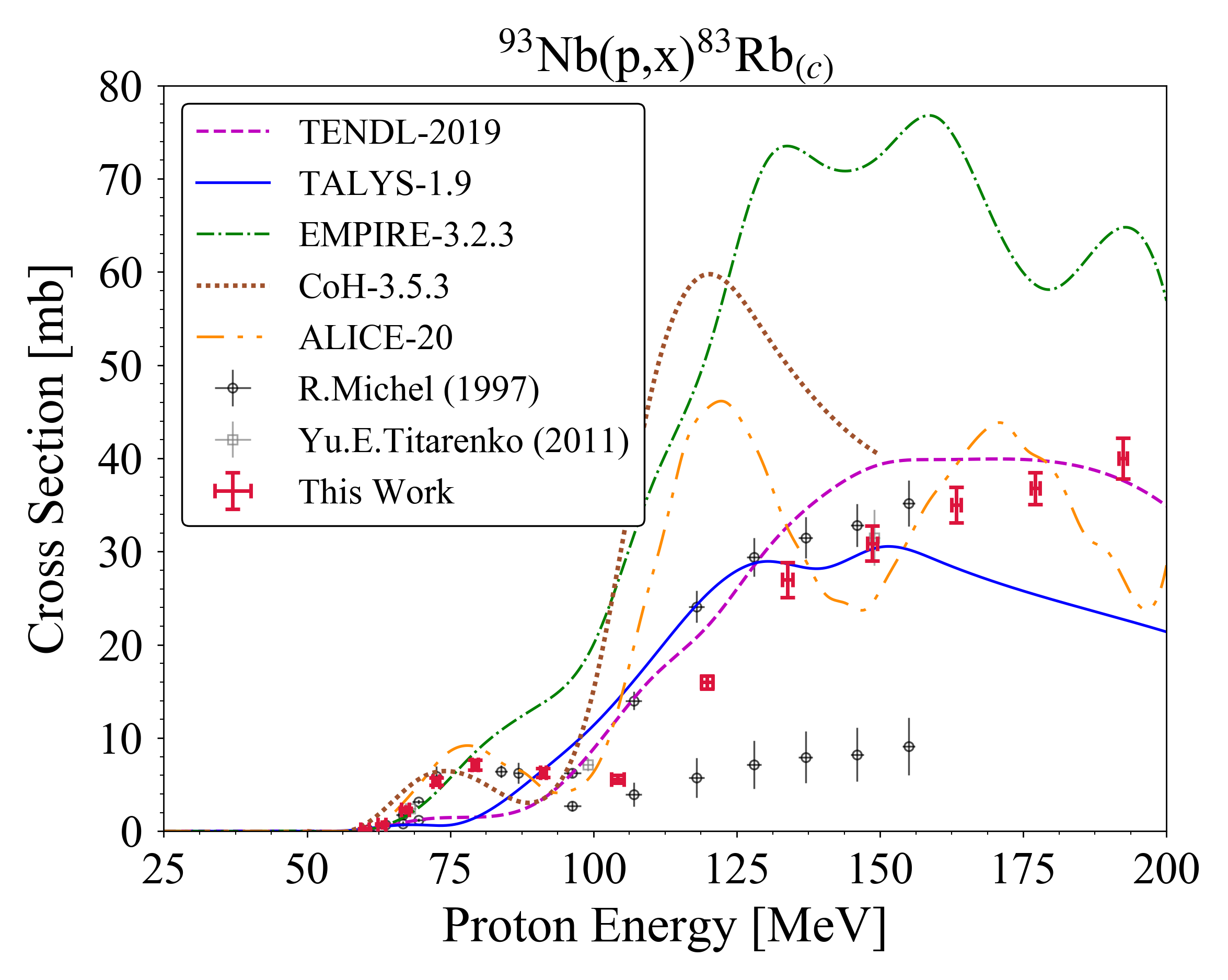}}
\vspace{-0.65cm}
\caption{Experimental and theoretical cross sections for $^{83}$Rb production.}\label{Nb_83Rb}
	{\includegraphics[width=1.0\columnwidth]{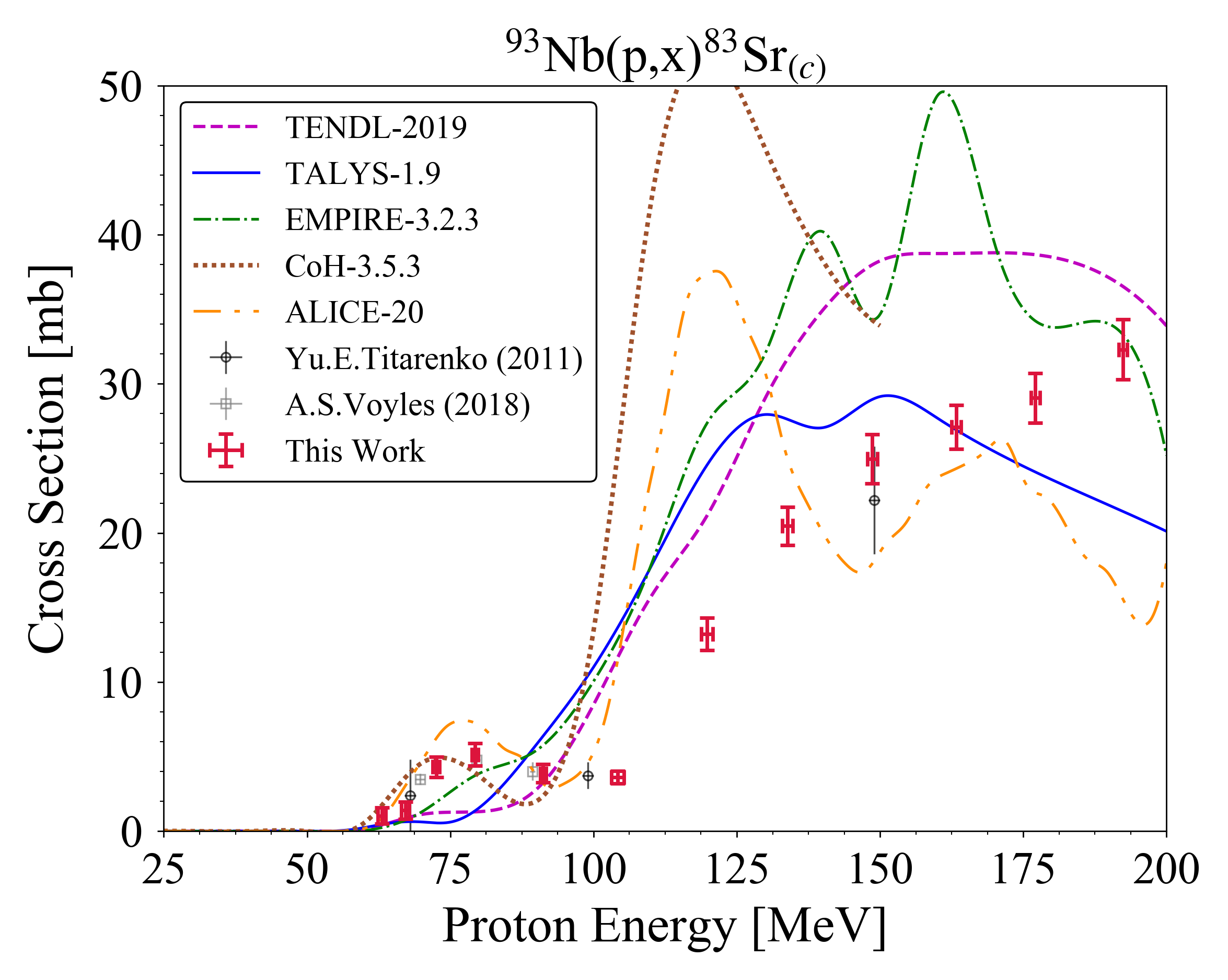}}
\vspace{-0.65cm}
\caption{Experimental and theoretical cross sections for $^{83}$Sr production.}\label{Nb_83Sr}
\end{figure}

\begin{figure}[H]
	{\includegraphics[width=1.0\columnwidth]{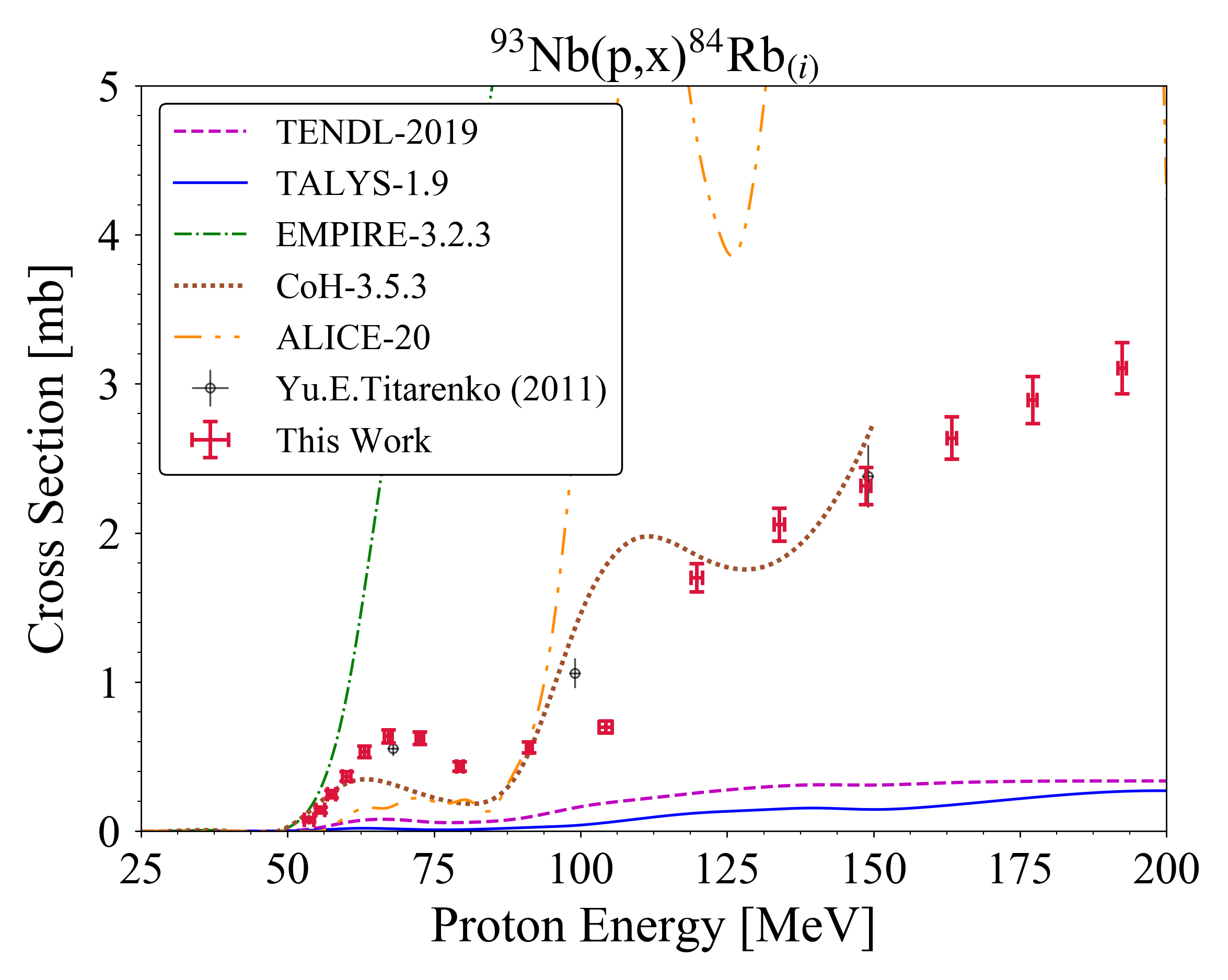}}
\vspace{-0.65cm}
\caption{Experimental and theoretical cross sections for $^{84}$Rb production.}\label{Nb_84Rb}
	{\includegraphics[width=1.0\columnwidth]{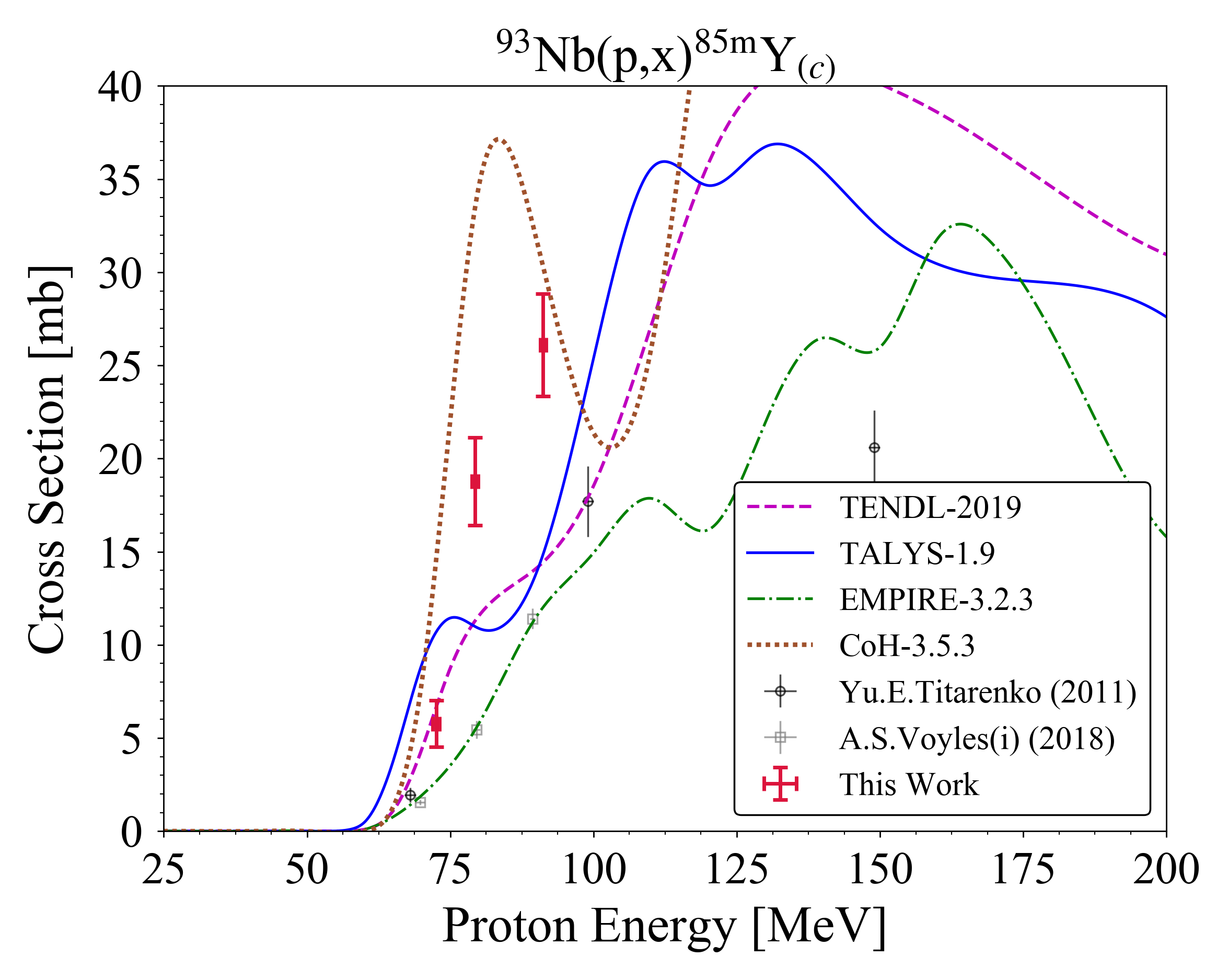}}
\vspace{-0.65cm}
\caption{Experimental and theoretical cross sections for $^{\textnormal{85m}}$Y production.}\label{Nb_85mY}
	{\includegraphics[width=1.0\columnwidth]{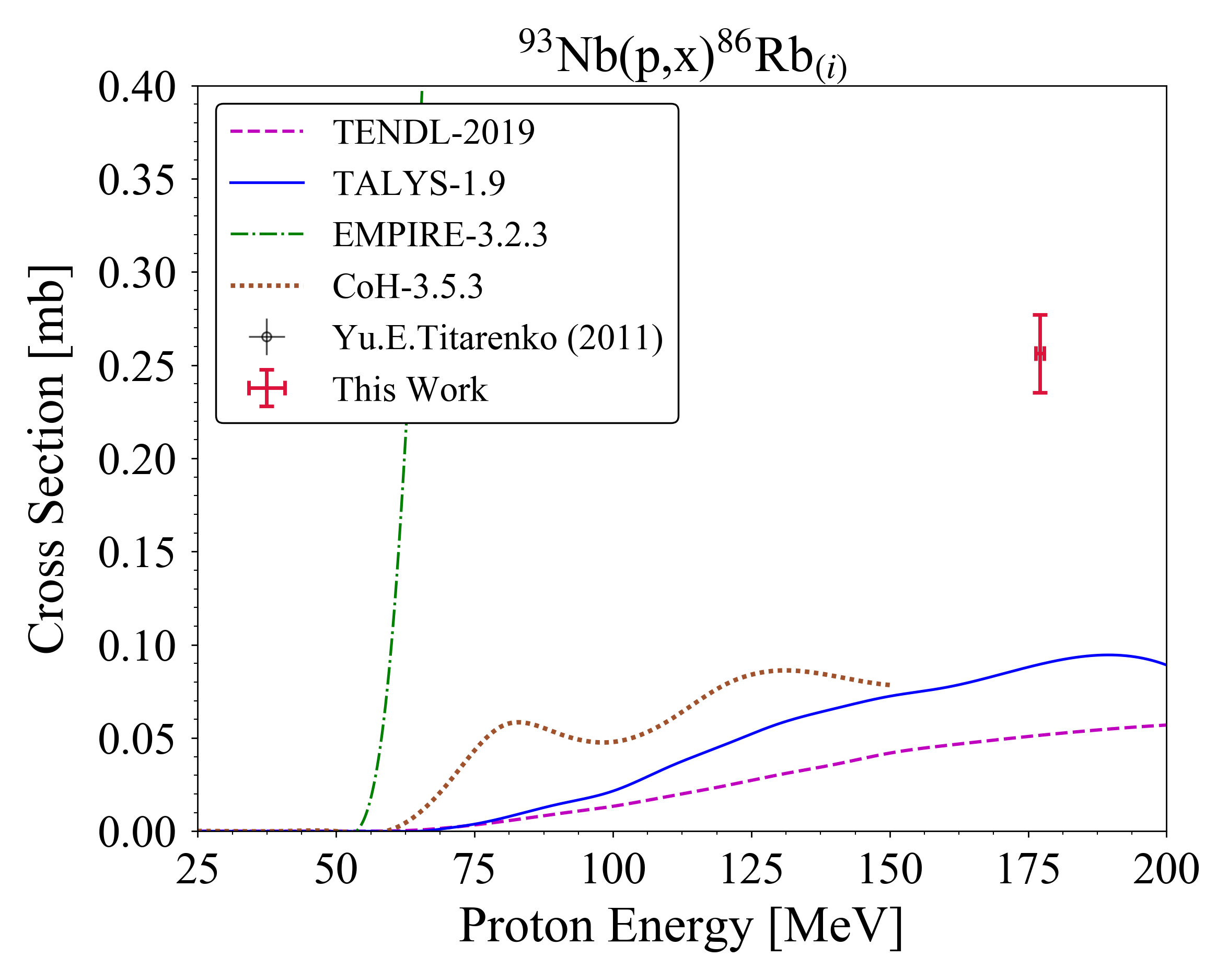}}
\vspace{-0.65cm}
\caption{Experimental and theoretical cross sections for $^{86}$Rb production.}\label{Nb_86Rb}
\end{figure}

\begin{figure}[H]
	{\includegraphics[width=1.0\columnwidth]{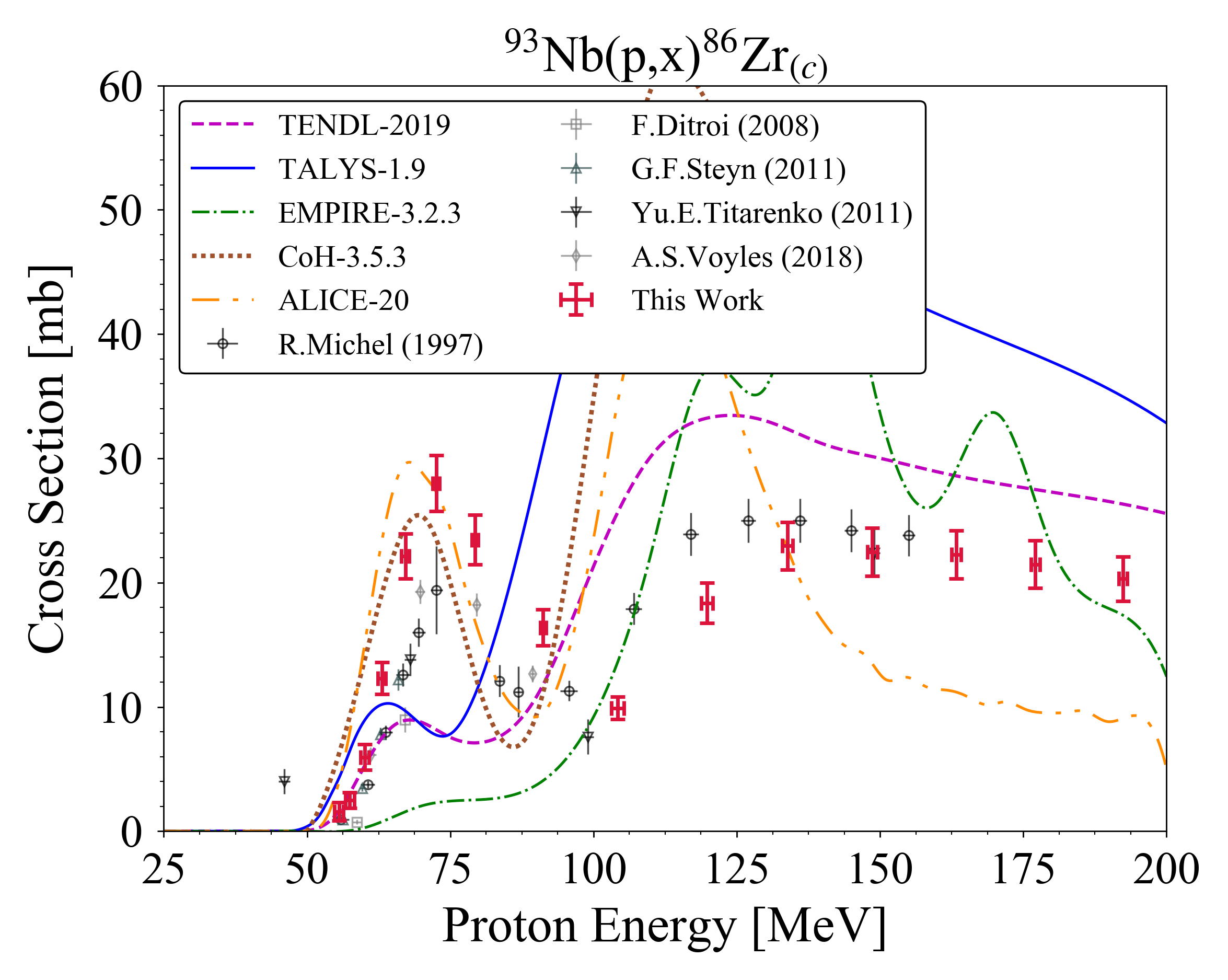}}
\vspace{-0.65cm}
\caption{Experimental and theoretical cross sections for $^{86}$Zr production.}\label{Nb_86Zr}
	{\includegraphics[width=1.0\columnwidth]{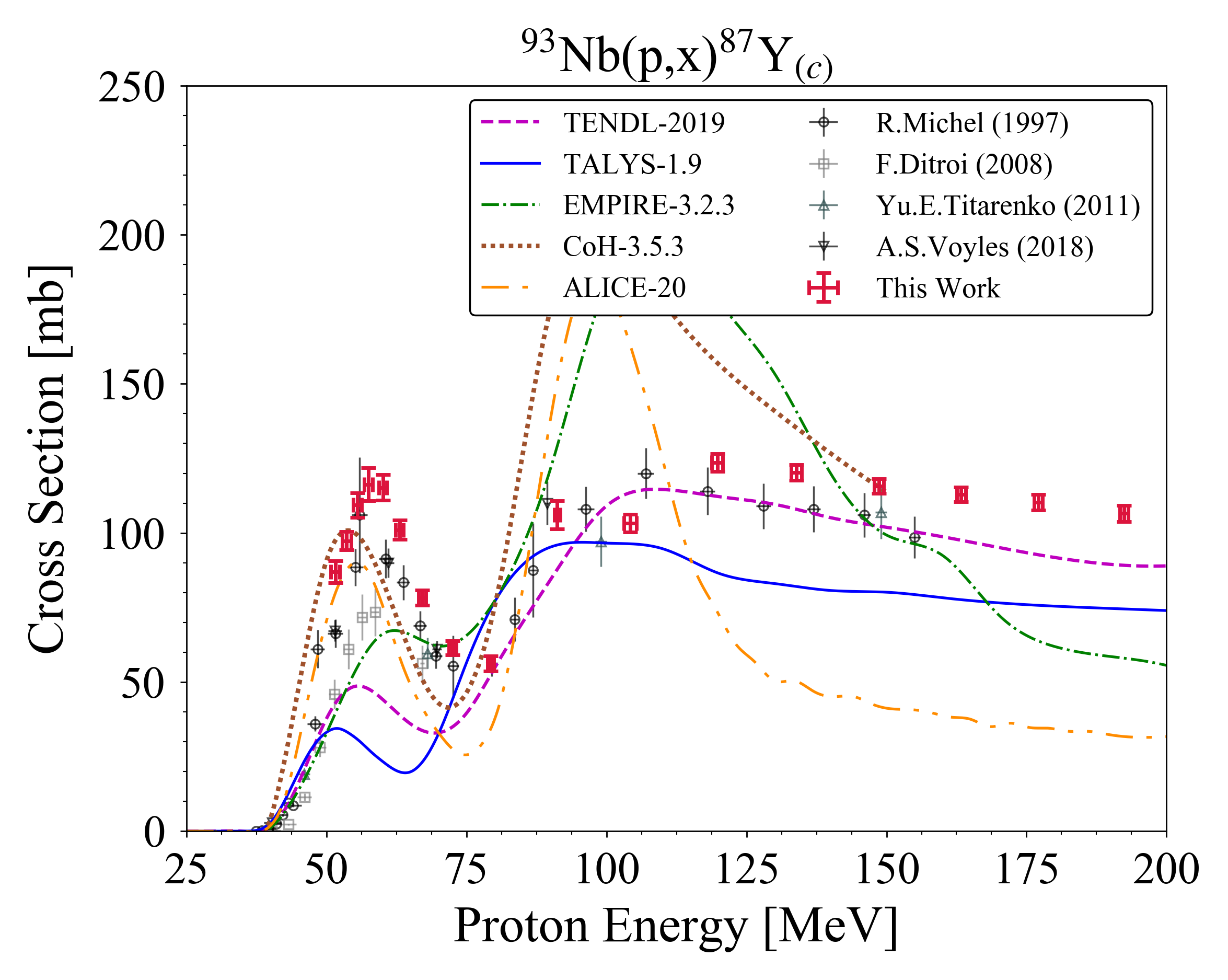}}
\vspace{-0.65cm}
\caption{Experimental and theoretical cross sections for $^{87}$Y production.}\label{Nb_87Y}
	{\includegraphics[width=1.0\columnwidth]{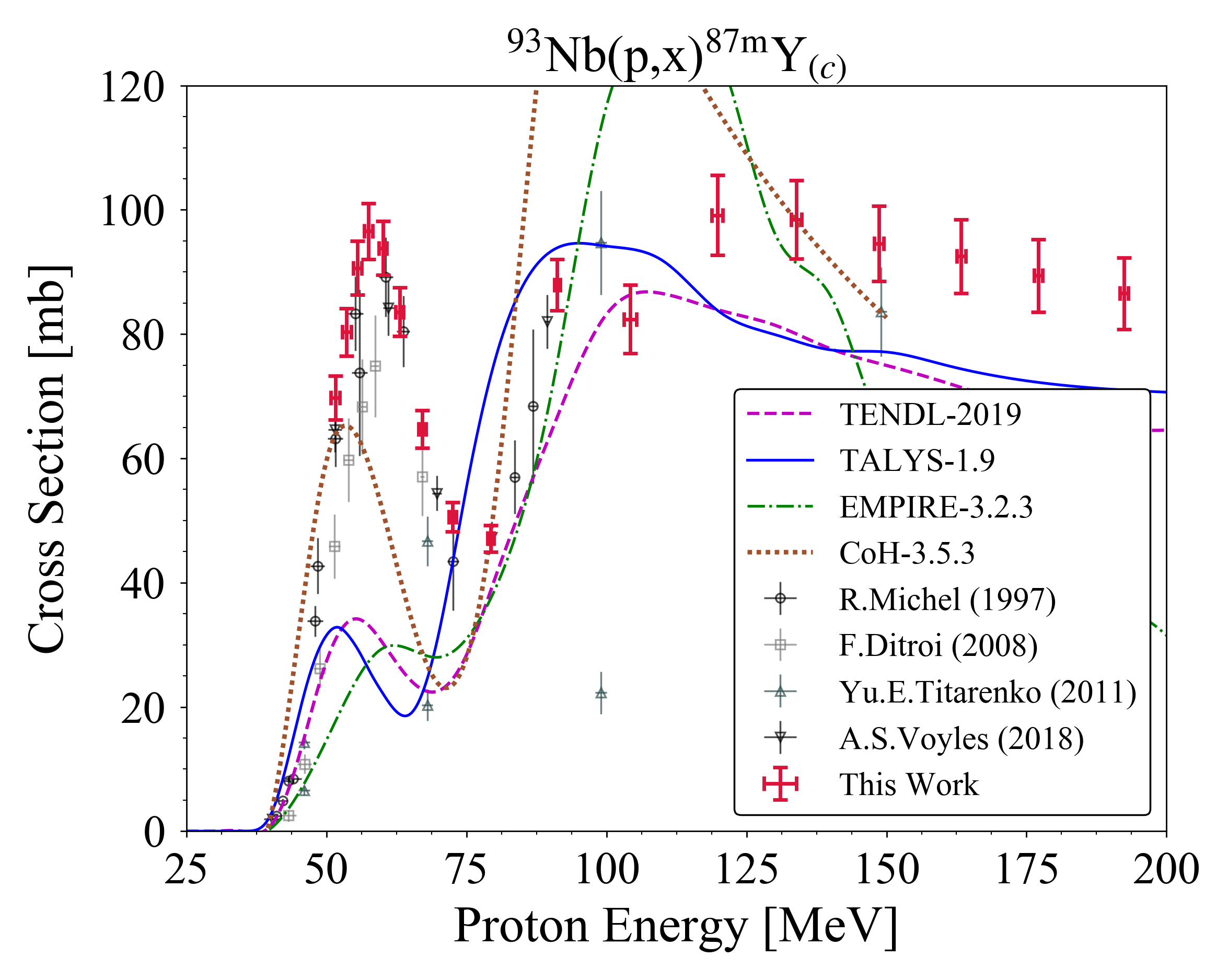}}
\vspace{-0.65cm}
\caption{Experimental and theoretical cross sections for $^{\textnormal{87m}}$Y production.}\label{Nb_87mY}
\end{figure}

\begin{figure}[H]
	{\includegraphics[width=1.0\columnwidth]{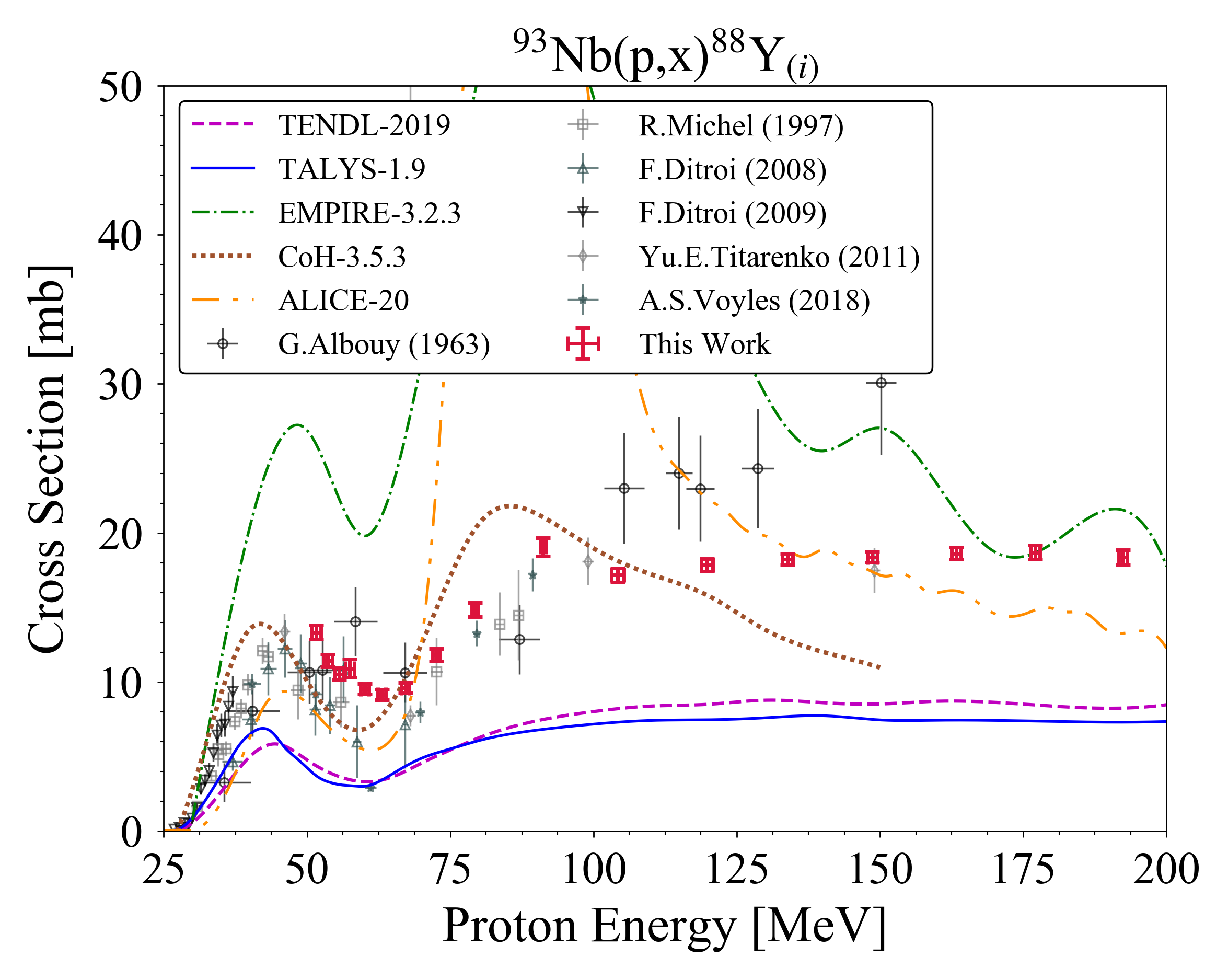}}
\vspace{-0.65cm}
\caption{Experimental and theoretical cross sections for $^{88}$Y production.}\label{Nb_88Y}
	{\includegraphics[width=1.0\columnwidth]{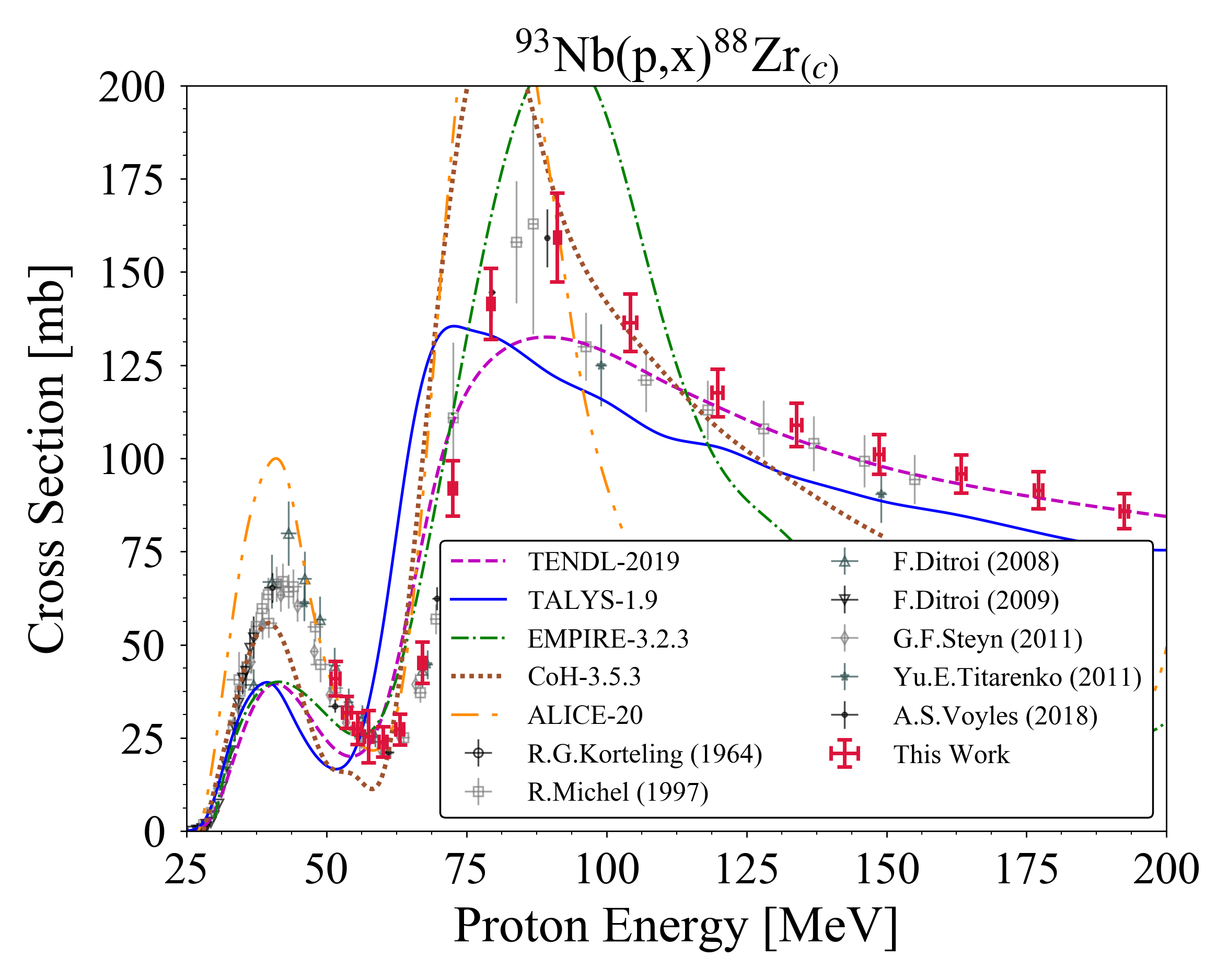}}
\vspace{-0.65cm}
\caption{Experimental and theoretical cross sections for $^{88}$Zr production.}\label{Nb_88Zr}
	{\includegraphics[width=1.0\columnwidth]{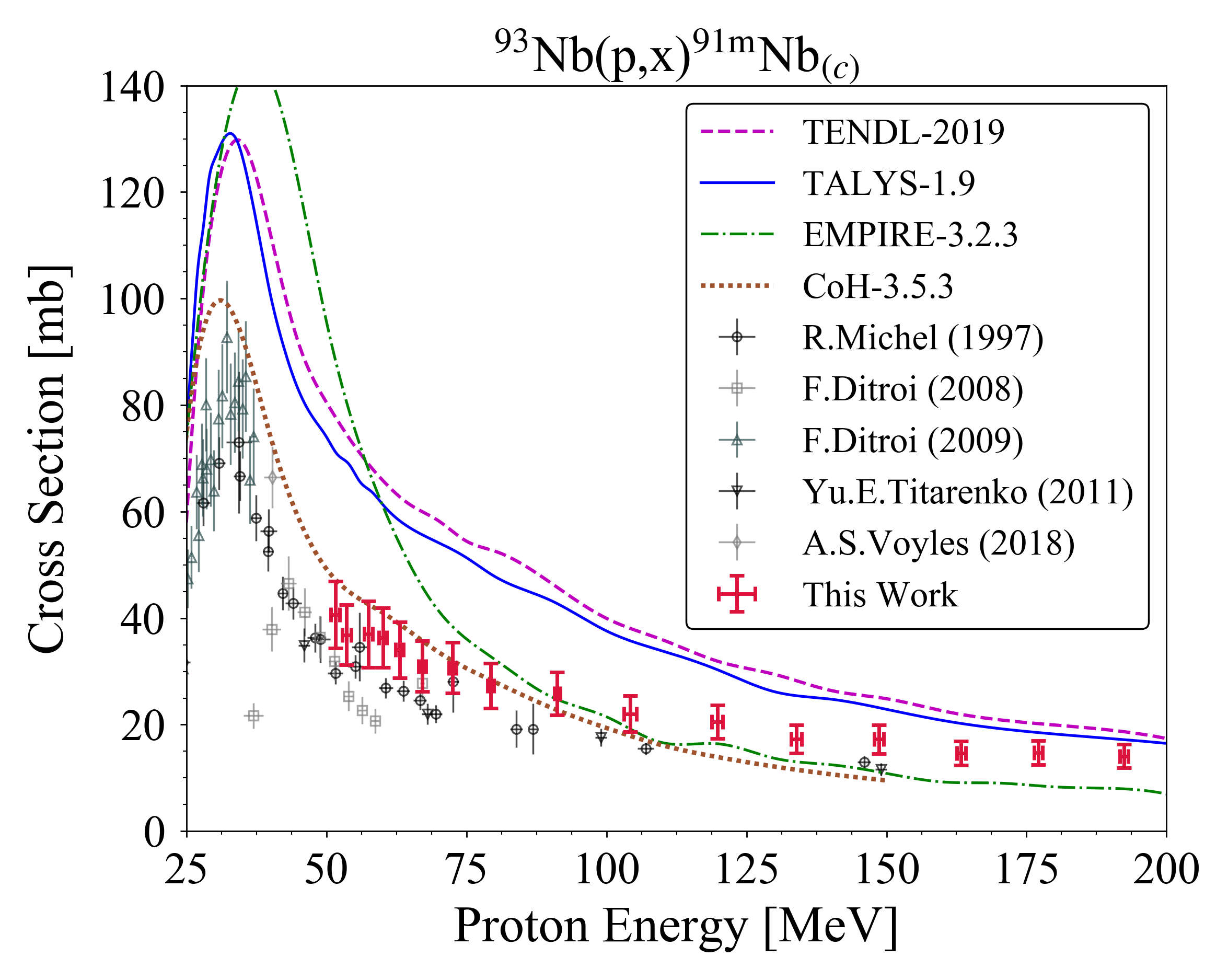}}
\vspace{-0.65cm}
\caption{Experimental and theoretical cross sections for $^{\textnormal{91m}}$Nb production.}\label{Nb_91mNb}
\end{figure}

\begin{figure}[H]
	{\includegraphics[width=1.0\columnwidth]{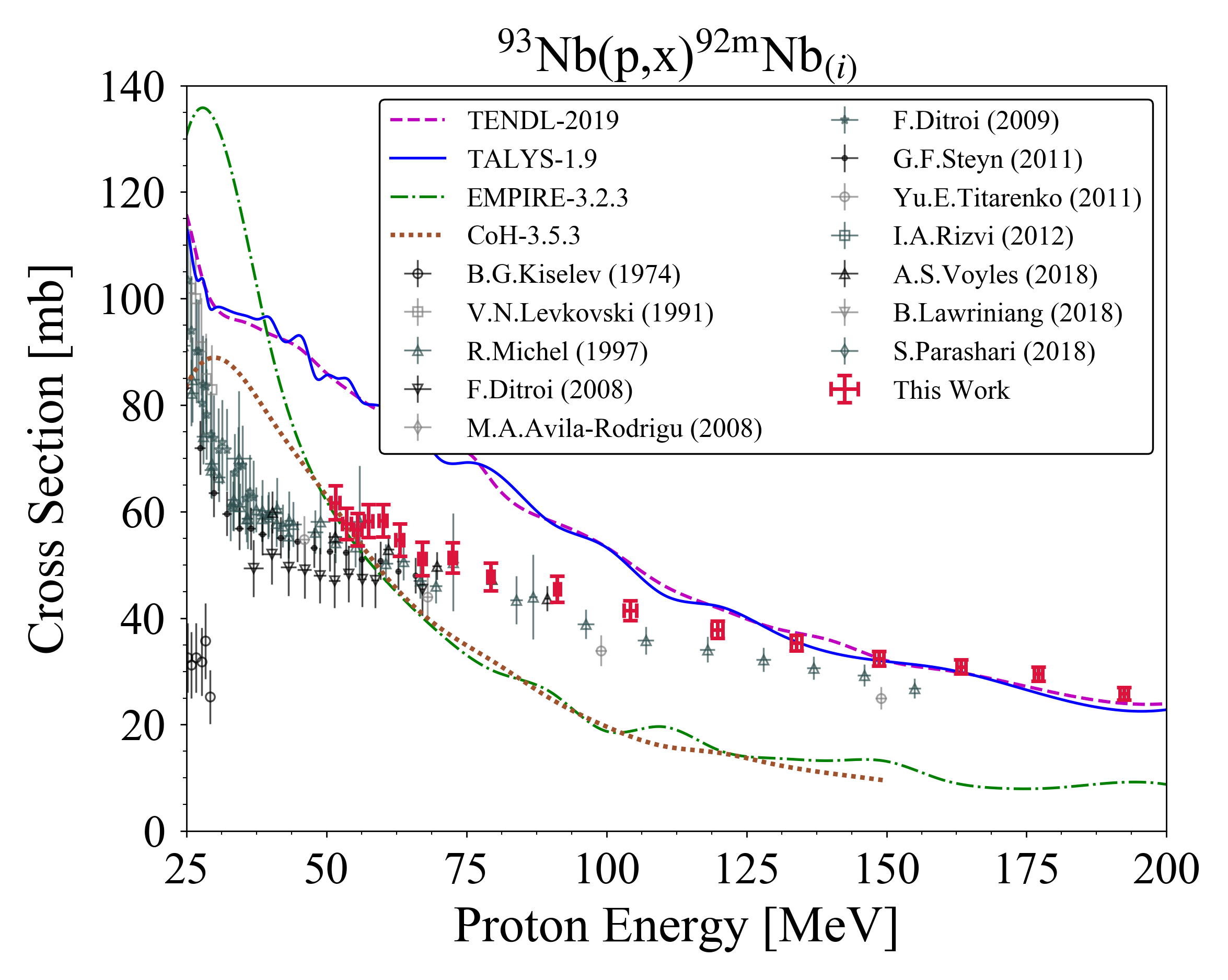}}
\vspace{-0.65cm}
\caption{Experimental and theoretical cross sections for $^{\textnormal{92m}}$Nb production.}\label{Nb_92mNb}
	{\includegraphics[width=1.0\columnwidth]{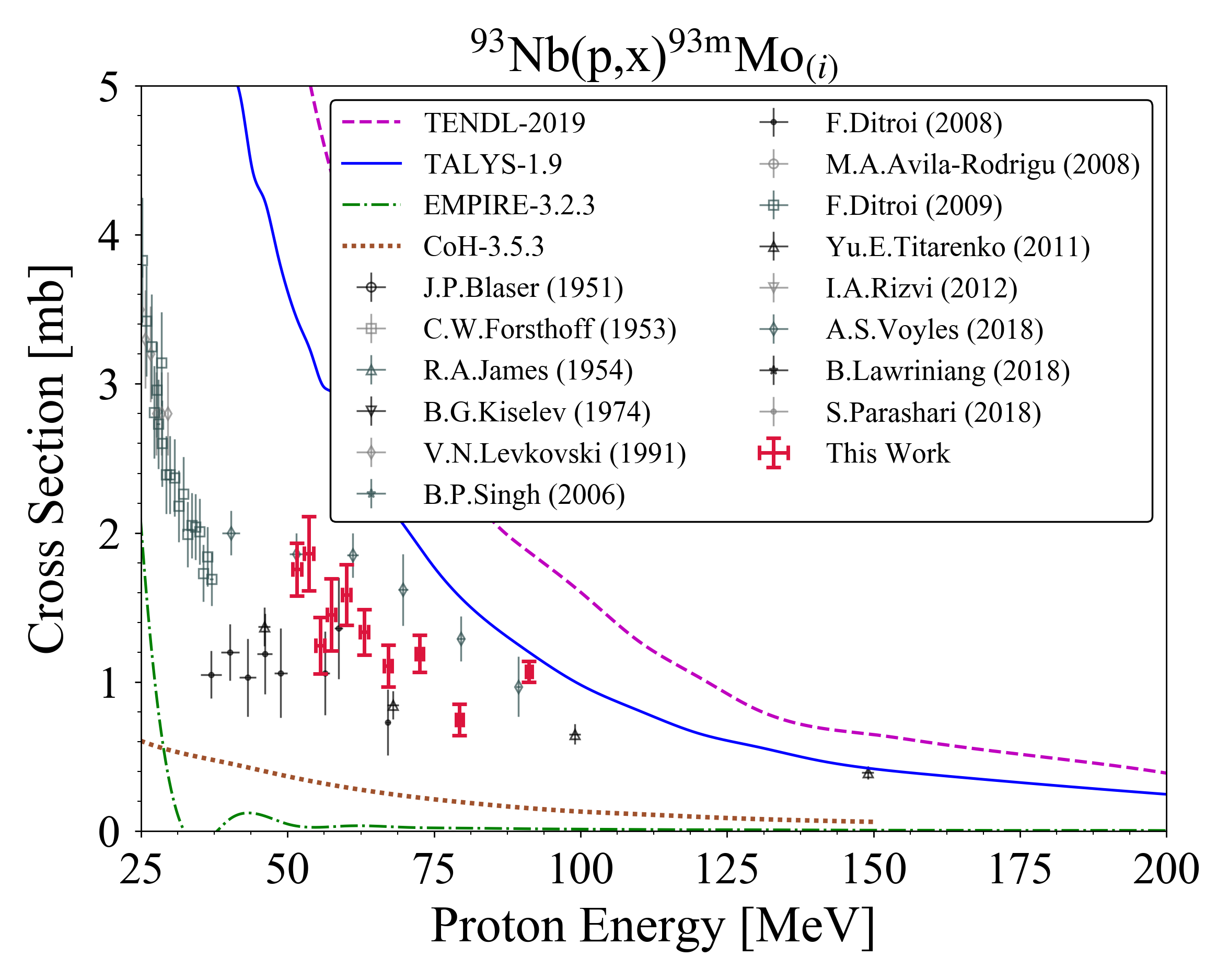}}
\vspace{-0.65cm}
\caption{Experimental and theoretical cross sections for $^{\textnormal{93m}}$Mo production.}\label{Nb_93mMo}
\end{figure}
\newpage

\section{\label{10Models}Non-Unique TALYS Parameter Adjustments}
Table \ref{NbParameterDetails} outlines the ambiguity surrounding TALYS parameter adjustments when modeling is based on a single excitation function.

\vspace{-0.1cm}
\begin{longtable}{P{2.85cm}P{4.5cm}P{1.2cm}}
\caption{Details of modeling cases used to reproduce similar behaviour for $^{93}$Nb(p,p3n)$^{90}$Nb reaction, shown in Figures \ref{NbTALYSModels} and \ref{NbParametersNewChannel}.}\label{NbParameterDetails}\\

    \hline\hline\\[-0.22cm]
    \multirow{1}{*}{Model Number}&\multirow{1}{*}{Parameter Adjustments}&\multirow{1}{*}{$\chi^2_\nu$}\\[0.1cm]
    \hline\\[-0.22cm]
    \endfirsthead
    
    \multicolumn{1}{l}{{\tablename\ \thetable{} -- cont.}} \\
    \hline\hline\\[-0.22cm]
    \multirow{1}{*}{Model Number}&\multirow{1}{*}{Parameter Adjustments}&\multirow{1}{*}{$\chi^2_\nu$}\\[0.1cm]
    \hline\\[-0.22cm]
    \endhead
    
    \hline Continued on next page&&
    \endfoot
    
    \hline\hline
    \endlastfoot
    
    \\[-0.22cm]
    \multirow{1}{*}{Default}&\multirow{1}{*}{---}&\multirow{1}{*}{57.9}\\[0.2cm]
    \multirow{3}{*}{1}&ldmodel 5 & \multirow{3}{*}{24.0}\\& strength 4\\& preeqmode 3\\[0.2cm]
    \multirow{5}{*}{2}&ldmodel 2 & \multirow{5}{*}{50.3}\\& strength 1\\& M2constant 1.8\\& avadjust p 0.85\\& rvadjust p 1.35\\[0.2cm]
    \multirow{5}{*}{3}&ldmodel 1 & \multirow{5}{*}{118.9}\\& strength 2\\& M2constant 3.0\\& M2shift 2.2\\& M2limit 2.0\\[0.2cm]
    \multirow{9}{*}{4}&ldmodel 3 & \multirow{9}{*}{298.4}\\& strength 2\\& M2constant 7.0\\& M2shift 0.1\\& M2limit 5.0\\& preeqmode 1\\& w1adjust p 1.5\\& v1adjust p 1.1\\& rvadjust p 1.33\\[0.2cm]
    \multirow{7}{*}{5}&ldmodel 6 & \multirow{7}{*}{34.5}\\& strength 8\\& M2constant 0.95\\& M2shift 0.95\\& M2limit 3.0\\& w1adjust p 1.4\\& ctable 41 90 0.15\\[0.2cm]
    \multirow{8}{*}{6}&ldmodel 4 & \multirow{8}{*}{57.8}\\& strength 5\\& M2constant 2.3\\& M2shift 0.6\\& M2limit 0.8\\& w1adjust p 1.3\\& rvadjust n 1.3\\& rvadjust a 0.85\\[0.2cm]
    \multirow{6}{*}{7}&ldmodel 1 & \multirow{6}{*}{46.9}\\& strength 2\\& M2constant 1.7\\& w1adjust p 1.2\\& v1adjust p 1.05\\& rvadjust p 1.25\\[0.2cm]
    \noalign{\penalty-2000}
    \multirow{3}{*}{8}&jlmomp y & \multirow{3}{*}{67.3}\\& preeqmode 3\\& lwadjust 1.08\\[0.2cm]
    \multirow{7}{*}{9}&ldmodel 1 & \multirow{7}{*}{45.1}\\& strength 2\\& M2constant 0.85\\&  localomp n\\& rvadjust n 0.85\\& v1adjust n 1.25\\& ctable 42 90 -1.0\\[0.2cm]
    \multirow{8}{*}{10}&ldmodel 5 & \multirow{8}{*}{23.5}\\& strength 4\\& M2constant 3.3\\&ctable 42 88 -1.2\\& ctable 42 87 -1.2\\& ctable 41 90 1.6\\& ctable 41 86 -1.0\\& ctable 40 86 -1.8\\[0.2cm]
\end{longtable}
\newpage

\section{\label{ParamsAdjustments}TALYS Parameter Adjustments From Fitting Procedure}
The derived parameter adjustments from the fitting procedure applied to the $^{93}$Nb(p,x) and $^{139}$La(p,x) data are listed in Tables \ref{NbAdjustedParams} and \ref{LaAdjustedParams}.

\vspace{-0.15cm}
\begin{table}[h]
\vspace{-0.3cm}
\caption{$^{93}$Nb(p,x) best fit parameter adjustments derived from proposed procedure. The \texttt{equidistant} keyword adjusts the width of excitation energy binning and will be a default in updated TALYS versions. The \texttt{strength} keyword selects the gamma-ray strength model and has little impact in this charged-particle investigation, so it is chosen as one of the available microscopic options.}
\label{NbAdjustedParams}
\begin{ruledtabular}
\begin{tabular}{ll}
\multirow{1}{*}{Parameter}&\multirow{1}{*}Value\\[0.1cm]
\hline\\[-0.22cm]
\multirow{3}{*}{ldmodel}& 4\\& 5 $^{94-86}$Nb\\& 5 $^{94}$Mo, $^{91-86}$Mo\\[0.1cm]
strength    &5\\[0.1cm]
equidistant    &y\\[0.1cm]
M2constant    &0.875\\[0.1cm]
M2limit    &4.5\\[0.1cm]
M2shift   &0.6\\[0.1cm]
w1adjust\ p    &2.2\\[0.1cm]
\multirow{4}{*}{ctable}& 39 86 -0.6\\& 40 86 -0.35\\& 40 87 -0.85\\& 42 90 -0.5\\[0.1cm]
ptable   &39 86 2.0\\[0.1cm]
\end{tabular}
\end{ruledtabular}
\end{table}

\begin{table}[h]
\vspace{-0.3cm}
\caption{$^{139}$La(p,x) best fit parameter adjustments derived from proposed procedure.}
\label{LaAdjustedParams}
\begin{ruledtabular}
\begin{tabular}{ll}
\multirow{1}{*}{Parameter}&\multirow{1}{*}Value\\[0.1cm]
\hline\\[-0.22cm]
ldmodel    &4\\[0.1cm]
strength    &5\\[0.1cm]
equidistant    &y\\[0.1cm]
M2constant    &0.85\\[0.1cm]
M2limit    &2.5\\[0.1cm]
M2shift   &0.9\\[0.1cm]
cstrip\ a    &2.0\\[0.1cm]
rvadjust\ p    &0.96\\[0.1cm]
ctable   &58 135 0.6\\[0.1cm]
\end{tabular}
\end{ruledtabular}
\end{table}
\clearpage

%

\end{document}